\documentclass[letter,apj]{emulateapj-rtx4}
\pdfoutput=1
\usepackage{graphicx}
\usepackage{enumerate}
\usepackage{amssymb, amsmath, amsfonts}
\usepackage{gensymb}
\usepackage{mathtools}
\usepackage{natbib}
\usepackage{color}
\usepackage{hyperref}
\usepackage{ulem}
\usepackage{soul}
\newcommand{\Lsun}{\mbox{$L_{\sun}$}}
\newcommand{\Lbol}{\mbox{$L_{\rm bol}$}}
\newcommand{\Msun}{\mbox{$M_{\sun}$}}

\newcommand{\Mjup}{\mbox{$M_{\rm Jup}$}}

\begin{document}

\altaffiltext{1}{Department of Physics, University of California, Santa Barbara, Santa Barbara, CA 93106, USA}
\altaffiltext{2}{Gemini Observatory, Northern Operations Center, 670 N.~Aohoku Place, Hilo, HI 96720, USA}
\altaffiltext{3}{Department of Astronomy, The University of Texas at Austin, Austin, TX 78712, USA}

\title{Precise Dynamical Masses of Directly Imaged Companions from Relative Astrometry, Radial Velocities, and Hipparcos--Gaia DR2 Accelerations}
\author{
Timothy D.~Brandt\altaffilmark{1}, 
Trent J.~Dupuy\altaffilmark{2}, and
Brendan P.~Bowler\altaffilmark{3}
}


\begin{abstract}
We measure dynamical masses for five objects---three ultracool dwarfs, one low-mass star, and one white dwarf---by fitting orbits to a combination of the {\it Hipparcos}-{\it Gaia} Catalog of Accelerations, literature radial velocities, and relative astrometry.  Our approach provides precise masses without any assumptions about the primary star, even though the observations typically cover only a small fraction of an orbit.  We also perform a uniform re-analysis of the host stars' ages.  Two of our objects, HD~4747B and HR~7672B, already have precise dynamical masses near the stellar/substellar boundary and are used to validate our approach.  For Gl~758B, we obtain a mass of $m=38.1_{-1.5}^{+1.7}$~$M_{\rm Jup}$, the most precise mass measurement of this companion to date.  Gl~758B is the coldest brown dwarf with a dynamical mass, and the combination of our low mass and slightly older host-star age resolves its previously noted discrepancy with substellar evolutionary models.  HD~68017B, a late-M dwarf, has a mass of $m = 0.147\pm 0.003$~$M_\odot$, consistent with stellar theory and previous empirical estimates based on its absolute magnitude.  The progenitor of the white dwarf Gl~86B has been debated in the literature, and our dynamical measurement of $m = 0.595 \pm 0.010$~$M_\odot$ is consistent with a higher progenitor mass and younger age for this planet-hosting binary system.  Overall, these case studies represent only five of the thousands of accelerating systems identified by combining {\it Hipparcos} and {\it Gaia}.  Our analysis could be repeated for many of them to build a large sample of companions with dynamical masses.
\end{abstract}

\maketitle

\section{Introduction} \label{sec:intro}

Dynamical masses represent one of the strongest observational cornerstones of stellar and substellar evolutionary models.  They are particularly important for objects that reside off of the main sequence and rapidly evolve in the Hertzsprung-Russell diagram, such as young stars, brown dwarfs, giant planets, and white dwarfs \citep[e.g.,][]{Hillenbrand+White_2004, 2006Natur.440..311S, 2013ApJ...779...21H, Dupuy+Liu_2017}. 
Combining masses with luminosities and ages (or radii and effective temperatures) offers a powerful way to test and calibrate cooling models of a given composition, providing direct constraints on the internal structure, atmospheric opacity, and radiative evolution of these objects.  For young directly imaged substellar companions in particular, dynamical masses can help distinguish among formation mechanisms by breaking degeneracies in hot-, warm-, and cold-start evolutionary models \citep{Marley+Fortney+Hubickyj_2007, Fortney+Marley+Saumon+etal_2008, Spiegel+Burrows_2012, Marleau+Cumming_2014}.

A variety of techniques can be used to measure dynamical masses.  Most require long-term orbit monitoring of resolved components of a binary system to determine total masses;
absolute astrometry or relative radial velocities (RVs) can then constrain the individual component masses \citep[][]{Dupuy+Liu+Ireland_2009a, Dupuy+Liu+Ireland_2009b, 2010ApJ...711.1087K, Montet+Bowler+Shkolnik+etal_2015, Bond+Bergeron+Bedard_2017}.  
Systems in which the primary star has a measurable radial velocity acceleration are particularly useful.  Assuming a known parallax and well-measured astrometric orbit, the primary's line-of-sight acceleration probes the mass of its companion.
These accelerating systems act as ``dynamical beacons'' and have been used to find and characterize both stellar and substellar companions \citep[e.g.,][]{2002ApJ...581L.115B, Crepp+Rice+Veicht+etal_2015, Cheetham+Segransan+Peretti+etal_2018, Bowler+Dupuy+Endl+etal_2018}.  However, the precision of masses for these wide companions is limited by their orbital periods---often decades to centuries---and requires long-baseline monitoring with radial velocities and high-contrast imaging to gradually improve mass constraints.

{\it Gaia} \citep{Gaia_General_2016} opens up the possibility of measuring the dynamical masses of systems across the sky.  {\it Hipparcos} and {\it Gaia} measured the motion of stars in an inertial reference frame, the ICRS, defined by distant quasars \citep{Ma+Arias+Eubanks+etal_1998, Fey+Gordon+Jacobs+etal_2015}.  Differences in the proper motions between {\it Hipparcos} and {\it Gaia} imply accelerations in an inertial frame; these may be used to constrain dynamical masses.  \cite{Calissendorff+Janson_2018} added such a proper motion difference to a study of Gl~758B \citep{Bowler+Dupuy+Endl+etal_2018} to refine its dynamical mass, obtaining a final constraint of $m = 42.4^{+5.6}_{-5.0}$~$M_{\rm Jup}$. Previous papers have also explored the use of the {\it Hipparcos} epoch astrometry in combination with radial velocity or relative astrometry to improve dynamical mass measurements \citep{Han+Black+Gatewood_2001,Sozzetti+Desidera_2010,Sahlmann+Segransan+Queloz_2011,Snellen+Brown_2018}.  

Previous studies have used catalog astrometry at face value, but \cite{Brandt_2018} has performed a cross-calibration of {\it Hipparcos} and {\it Gaia} DR2, the {\it Hipparcos}--{\it Gaia} Catalog of Accelerations (HGCA), that accounts for systematics as a function of position on the sky. The HGCA adopts a 60/40 linear combination of the two {\it Hipparcos} reductions \citep{ESA_1997,vanLeeuwen_2007}, inflates uncertainties of both {\it Hipparcos} and {\it Gaia} DR2, and applies locally variable frame rotations of $\sim$0.5~mas\,yr$^{-1}$ to {\it Hipparcos} and $\sim$0.2~mas\,yr$^{-1}$ to {\it Gaia}.  All of the resulting astrometry is then placed in the DR2 reference frame. Figure 9 of \cite{Brandt_2018} demonstrates the Gaussianity of the residuals between the HGCA's three proper motion measurements. This Gaussianity makes the catalog well-suited for use in orbit fitting; we employ it here to validate its accuracy and improve on previous dynamical mass estimates.

Combining {\it Gaia} and {\it Hipparcos} proper motions provides an acceleration in the plane of the sky.  Adding a radial velocity trend gives a full three-dimensional acceleration.  Together with a projected separation from direct imaging, this is sufficient to determine a dynamical mass {\it without observing a substantial fraction of an orbit.}  This opens up long-period systems, where the components are well-separated, to dynamical mass measurements.  It also reduces the observational effort needed to obtain these masses to intermittent radial velocity monitoring and a few direct imaging snapshots.  While this technique can constrain masses to high precision, it does a poorer job of measuring the other orbital parameters.  {\it Gaia} and {\it Hipparcos} are well-suited to measuring only one parameter, but this parameter is the most physically significant one.  

We demonstrate the use of absolute stellar astrometry to constrain masses using a sample of five binaries: Gl~758 \citep{Thalmann+Carson+Janson+etal_2009,Janson+Carson+Thalmann+etal_2011}, Gl~86 \citep{Els+Sterzik+Marchis+etal_2001,Lagrange+Beust+Udry+etal_2006}, HR~7672 \citep{Liu+Fischer+Graham+etal_2002}, HD~4747 \citep{Crepp+Gonzales+Bechter+etal_2016,Crepp+Principe+Wolff+etal_2018}, and HD~68017 \citep{Crepp+Johnson+Howard+etal_2012}.  This sample includes one $\sim$40~$M_{\rm Jup}$ brown dwarf (Gl~758B), two higher-mass brown dwarfs near the stellar/substellar boundary (HR~7672B and HD~4747B), one low-mass M-dwarf (HD~68017B), and one white dwarf (Gl~86B).  All of the companions are optically faint, which minimizes their effects on the {\it Hipparcos} and {\it Gaia} astrometry.  We have not selected this sample systematically, but rather choose five stars with a range of companion properties, well-measured radial velocities, and high signal-to-noise ratio accelerations between the {\it Hipparcos} and {\it Gaia} epochs.

The paper is organized as follows.  
Section \ref{sec:ages_masses} contains a uniform re-analysis of the host stars' ages and basic physical parameters using both activity-age relations and isochrone fitting.  Section \ref{sec:rv_imaging} describes the archival direct imaging and radial velocity measurements that we make use of.  Section \ref{sec:gaia_astrometry} summarizes the stellar astrometry, derived from {\it Hipparcos} and {\it Gaia} DR2 and cross-calibrated by \cite{Brandt_2018}.  Section \ref{sec:singleepochmasses} shows how our approach can provide companion masses to high precision even for short orbital arcs, while Section \ref{sec:orbitfitting} describes how we actually fit orbits and derive masses.  Section \ref{sec:results} discusses our results, and we conclude in Section \ref{sec:conclusions}.

\section{Stellar Ages and Masses} \label{sec:ages_masses}

\begin{deluxetable*}{lcccccccccccr}
\tablewidth{0pt}
\tablecaption{Adopted Stellar Parameters\tablenotemark{*}}
\tablehead{
    \colhead{Star} &
    \colhead{HIP ID} &
    \colhead{$\varpi$ (mas)} &
    \colhead{$\sigma[\varpi]$} &
    \colhead{$V_T$ (mag)} &
    \colhead{$\sigma[V_T]$\tablenotemark{$\dagger$}} & 
    \colhead{$K_s$ (mag)} & 
    \colhead{$\sigma[K_s]$} &
    \colhead{$R_X$} &
    \colhead{$R^\prime_{\rm HK}$} &
    \colhead{$P_{\rm rot}$ (d)} &
    \colhead{$T_{\rm eff}$ (K)} & 
    \colhead{[Fe/H]}
}   
\startdata
   HD 4747 &   3850 &  53.18 & 0.13 &  7.226 & 0.02 & 5.305 & 0.029 &  $-5.39$ & $-4.79$ &  27.7 & $5390$ & $-0.21$ \\
    Gl 86 &  10138 &  92.70 & 0.05 &  6.209 & 0.02 & 4.125 & 0.036 &  $-5.42$ & $-4.75$ & \ldots & $5190$ & $-22$ \\
  HD 68017 &  40118 &  46.33 & 0.06 &  6.859 & 0.02 & 5.090 & 0.02\tablenotemark{$\dagger$} & $<-5.32$ & $-4.89$ & \ldots & $5531$ & $-0.44$ \\
    GJ 758 &  95319 &  64.06 & 0.02 &  6.447 & 0.02 & 4.493 & 0.036 & $<-5.04$ & $-5.05$ & \ldots & $5426$ & $0.21$ \\
   HR 7672 &  98819 &  56.43 & 0.07 &  5.857 & 0.02 & 4.388 & 0.027 &  $-5.88$ & $-4.77$\tablenotemark{$\dagger\dagger$} &  13.94 & $5921$ & $0.05$
\enddata
\tablenotetext{*}{References: $\varpi$ from {\it Gaia} DR2 \citep{Gaia_Astrometry_2018}, $V_T$ from {\it Tycho}-2 \citep{Hog+Fabricius+Makarov+etal_2000}, $K_s$ from 2MASS \citep{Cutri+Skrutskie+vanDyk+etal_2003}, $R_X$ from {\it ROSAT} \citep{Voges+Aschenbach+Boller+etal_1999}, $R^\prime_{\rm HK}$ from \cite{Pace_2013} and references therein (we have corrected several errors in the catalog), $T_{\rm eff}$ and [Fe/H] from \cite{Soubiran+Campion+Brouillet+etal_2016} and references therein, rotation periods: HR~7672 \citep{Wright+Drake+Mamajek+etal_2011}; HD~4747 \citep{Peretti+Segransan+Lavie+etal_2018}.}
\tablenotetext{$\dagger$}{We have inflated the formal uncertainties to limit the impact of stellar atmospheric modeling and bandpass corrections.}
\tablenotetext{$\dagger\dagger$}{Multi-decade Mt.~Wilson measurement \citep{Baliulnas+Donahue+Soon+etal_1995}}
\label{tab:stellarparam}
\end{deluxetable*}

Our targets consist of five field G and K dwarfs, each with a long history of observations including high-resolution spectroscopy.  In this section we revisit their fundamental properties of age and mass.  The ages, in particular, are needed to constrain models of their companions.  We use both the Bayesian activity age measurement described in \cite{Brandt+Kuzuhara+McElwain+etal_2014} and a comparison to the PARSEC stellar models \citep{Bressan+Marigo+Girardi+etal_2012}.  One of our stars, the low-metallicity G~dwarf HD~68017A, is not fit by these stellar models at any age or mass.  

\subsection{Ages from Stellar Activity and Rotation} \label{subsec:activity_rotation}

Age dating of Sun-like stars by their chromospheric and coronal activity is possible because late-type stars have convective outer envelopes that support magnetic dynamos.  As these stars launch a wind, the wind rotates at the same angular velocity as the photosphere out to the Alfv\'en radius \citep{Mestel_1968}.  The star imparts its angular momentum to the magnetized wind and spins down \citep{Noyes+Hartmann+Baliunas+etal_1984}.  This has been recognized as a possible ``clock'' for many years \citep{Barnes_2003}, albeit with large uncertainties, and has been calibrated using clusters and field binaries \citep[e.g.][]{Mamajek+Hillenbrand_2008}.  The magnetic dynamo also heats the chromosphere and corona, resulting in narrow Ca\,{\sc ii}~HK emission lines and X-ray emission.  Both decline as the star spins down, and are much weaker in old field stars than in young clusters \citep[][and references therein]{Soderblom_2010}.  At very old ages, gyrochronology may be less useful as an age diagnostic due to a weakening of the stellar dynamo \citep{vanSaders+Ceillier+Metcalfe+etal_2016}.  The activity relations have also only been calibrated to activity levels of $R^\prime_{\rm HK} \geq -5.0$ corresponding to a Rossby number of 2.2 \citep{Mamajek+Hillenbrand_2008}.  This is only slightly weaker than the Solar activity level.  

\cite{Brandt+Kuzuhara+McElwain+etal_2014} have developed a Bayesian method using Ca\,{\sc ii}~HK and X-ray emission to constrain the Rossby number and, in turn, the stellar age.  This method accounts for uncertainties at old ages by treating all Rossby number measurements above 2.2 as lower limits, resulting in a long tail to old ages for quiescent stars like Gl~758A. It also includes a 5\% chance of a star being a catastrophic outlier in the sense that its activity does not reflect its age.  We scale a uniform distribution between 0 and 13~Gyr to account for this possibility.  We refer to \cite{Brandt+Kuzuhara+McElwain+etal_2014} for a detailed discussion of the approach.

We collect activity diagnostics from the literature, Ca\,{\sc ii} $S$-indices from the catalog of \cite{Pace_2013} and references therein, and X-ray activities from {\it ROSAT} \citep{Voges+Aschenbach+Boller+etal_1999}.  The \citeauthor{Pace_2013} catalog has many erroneous values that we have corrected.  We adopt the average of the maximum and minimum $S$-indices reported in the literature and transform this to the Mt.~Wilson $R^\prime_{\rm HK}$, roughly the log of the ratio of emission in the HK lines to the intensity of the underlying photospheric continuum, using the relations given in \cite{Noyes+Hartmann+Baliunas+etal_1984}.  For stars that are not detected in {\it ROSAT}, we use nearby detections to estimate the level of X-ray emission that would result in a 5-$\sigma$ detection and use this level as an upper limit.  

Table \ref{tab:stellarparam} includes our $R^\prime_{\rm HK}$ measurements and $R_X$ values (the log of the ratio of X-ray to bolometric luminosity) for each target.  One of our stars, HR~7672A, has multi-decade Mt.~Wilson measurements from \cite{Baliulnas+Donahue+Soon+etal_1995} that we adopt in lieu of an average of literature values.  
Two of our stars, HD~4747A and HR~7672A, also have rotation periods measured in the literature.  These provide a more direct probe of the Rossby number and give smaller errors on the resulting ages \citep{Mamajek+Hillenbrand_2008}.

\subsection{Stellar Model Fitting} \label{sec:isochrones}

We also fit stellar models as an independent measure of the age and stellar mass.  We perform the analysis using both a uniform age prior and the age distribution inferred from activity as described in the preceding section.  We adopt the PARSEC stellar models \citep{Bressan+Marigo+Girardi+etal_2012}.

We perform our fits using apparent magnitude, color, distance, and spectroscopic effective temperature and metallicity.  We take the measured values from {\it Tycho} \citep{Hog+Fabricius+Makarov+etal_2000} and 2MASS \citep{Cutri+Skrutskie+vanDyk+etal_2003} as our magnitudes, and transform apparent magnitudes into absolute magnitudes using the parallax measured by {\it Gaia} DR2 \citep{Gaia_General_2016,Gaia_Astrometry_2018}.  We adopt a floor of 0.02~mag on our photometric errors to account for possible systematics in the colors of the stellar models. 

For our spectroscopic $T_{\rm eff}$ and metallicity, we adopt the median values reported in the literature as tabulated in the PASTEL catalog \citep{Soubiran+Campion+Brouillet+etal_2016}.  While these measurements are all derived from high-resolution, high signal-to-noise spectra, they show significant scatter.  We adopt (Gaussian) uncertainties of 75~K in $T_{\rm eff}$ and 0.05 in [Fe/H] for all of our stars; this roughly reflects the range of values reported in the literature.

We show results for our stars with three combinations of photometry: $V_T$ only (sampling the peak of the intensity distribution), $K_s$ only (sampling the Rayleigh-Jeans tail), and both $V_T$ and $K_s$ (providing a photometric $T_{\rm eff}$).  The results are generally consistent, but show significant variation for Gl~758A.  One of our G dwarfs, HD~68017A, is incompatible with all of the PARSEC models at high significance.

We perform our analysis on grids of isochrones downloaded from the PARSEC web server\footnote{http://stev.oapd.inaf.it/cgi-bin/cmd} on the native {\it Tycho} and 2MASS photometric systems.  We interpolate $T_{\rm eff}$ and magnitudes onto a fine grid of masses and use a Salpeter initial mass function, $dN/dM \propto M^{-2.35}$ \citep{Salpeter_1955}, as our mass prior.  We then adopt either a uniform prior in age or the prior recovered from the activity-age relation.  We calculate our weights as 
\begin{equation}
{\rm weight} = p[{\rm age}] \times \exp \left[-\chi^2/2 \right],    
\end{equation}
where $p[{\rm age}]$ is the age prior and 
\begin{align}
    \chi^2 = &\frac{(V_T - V_{T,\,{\rm obs}})^2}{\sigma_V^2}
    + \frac{(K_s - K_{s,\,{\rm obs}})^2}{\sigma_K^2} \nonumber \\
    &+ \frac{(T_{\rm eff} - T_{\rm eff,\,obs})^2}{(75\,{\rm K})^2} + \frac{(Z - Z_{\rm obs})^2}{\sigma_Z^2}.
\end{align}
We compute these weights for all models and integrate to obtain marginalized posterior probability distributions.

\subsection{Results and Notes on Individual Stars}

\begin{figure}
    \centering
    \includegraphics[width=\linewidth]{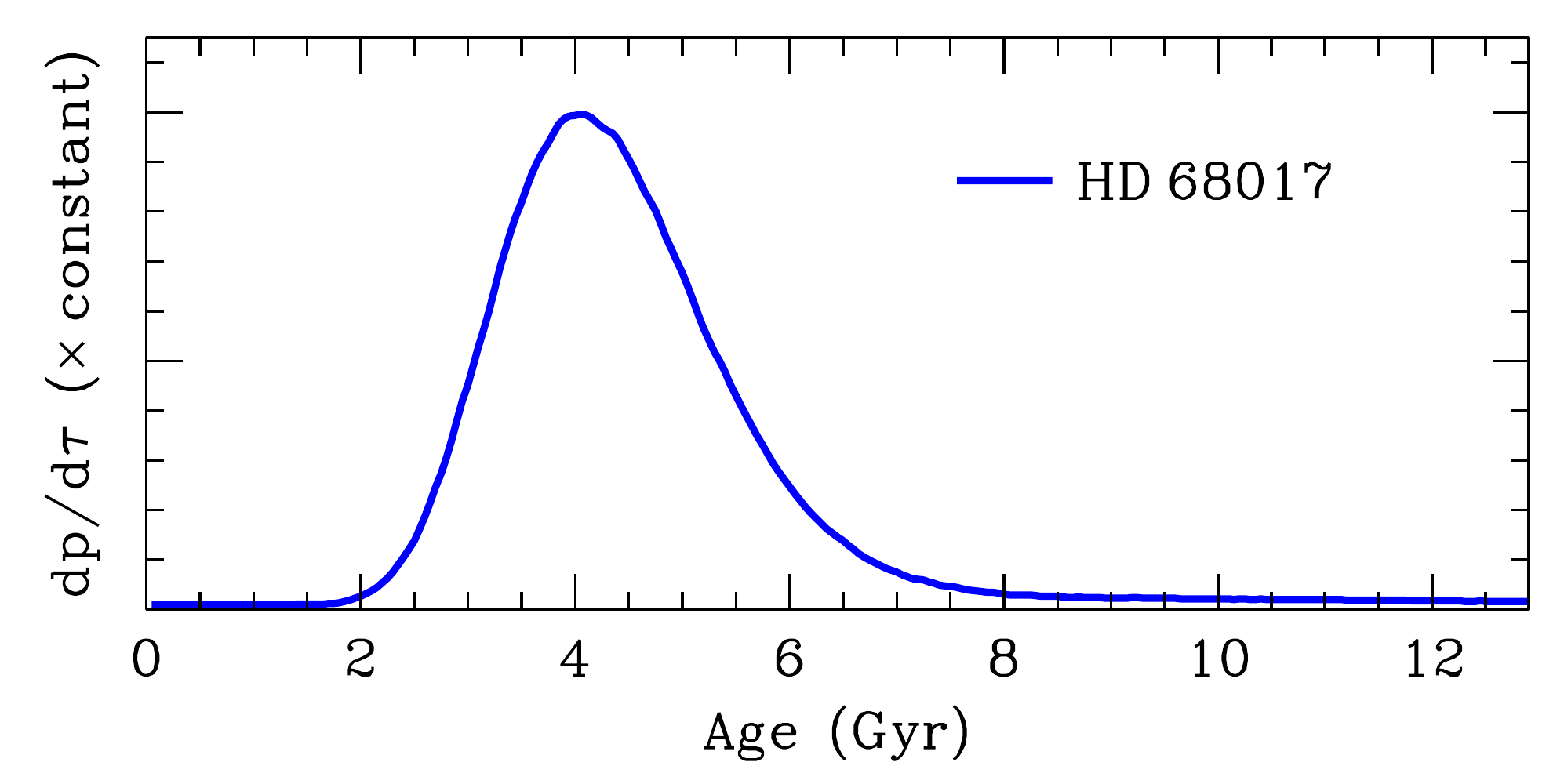}
    \caption{Age posterior for the G dwarf HD~68017A using the Bayesian technique of \cite{Brandt+Kuzuhara+McElwain+etal_2014}.  Our method uses both X-ray and chromospheric activity indicators to infer a Rossby number and convert this to an age using the calibration of \cite{Mamajek+Hillenbrand_2008}, adding a uniform distribution weighted by 5\% to account for the possibility that the activity age is pathological.  To provide a satisfactory fit to HD~68017 with PARSEC stellar models, we would either have to inflate our photometric errors by factor of $\sim$10 or increase its assumed metallicity to $\sim$$Z_\odot$ and increase its temperature by $\sim$100~K.}
    \label{fig:b14_bayesage}
\end{figure}

\begin{figure*}
\begin{center}
    \includegraphics[height=0.29\linewidth]{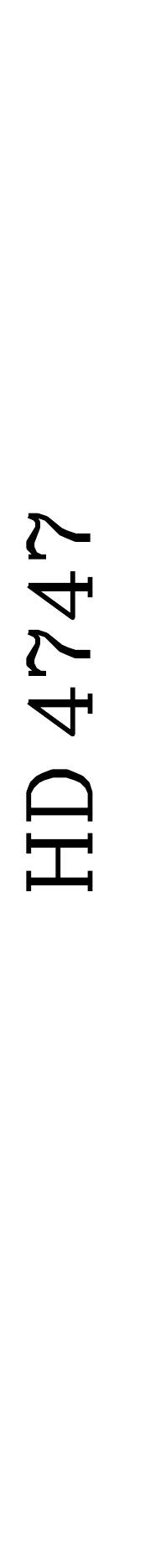}
    \includegraphics[height=0.29\linewidth]{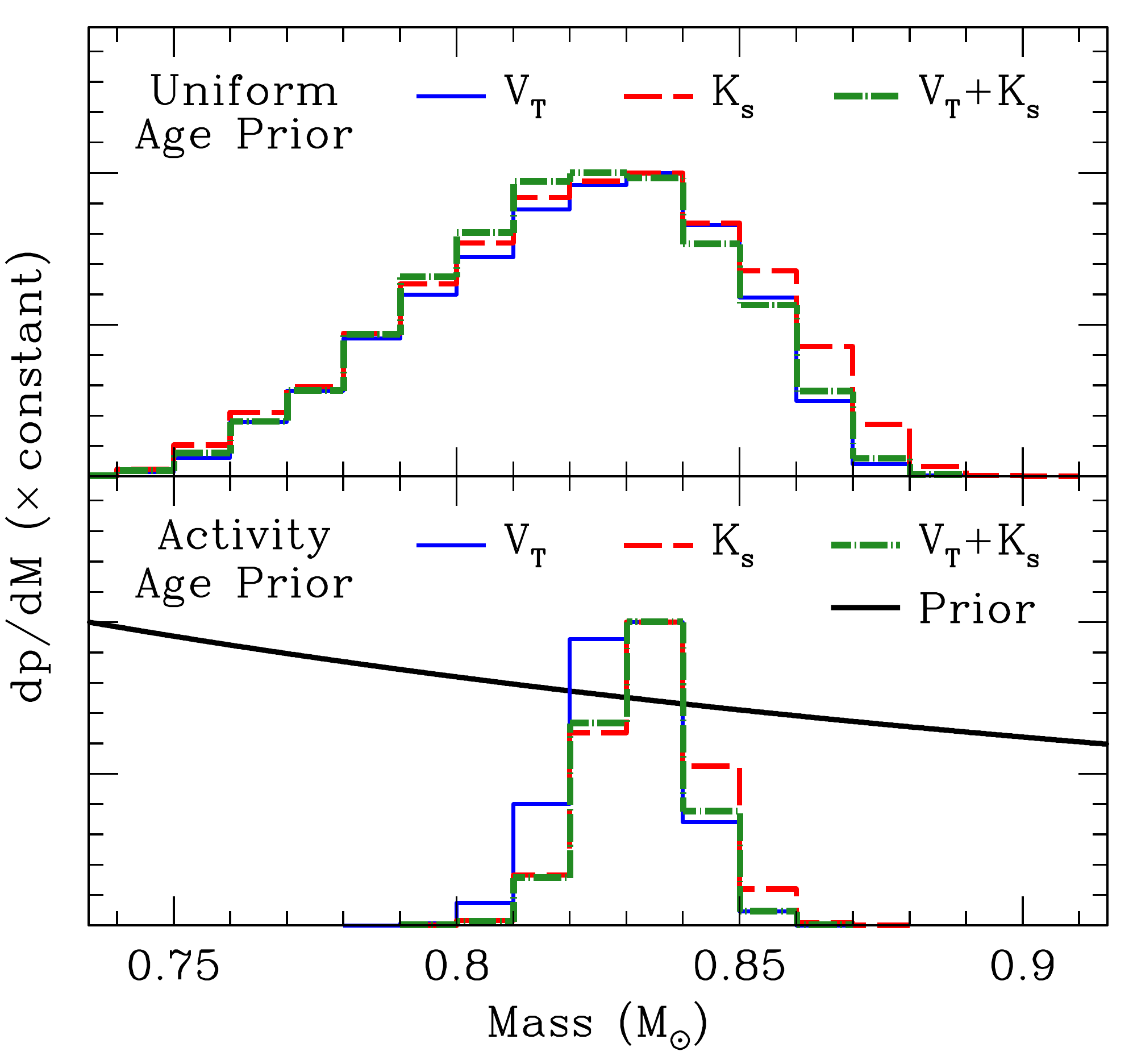}
    \includegraphics[height=0.29\linewidth]{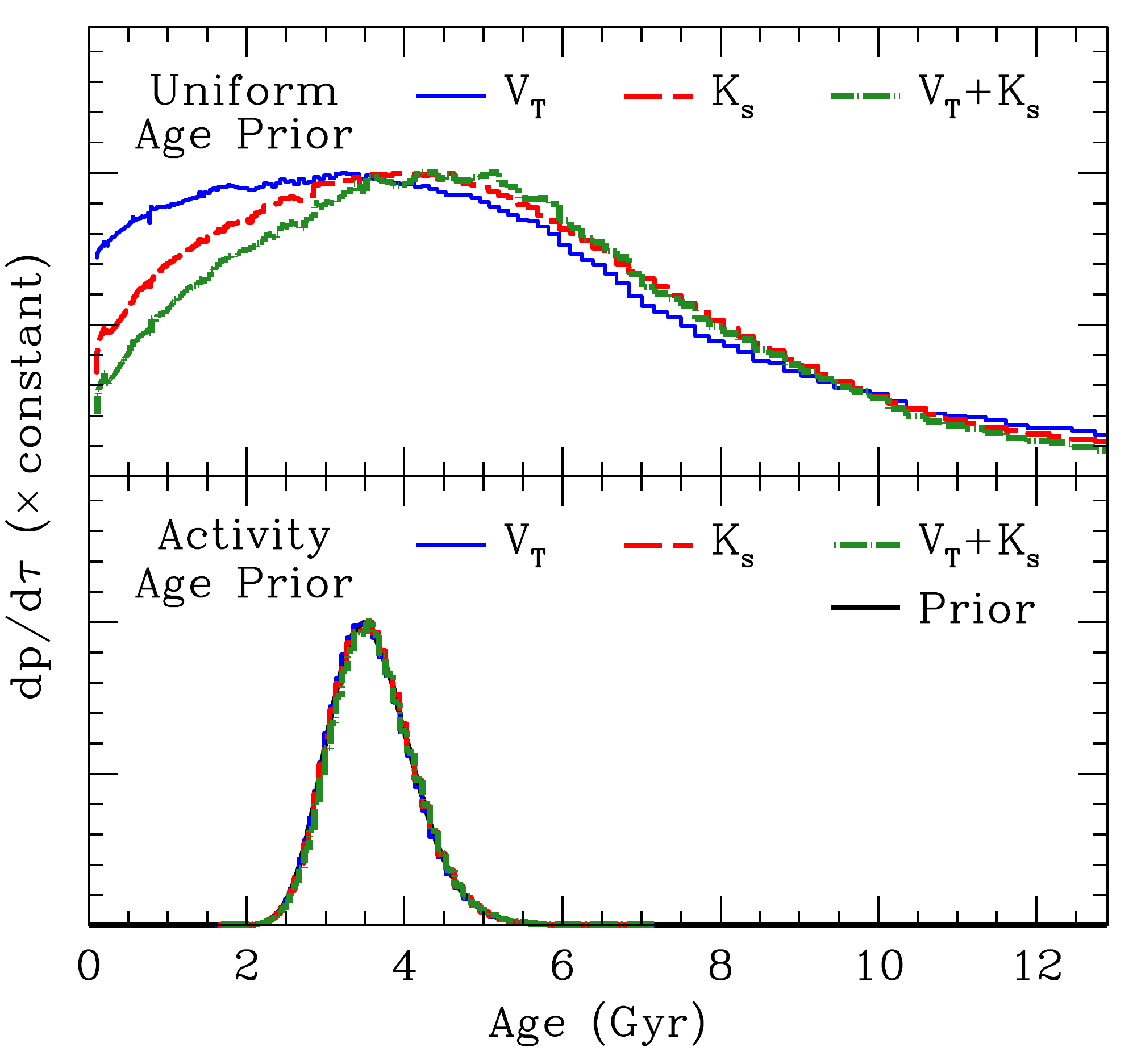}
    \includegraphics[height=0.29\linewidth]{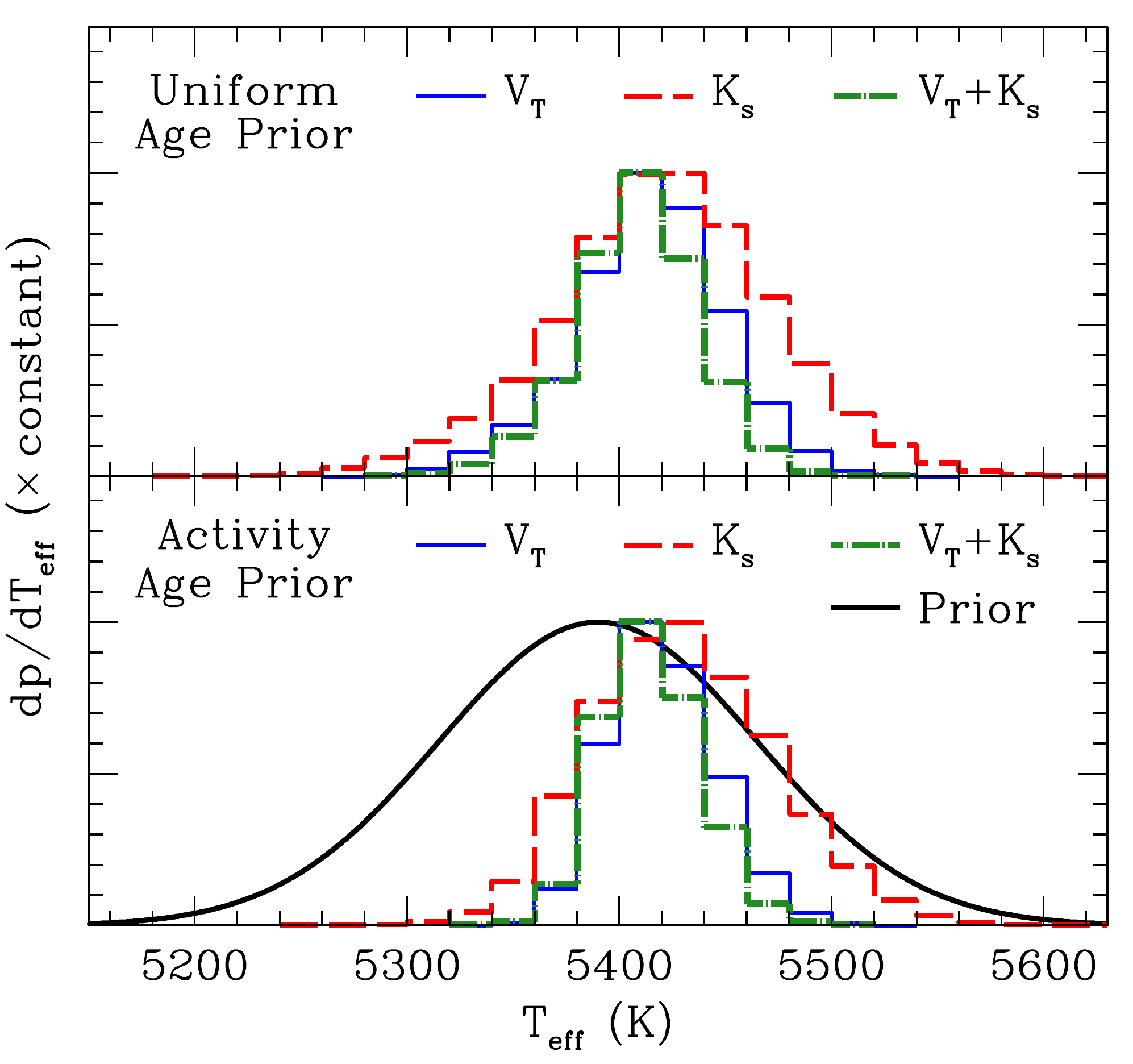}
    \includegraphics[height=0.29\linewidth]{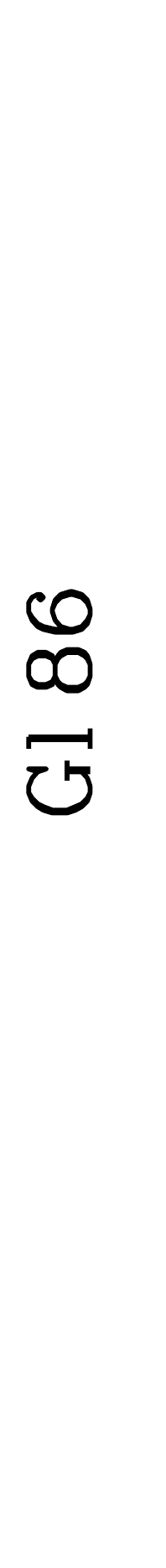}
    \includegraphics[height=0.29\linewidth]{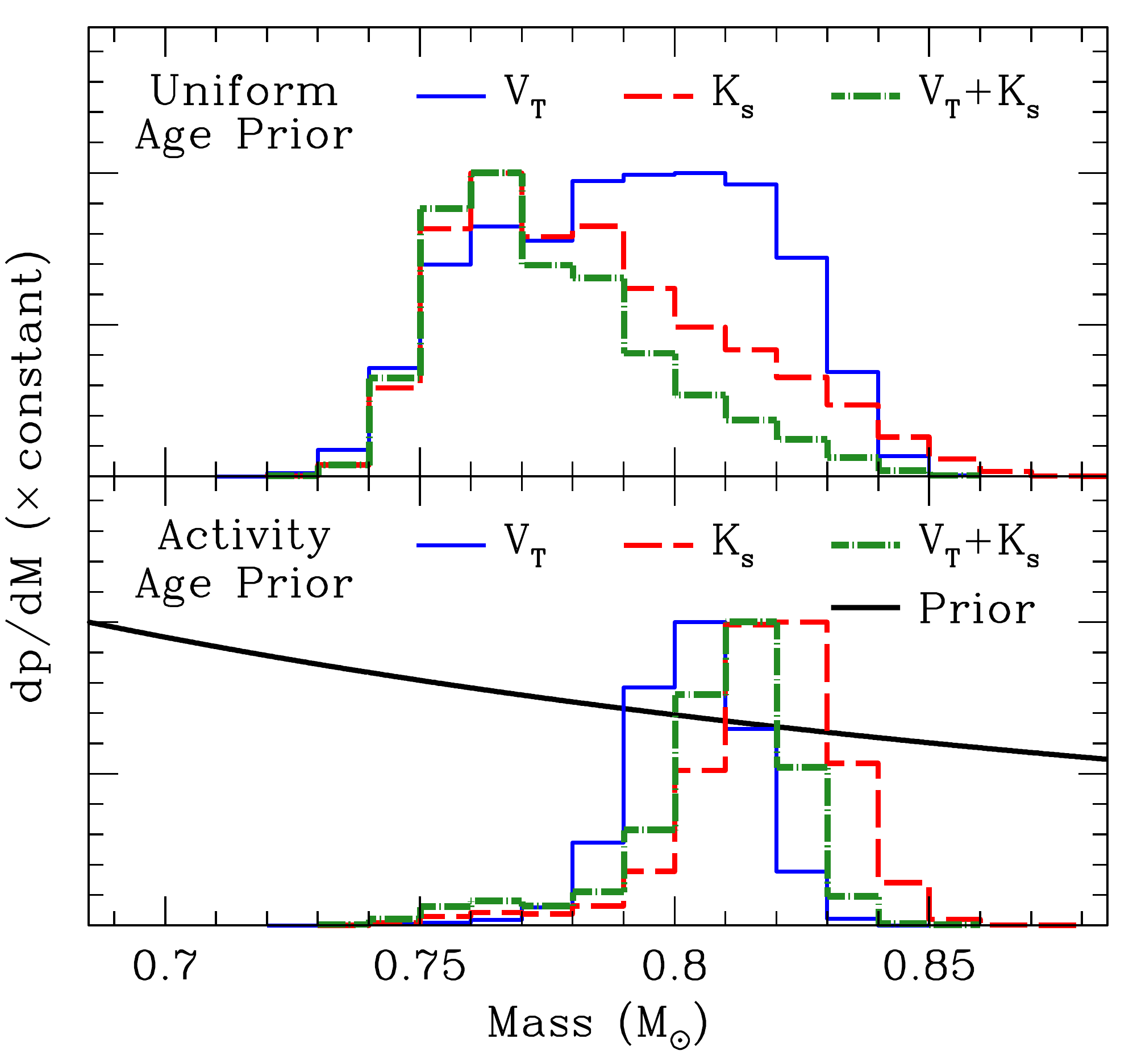}
    \includegraphics[height=0.29\linewidth]{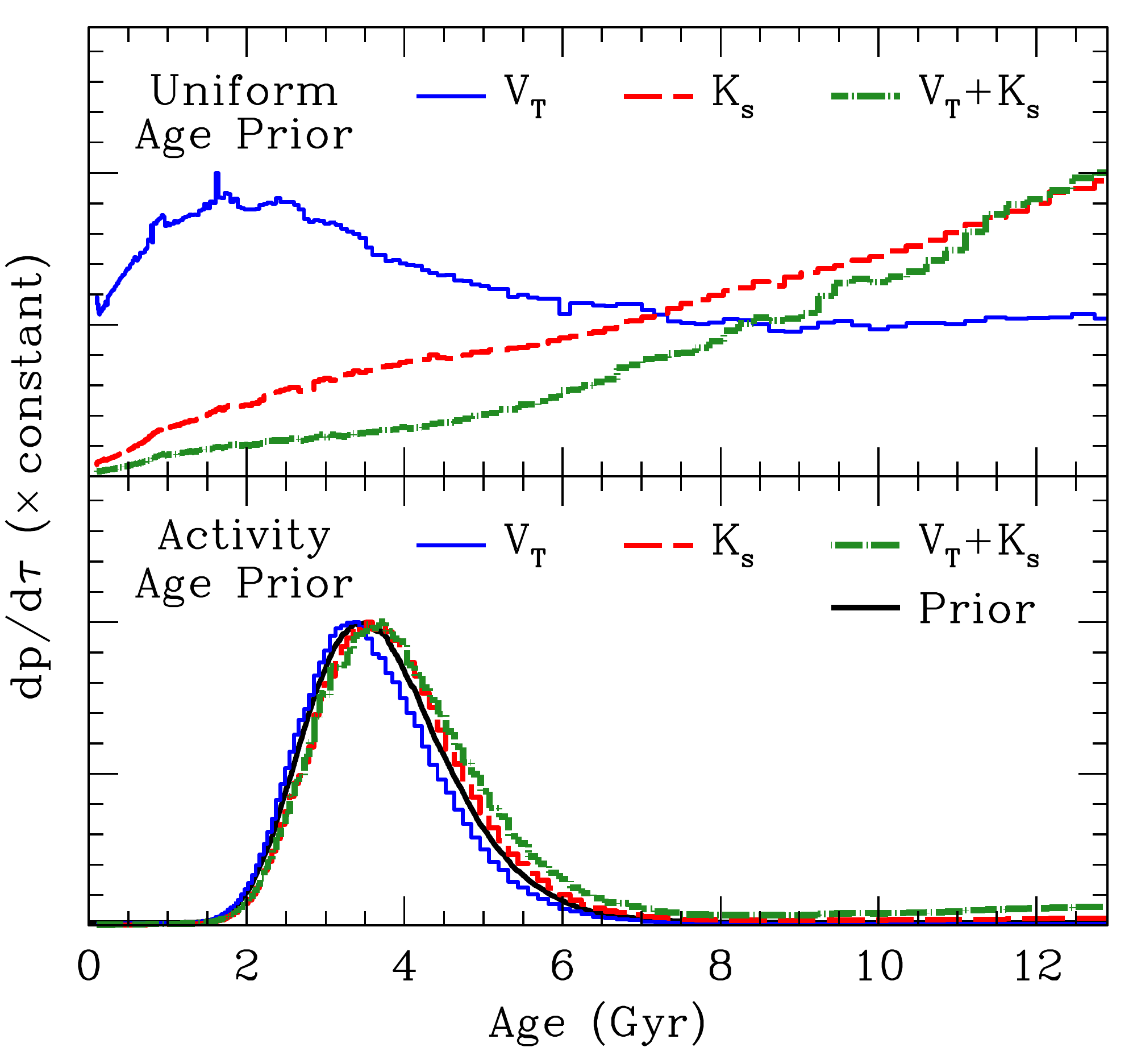}
    \includegraphics[height=0.29\linewidth]{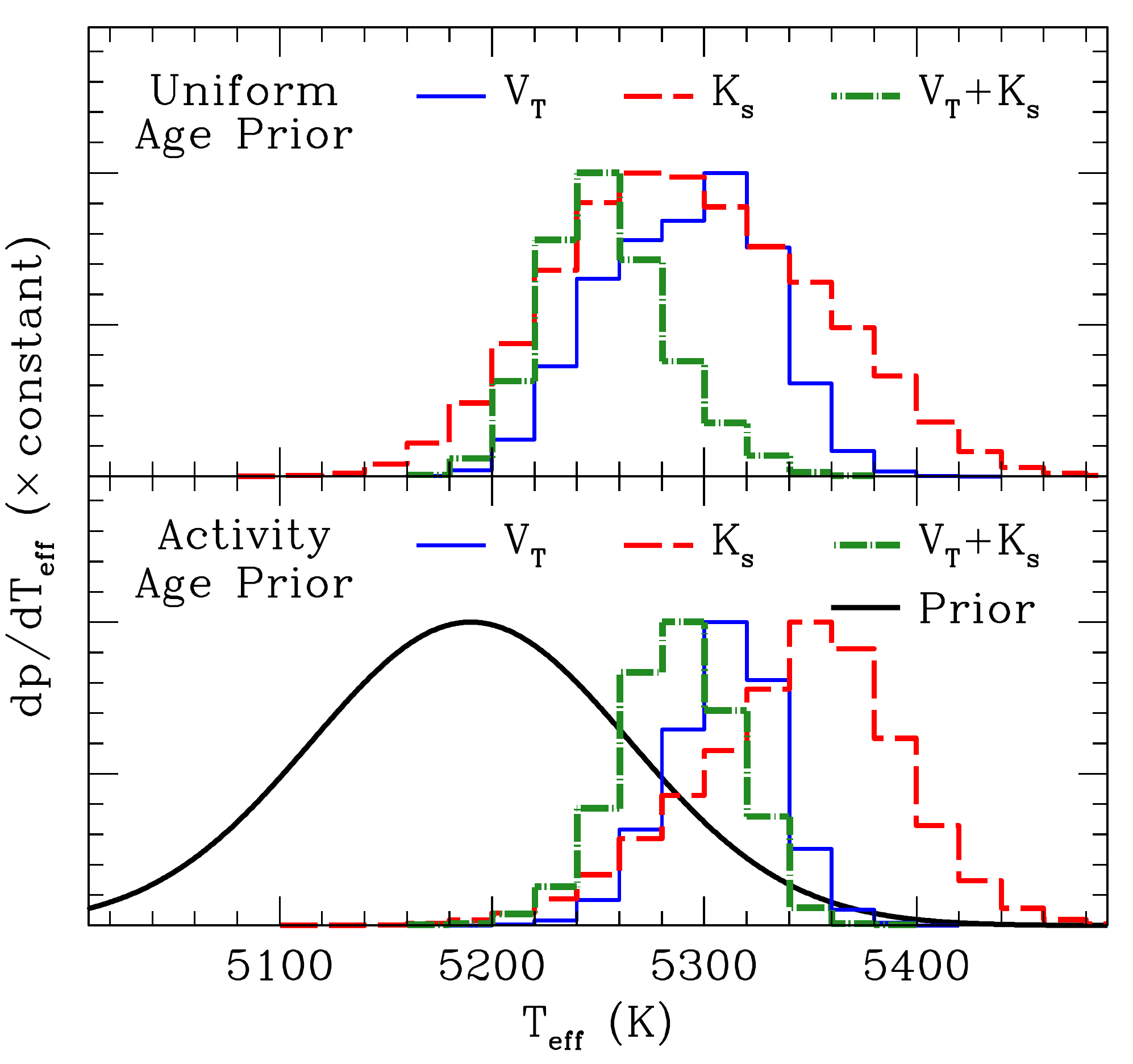}
    \includegraphics[height=0.29\linewidth]{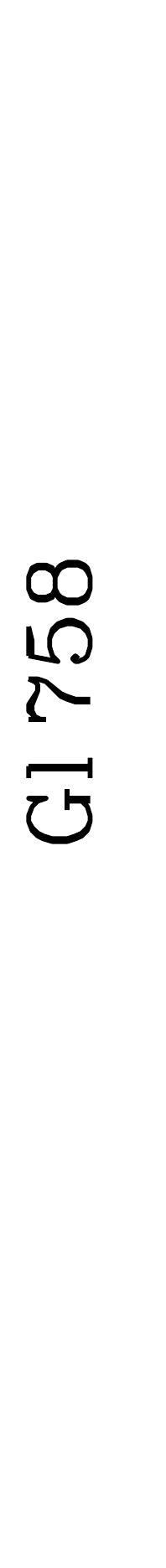}
    \includegraphics[height=0.29\linewidth]{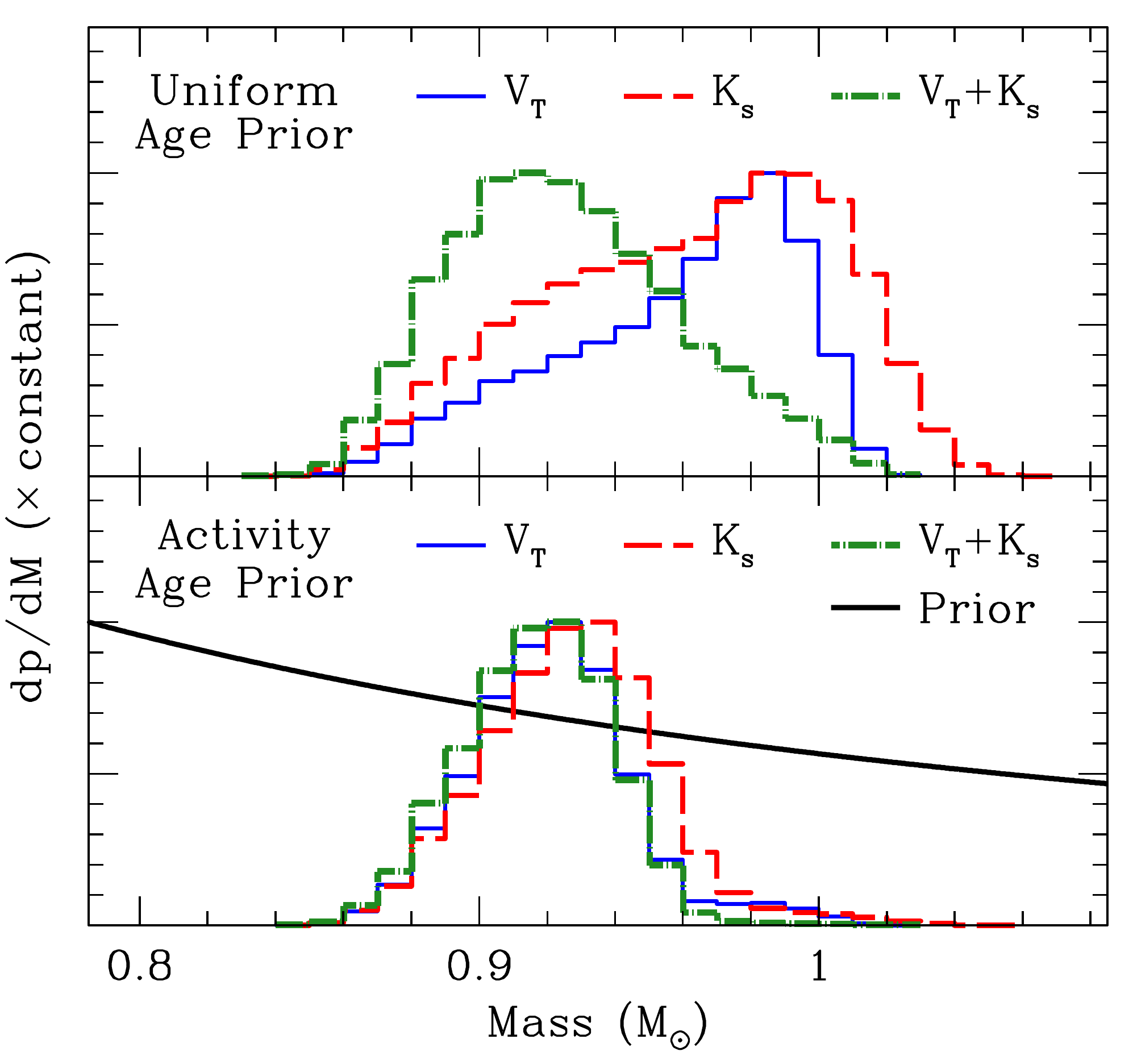}
    \includegraphics[height=0.29\linewidth]{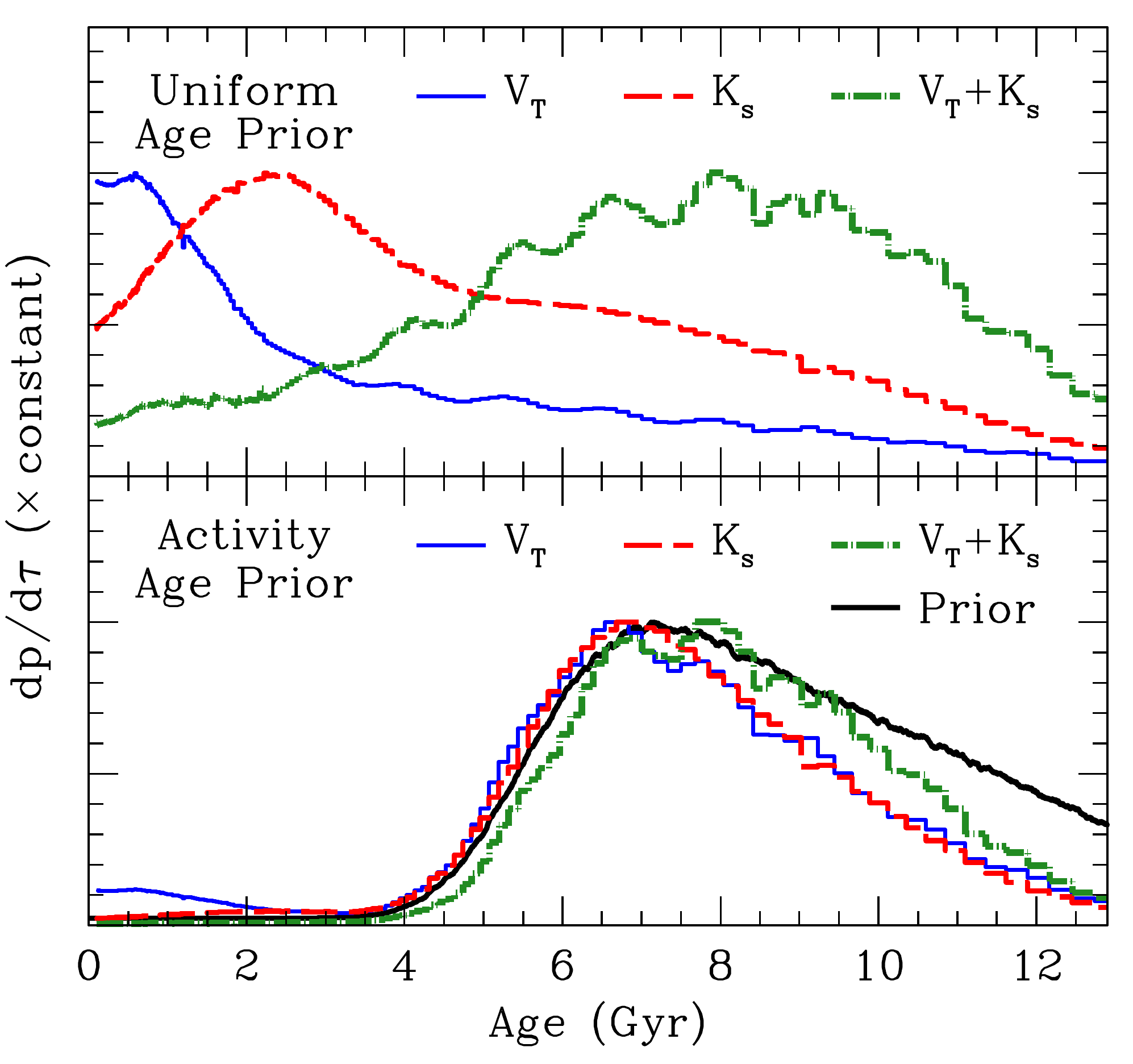}
    \includegraphics[height=0.29\linewidth]{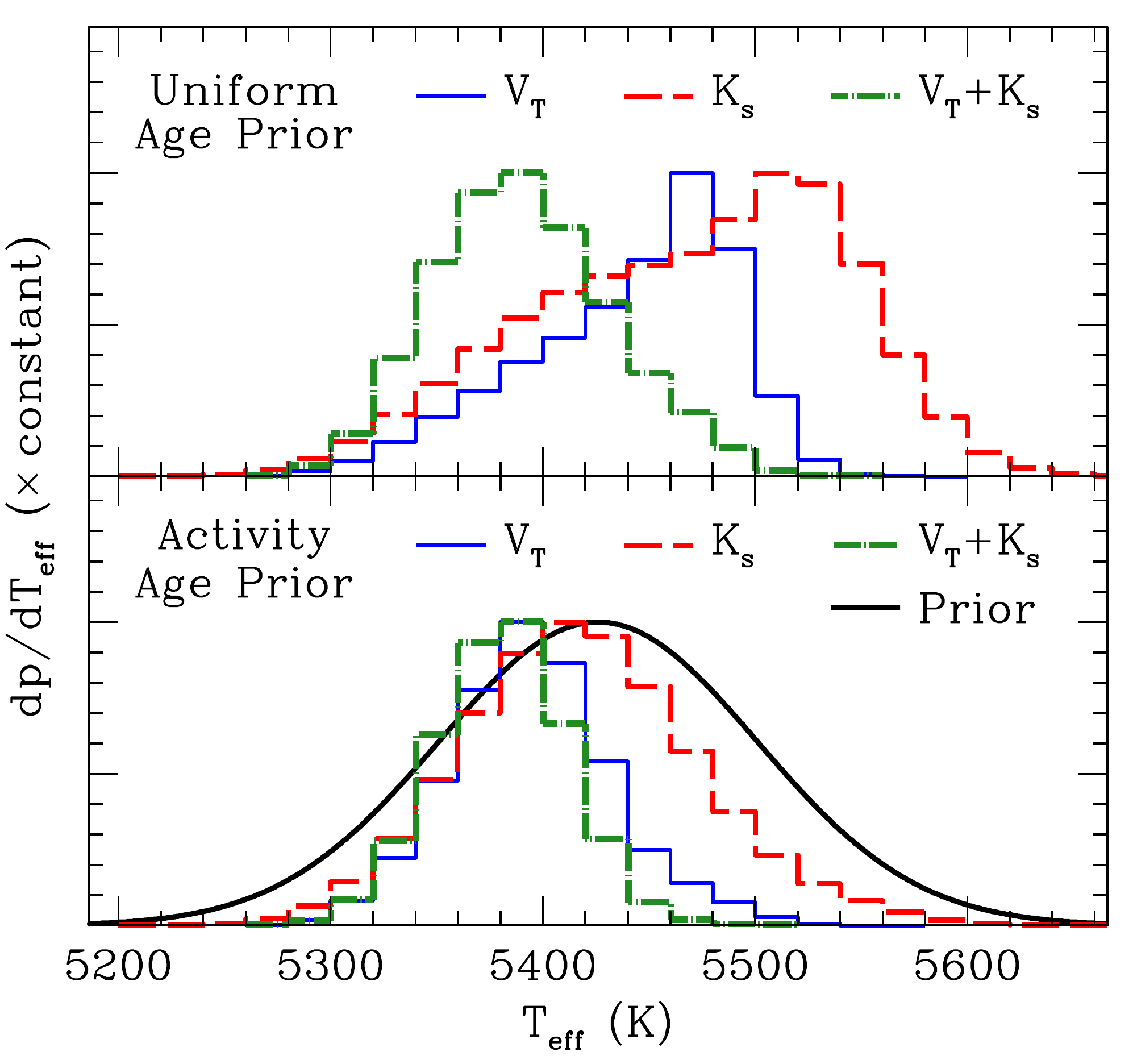}
    \includegraphics[height=0.29\linewidth]{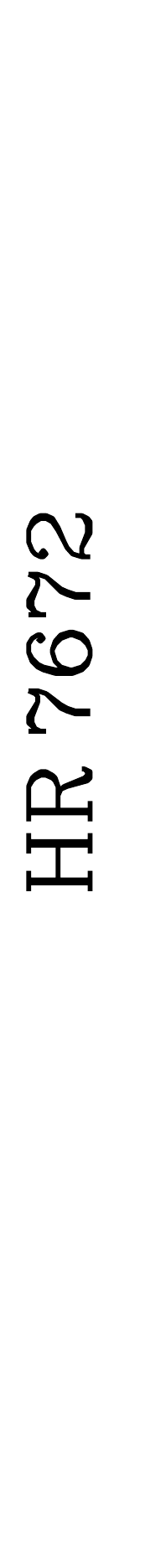}
    \includegraphics[height=0.29\linewidth]{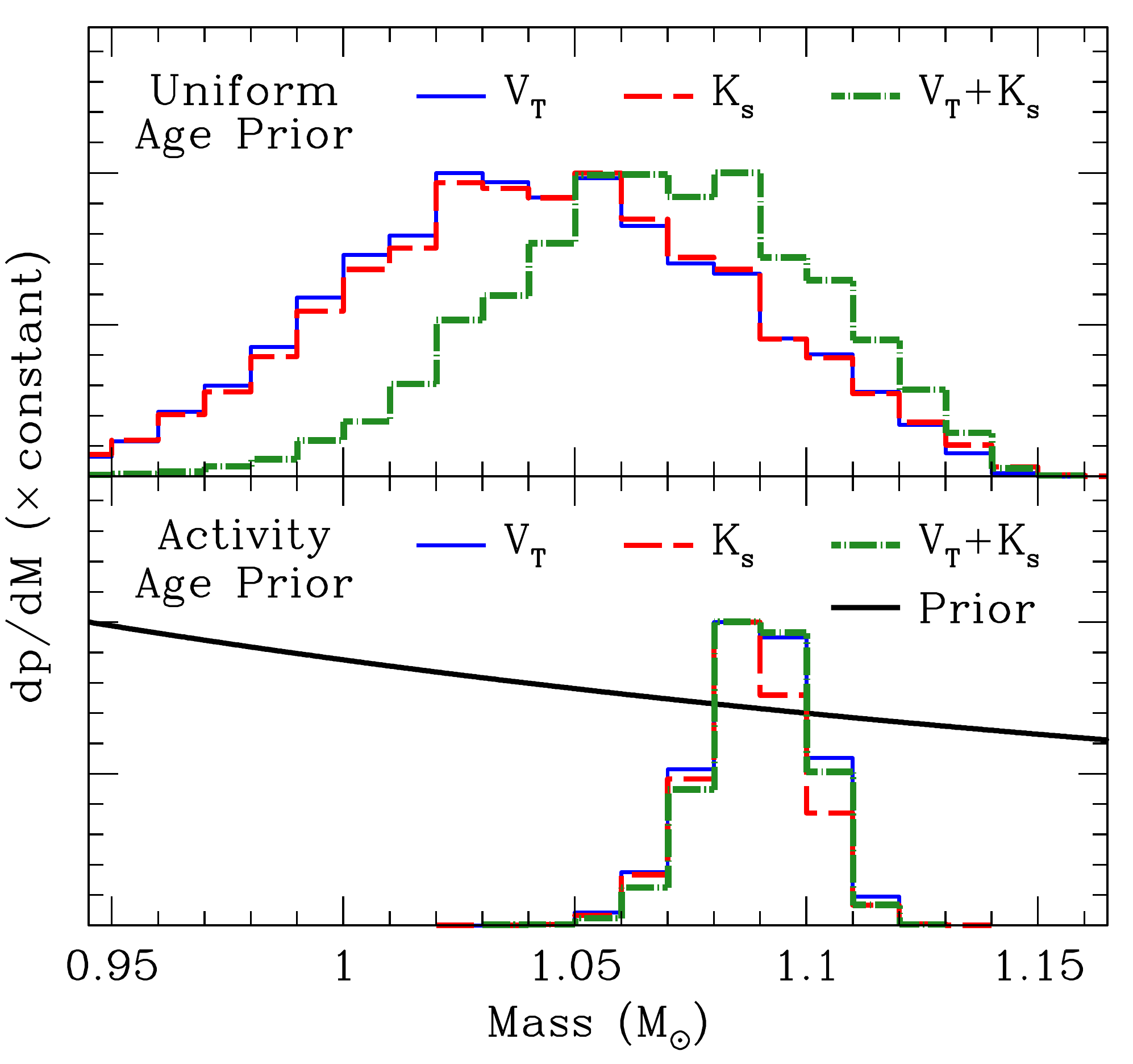}
    \includegraphics[height=0.29\linewidth]{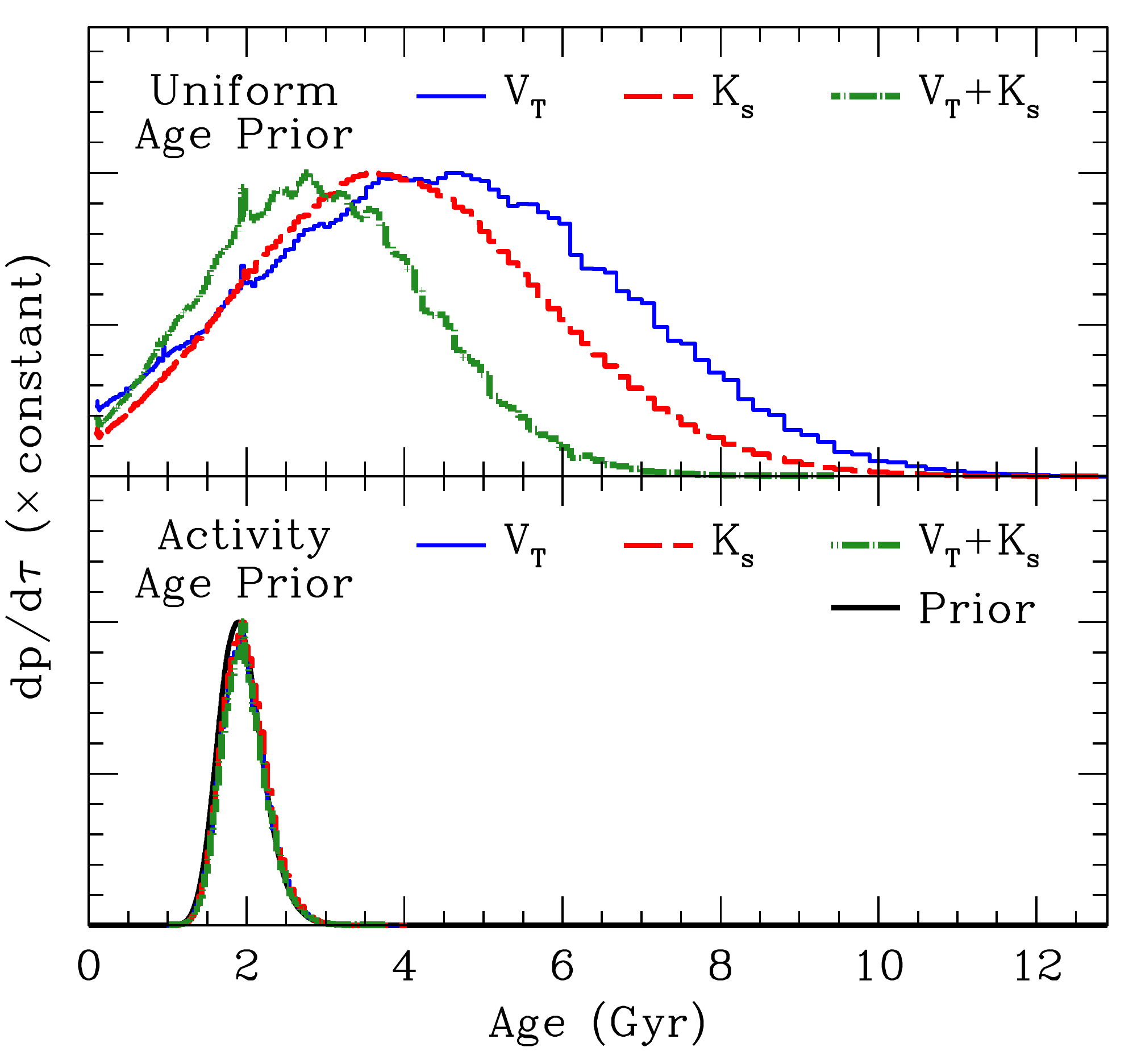}
    \includegraphics[height=0.29\linewidth]{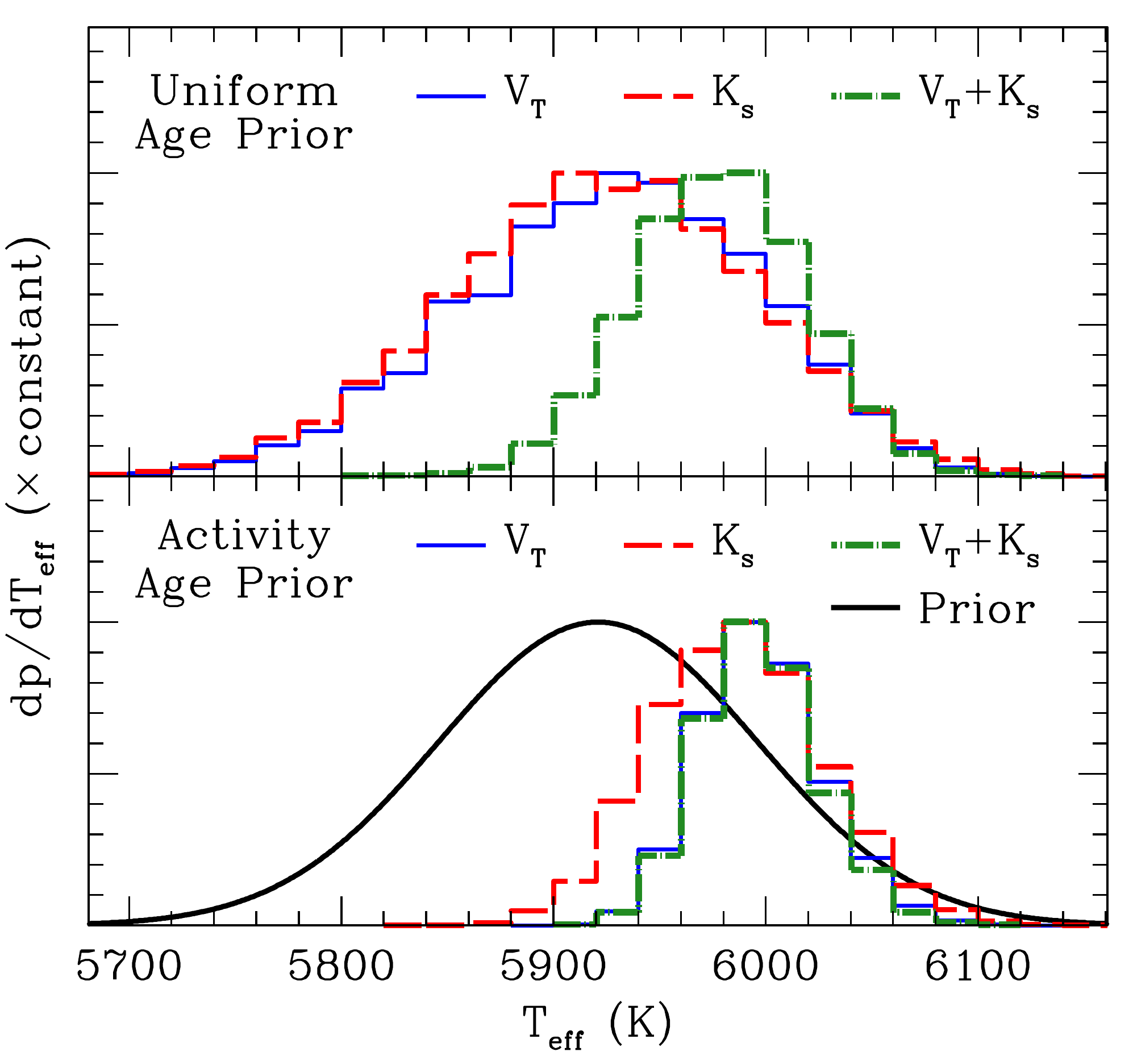}
    \end{center}
    \caption{Mass, age, and $T_{\rm eff}$ posterior probability distributions for our G and K dwarfs computed using the PARSEC isochrones \citep{Bressan+Marigo+Girardi+etal_2012} with a Salpeter mass prior and either a uniform age prior or one derived from stellar activity according to the method described in \cite{Brandt+Kuzuhara+McElwain+etal_2014}.  We do not show HD~68017, which the PARSEC models are unable to fit.  Our adopted stellar parameters are given in Table \ref{tab:stellarparam}.  The isochrone ages are in good agreement with the activity ages for all of these stars.  Gl~758 has posteriors that differ significantly depending on which photometric band(s) we use; this likely reflects a mismatch between our spectroscopic $T_{\rm eff}$ and the photometric $T_{\rm eff}$ (as measured across the $V_T$ and $K_s$ wavelength baseline) implicit in the PARSEC isochrones.  }
    \label{fig:age_mass_posteriors}
\end{figure*}

We perform our Bayesian activity-based age dating and our isochrone analysis to all of our stars.  Figure~\ref{fig:b14_bayesage} shows the activity-based age for HD~68017A.  As we discuss in more detail below, no PARSEC model provides an acceptable fit to this star.  Figure~\ref{fig:age_mass_posteriors} shows the results of our stellar model fitting for the rest of the sample, with one row of plots for each star.  The left column of plots shows the posterior distributions for mass, the middle plots show the posteriors for age, and the right plots show the posteriors for effective temperature.  The upper plots for each object assume a uniform prior for age, while the lower plots adopt the posteriors of the activity age analysis (Section~\ref{subsec:activity_rotation}) as their age priors.  The thick black lines on the lower age panels show these activity age posteriors. \\

\noindent {\it HD~4747A}---This early-K/late-G star \citep{Houk+Smith-Moore_1988} has measured X-ray and Ca\,{\sc ii} HK emission together with a photometric rotation period of just under 28 days \citep{Peretti+Segransan+Lavie+etal_2018}.  These measurements combine to give a relatively precise activity-based age peaked at just under 4~Gyr.  The activity-based age is fully consistent with an age based on stellar models, and agrees well with the spectroscopic effective temperature.  \\

\noindent {\it Gl~86A}--- The activity--age indicators yield an age of $\sim$3--4~Gyr for this star, in moderate tension with stellar models using both the $V_T$ and $K_s$ bands.  The star's kinematics indicate that it is not a member of the thin disk, and suggest an old age \citep{Rocha-Pinto+Castilho+Maciel_2002}, again in tension with stellar activity.  More recently, \cite{Fuhrmann+Chini+Buda+etal_2014} have found that the star has a high ratio of magnesium to iron.  This provides a chemical association intermediate between the thin disk and the old, thick disk, corroborating the kinematic evidence.  \citeauthor{Fuhrmann+Chini+Buda+etal_2014} conjecture that mass transfer from the secondary's wind added angular momentum to the primary and accounts for its relatively strong activity.  

\cite{Farihi+Bond+Dufour+etal_2013} infer a mass of $0.59 \pm 0.01 $~$M_\odot$ for the white dwarf companion.  This is based on angular diameter, parallax, and theoretical mass-radius relations \citep{Fontaine+Brassard+Bergeron_2001}.  The angular diameter is computed from the white dwarf's effective temperature, measured by \cite{Farihi+Bond+Dufour+etal_2013} both spectroscopically and from broad-band photometry.  
The {\it Hipparcos} parallax of $92.74\pm 0.32$~mas adopted by \cite{Farihi+Bond+Dufour+etal_2013} is consistent with the more precise {\it Gaia} DR2 parallax of $92.704 \pm 0.045$~mas.  The angular diameter and parallax combine to give a physical radius and a resulting mass of $0.59 \pm 0.01$~$M_\odot$.  Assuming that a white dwarf of this mass descends from a $\sim$2~$M_\odot$ star, this implies a total system age of $\lesssim$3~Gyr \citep{Farihi+Bond+Dufour+etal_2013}.  A very old age for Gl~86 implies a relatively low-mass progenitor for its white dwarf.  This conclusion is consistent with the analysis of \cite{Lagrange+Beust+Udry+etal_2006}, but is in tension with a remnant mass of $\sim$0.6~$M_\odot$. \\

\noindent {\it HD~68017A}---This G3 star \citep{Gray+Corbally+Garrison+etal_2003} has an upper limit on its X-ray flux but a detection of chromospheric activity slightly above the Solar value.  This points to a star very similar to the Sun in both mass and age.  However, no PARSEC model at the star's spectroscopic metallicity of $-0.44$ can match the observed photometry and {\it Gaia} parallax.  This metallicity is not from a single, isolated analysis: \cite{Fuhrmann_2004, Mishenina+Soubiran+Kovtyukh+etal_2004, Valenti+Fischer_2005, Ramirez+Allende_Prieto+Lambert_2007,Takeda+Kawanomoto+Honda+etal_2007}, and \cite{Ramirez+Allende_Prieto+Lambert_2013} all use high-resolution spectra to infer $[{\rm Fe/H}]$ values from $-0.40$ to $-0.47$.  The best-fitting PARSEC models are at very old ages, but these models are not good fits.  We do not show the formal posterior probability distributions from isochrone fitting; the poor goodness-of-fit metric renders these distributions meaningless.

We do not have an explanation for the failure of the stellar models to fit the spectroscopic and photometric constraints at any mass and age.  The star's $B-V$ color of 0.69\,mag \citep{Ducati_2002} is similar to that of the Sun, while its absolute $V$-band magnitude of 5.14\,mag (computed using the {\it Gaia} parallax) is slightly less luminous than the Sun.  The star does have a close M-dwarf companion, but at visible through near-infrared wavelengths the contrast is $>$100 \citep{Crepp+Johnson+Howard+etal_2012}.  The flux added by the companion is smaller than the photometric errors, and should also have a negligible effect on stellar parameters inferred from optical spectroscopy.  In the absence of a constraint from stellar models, Figure \ref{fig:b14_bayesage} shows the age inferred from stellar activity alone. \\

\noindent {\it Gl~758A}--- Our analysis of Gl~758A favors an old age, $\gtrsim$6~Gyr, based largely on its low levels of chromospheric and coronal activity.  These are in mild tension with the PARSEC isochrone ages.  \cite{Vigan+Bonnefoy+Ginski+etal_2016} used Gl~758A's $V$-band magnitude, spectroscopic metallicity, and effective temperature to derive an age of $2.2 \pm 1.4$~Gyr based on the same PARSEC models that we use.  This analysis corresponds to the blue solid curve in Figure \ref{fig:age_mass_posteriors}, with a lower limit of 600~Myr based on a non-detection of lithium \citep{Janson+Carson+Thalmann+etal_2011,Vigan+Bonnefoy+Ginski+etal_2016}.

When only using $V_T$ and adopting a uniform age prior, our modeling using the PARSEC isochrones favor a younger age and a higher mass.  This is also true, though slightly less so, when using only $K_s$.  In both cases the age posteriors are consistent with the age inferred from activity; they do not exclude an old age.  Using both $V_T$ and $K_s$ gives a photometric constraint on effective temperature.  The right panel of Figure \ref{fig:age_mass_posteriors} shows that these photometric $T_{\rm eff}$, which are implicit in the models, are in mild tension with the spectroscopic effective temperature.  However, the large, homogeneous, and calibrated analysis of \cite{Brewer+Fischer+Valenti+etal_2016} produces a very low $T_{\rm eff}$ of 5358~K even as those authors note strong systematic uncertainties in spectroscopic determinations of $T_{\rm eff}$ and their adoption of an empirical offset.  Adopting the \cite{Brewer+Fischer+Valenti+etal_2016} value would push the tension with the photometric $T_{\rm eff}$ in the other direction.  Because of the risks in relying on a precise value of $T_{\rm eff}$, we favor a relatively broad prior on $T_{\rm eff}$ and the use of an independently calibrated activity age prior.  

Regardless of which photometric band(s) we adopt, including the activity age as our prior conclusively favors an old age for the system.  These older ages of $\sim$6--10~Gyr give a stellar mass between 0.89 and 0.97~$M_\odot$ at 90\% confidence (the exact range depends on the choice of photometric bands).  An old age also eases the tension between the observed luminosity of Gl~758B and models of substellar evolution \citep{Bowler+Dupuy+Endl+etal_2018}.  Finally, we note that other authors have derived masses and ages for Gl~758A from different stellar evolutionary models.  \cite{Brewer+Fischer+Valenti+etal_2016}, for example, obtain a mass of $0.92 \pm 0.03$~$M_\odot$ and an age of 4.6 to 10.4~Gyr by combining the Yonsei-Yale models \citep{Demarque+Woo+Kim+etal_2004,Spada+Demarque+Kim+etal_2013} with their spectroscopic measurements.  These age constraints, which were derived using a lower value of $T_{\rm eff}$, agree well with our activity age. \\

\noindent {\it HR~7672A}---This G0 star \citep{Gray+Corbally+Garrison+etal_2006} has Ca\,{\sc ii} HK emission measured from multi-decade Mt.~Wilson data \citep{Baliulnas+Donahue+Soon+etal_1995} and a photometric rotation period \citep{Wright+Drake+Mamajek+etal_2011}.   This provides the best activity-based age of any star in our sample, centered at 2~Gyr.  

HR 7672A also has an angular diameter of $0.584 \pm 0.010$~mas measured interferometrically \citep{Crepp+Johnson+Howard+etal_2012}, or a radius of $1.113 \pm 0.019$~$R_\odot$ adopting the {\it Gaia} DR2 parallax.  \cite{Crepp+Johnson+Fischer+etal_2012} used this radius to obtain an isochrone-based age of $2.5 \pm 1.8$~Gyr.  The activity-based age is consistent with the isochrone-based ages, but is much more precise.  It favors a slightly higher effective temperature than the spectroscopic value, though well within our adopted uncertainty.

\section{Radial Velocity and Direct Imaging Astrometry} \label{sec:rv_imaging}

All of our targets have both direct imaging and radial velocity data available in the literature.  In this section, we provide a brief summary of the observations for each star.  The radial velocity time series are mostly taken with the HIRES instrument on Keck \citep{Vogt+Allen+Bigelow+etal_1994}, as published by \cite{Butler+Vogt+Laughlin+etal_2017}.  Gl~86 is a southern target with a radial velocity time series from the UCLES \'echelle spectrograph \citep{Diego+Charalambous+Fish+etal_1990} on the Anglo-Australian Telescope.  Its data were published in \cite{Butler+Wright+Marcy+etal_2006}.  The direct imaging astrometry comes from a variety of sources; they are listed in Table \ref{tab:companion_astrometry}.

\subsection{HD~4747}

The radial velocity curve for HD~4747A is from HIRES, as published by \cite{Butler+Vogt+Laughlin+etal_2017}.  As for our other targets with HIRES data, the spectra were reduced and calibrated using the same techniques as the California Planet Survey \citep{Howard+Johnson+Marcy+etal_2010}.  The radial velocity time series for HD~4747A runs from 1996 through 2013, covering nearly 17 years with 49 measurements.  The median uncertainty of the radial velocities is 1.65~m\,s$^{-1}$, and the radial velocity curve contains a significant amount of orbital information beyond a simple linear trend.

Direct imaging observations of HD~4747 include four different instruments spanning nearly ten years.  High-contrast imaging was obtained in 2008 using NACO \citep{Lenzen+Hartung+Brandner+etal_2003,Rousset+Lacombe+Puget+etal_2003} on VLT.  Initially a non-detection, the data were re-reduced by \cite{Peretti+Segransan+Lavie+etal_2018}.  These authors constructed $\chi^2$ maps at different roll angles to obtain astrometric uncertainties, and added a small additional amount of error in quadrature.  More recently, the system has been observed by NIRC2 on Keck \citep{Crepp+Gonzales+Bechter+etal_2016}, by GPI \citep{Crepp+Principe+Wolff+etal_2018}, and by SPHERE \citep{Peretti+Segransan+Lavie+etal_2018}.  The NIRC2 observations in 2014 and 2015 were taken as part of the TRENDS survey, and were reduced and presented in \citep{Crepp+Gonzales+Bechter+etal_2016}.  
The more recent SPHERE observations have the lowest claimed uncertainties, particularly in separation, where the fractional error is 1\%.  

The two GPI position angle measurements by \cite{Crepp+Principe+Wolff+etal_2018} are $\sim$2$\sigma$ discrepant with one another.  The first of the two GPI measurements also has a position angle $\sim$4$\sigma$ discrepant from the trend favored by the rest of the astrometry.  We therefore omit these measurements from our fit, adopting the same astrometry as \cite{Peretti+Segransan+Lavie+etal_2018}.  \cite{Bowler+Dupuy+Endl+etal_2018} encountered similar difficulties in combining astrometry of Gl~758B from different instruments; those authors finally restricted their analysis to NIRC2.

The combination of ten years of companion astrometry, twenty years of radial velocities, and a model isochrone-derived host-star mass of $0.856\pm0.014$\,\Msun\ enabled \cite{Peretti+Segransan+Lavie+etal_2018} to derive a companion mass of $m = 70.2 \pm 1.6$~$M_{\rm Jup}$. This is somewhat higher than the companion mass of $m=65.3^{+4.4}_{-3.3}$\,\Mjup\ derived by \cite{Crepp+Principe+Wolff+etal_2018} using a host star mass of $0.82\pm0.04$\,\Msun.

\subsection{Gl~86}

Gl~86 is a southern target ($\delta \approx -51^\circ$) and is not accessible from Keck.  The radial velocity time series for this object was taken with the UCLES \'echelle spectrograph \citep{Diego+Charalambous+Fish+etal_1990} on the Anglo-Australian Telescope.  \cite{Butler+Wright+Marcy+etal_2006} published 42 radial velocity measurements taken between 1998 and 2005 with a median uncertainty of 4.2~m\,s$^{-1}$.  The radial velocities show both a strong linear trend and a periodic signal from an inner companion (Gl~86~b) on a 15.8 day orbit \citep{Queloz+Mayor+Weber+etal_2000}.  \cite{Han+Black+Gatewood_2001} used the {\it Hipparcos} epoch astrometry together with the radial velocity amplitude to obtain a mass of $m_b \sim 15$~$M_{\rm Jup}$.  However, \cite{Pourbaix+Arenou_2001} showed that the {\it Hipparcos} precision is insufficient to establish the true mass of the inner planet.  A full astrometric analysis may be possible once the {\it Gaia} epoch astrometry is published.

Relative astrometry for the companion extends over more than a decade, beginning with the first detection by \cite{Els+Sterzik+Marchis+etal_2001} using ADONIS-SHARPII \citep{1997ExA.....7..285B}.  Those authors detected the companion at three observational epochs in late 2000, but their formal astrometric errors render the astrometric measurements inconsistent with one another.  Their tabulated astrometric errors also do not account for the $\sim$25~mas uncertainty they quote for the position of the bright Gl~86A.  In the text, \citeauthor{Els+Sterzik+Marchis+etal_2001} give a separation of $1.\!\!''73 \pm 0.\!\!''03$ and a position angle of $119^\circ \pm 1^\circ$.  We adopt these values at 2000.82, the mean of their three observational epochs.

\cite{Lagrange+Beust+Udry+etal_2006} used VLT/NACO to obtain additional astrometric measurements from 2003 to 2005 and also performed an orbital fit, finding the companion to be a $\sim$0.5~$M_\odot$ white dwarf.  \cite{Mugrauer+Neuhauser_2005} obtained another astrometric measurement in January of 2005, also with VLT/NACO, and independently confirmed the white dwarf identity of the companion.  The \citeauthor{Mugrauer+Neuhauser_2005} data were taken with NACO in spectral differential imaging (SDI) mode.  They used a {\it Hipparcos} binary to calibrate the plate scale.  

The \citeauthor{Lagrange+Beust+Udry+etal_2006} measurements were taken using both the wide-field camera S27 and the narrow-field camera S13.  In the case of the 2004 observation using S13, the quoted uncertainty in separation, $0.\!\!''014$, is no larger than that implied by the quoted uncertainty in the plate scale ($13.25 \pm 0.10$\,mas\,pixel$^{-1}$) at Gl~86B's separation.  Astrometric calibration for all images, both those taken with S13 and S27, was done using the $\Theta_1$~Ori~C system.  However, the S27 camera in particular has nonlinear distortion averaging a few tenths of a pixel, or $\sim$5~mas, and this distortion varies strongly across the field-of-view \citep{Plewa+Gillessen+Eisenhauer_2015}.  In many places, even near the center of the field-of-view, it is $\gtrsim$10~mas.  Though the distortion correction varies with time \citep{Plewa+Gillessen+Eisenhauer_2015}, it may be possible to improve the astrometry with a re-reduction of these archival data.  For our analysis here, we add 10~mas of astrometric uncertainty in quadrature with the published values of \cite{Lagrange+Beust+Udry+etal_2006}.  There does not appear to be a published distortion correction for the SDI camera on NACO; in the absence of one, we add the same 10~mas in uncertainty to these measurements.  \cite{Mugrauer+Neuhauser_2005} only provided one significant figure for their separation.  As a result, we round our separation uncertainty up to 20~mas.  

The four VLT/NACO data points that we use derive from three separate cameras.  Field distortion could have induced similar errors to more than one data point (depending on the location and orientation of the calibration field and science imaging), potentially introducing covariance between astrometric measurements.  

\cite{Farihi+Bond+Dufour+etal_2013} obtained precise {\it HST} astrometry in 2012, giving 
an 11 year baseline between the earliest and latest astrometry.  Even after our error inflation, the four VLT/NACO points from 2003-5 are only marginally consistent with the slope of the separation vs.~time relation of 57\,mas\,yr$^{-1}$ implied by the earliest and latest astrometry ($\chi^2 = 5.1$ for three degrees of freedom).  This suggests that our uncertainties for the NACO data might remain underestimated.  One additional piece of circumstantial evidence for underestimated errors in NACO astrometry comes from \cite{Peretti+Segransan+Lavie+etal_2018}.  Those authors re-reduced VLT/NACO imaging taken in SDI mode for HD~4747B and derived a separation uncertainty of 11~mas.  However, this point remained nearly $2\sigma$ discrepant with their orbital fit.

\subsection{HD~68017}

HD~68017A is a G~dwarf with an M~dwarf companion indicated by a nonlinear radial velocity trend and later discovered by direct imaging \citep{Crepp+Johnson+Howard+etal_2012}.  We use the radial velocity time series obtained with HIRES on Keck and published by \cite{Butler+Vogt+Laughlin+etal_2017}.  These measurements extend from 1997 to 2014, comprising 182 radial velocities with a median precision of 2.6~m\,s$^{-1}$. (We exclude one highly discrepant radial velocity measurement with very low counts from 2454547.8~JD.) The RVs show a strong trend with substantial curvature.  \cite{Crepp+Johnson+Howard+etal_2012} reported a photometric mass estimate of $0.15\pm0.01$\,\Msun based on the empirical mass-luminosity relation of \cite{Delfosse+Forveille+Segransan+etal_2000}, consistent with the lower limit of $>$0.08\,\Msun\ implied by the radial velocity amplitude.

There are only two relative astrometry epochs available for HD~68017, both from 2011--2012, and taken using NIRC2 on Keck \citep{Crepp+Johnson+Howard+etal_2012}.  The companion is relatively bright in the $H$~band, giving very small astrometric errors, though \cite{Crepp+Johnson+Howard+etal_2012} did not perform a full orbital analysis.  Perhaps due to the stellar nature of the companion it has not been targeted for subsequent astrometric follow-up.  

\cite{Bowler+Dupuy+Endl+etal_2018}, also using NIRC2, added 1~mas to the positional errors of both Gl~758A and its companion to account for uncertainties in the distortion correction.  We repeat that procedure here, adding in quadrature 1.4~mas to the separations and 0.\!\!\degree14\ to the PAs reported by \cite{Crepp+Johnson+Howard+etal_2012}, with Table~\ref{tab:companion_astrometry} giving the errors as used in our analysis.  \cite{Bowler+Dupuy+Endl+etal_2018} added another 4.3~mas in quadrature to the separation uncertainties for Gl~758B to achieve an acceptable reduced $\chi^2$ in their orbital fit.  We do not repeat that step here, though further error inflation may be warranted as the data were collected with the same instrument (NIRC2) as for Gl~758B.  

\subsection{Gl~758}

All of our data for Gl~758, apart from the {\it Gaia} and {\it Hipparcos} astrometry, were published in \cite{Bowler+Dupuy+Endl+etal_2018}.  Following that paper, we combine the HIRES radial velocity time series for Gl~758A \citep{Butler+Vogt+Laughlin+etal_2017} with data from the McDonald observatory and the Automated Planet Finder at Lick Observatory.  The McDonald measurements used the Tull \'echelle spectrograph \citep{Tull+MacQueen+Sneden+etal_1995} as part of a radial velocity planet search \citep{Cochran+Hatzes+Butler+etal_1997,Endl+Brugamver+Cochran+etal_2016}; the data for Gl~758A span 19 years with a typical uncertainty of 4.6~m\,s$^{-1}$.  Since 2013, Gl~758A has been observed with the Automated Planet Finder at Lick Observatory \citep[APF,][]{Vogt+Radovan+Kibrick+etal_2014} as part of a search for rocky planets \citep{Fulton+Weiss+Sinukoff+etal_2015}.  This program collected 250 spectra with a typical uncertainty of 1.4~m\,s$^{-1}$.  The Keck HIRES time series encompasses 262 spectra with a median uncertainty of 1.2~m\,s$^{-1}$.

The relative astrometry used by \cite{Bowler+Dupuy+Endl+etal_2018} is exclusively from the NIRC2 camera on the Keck telescope with natural guide star adaptive optics \citep{Wizinowich+Acton+Shelton+etal_2000}.  These observations comprise four epochs spanning 7.4 years, from May 2010 through October 2017.  \cite{Bowler+Dupuy+Endl+etal_2018} processed the data using the locally optimized combination of images algorithm \citep[LOCI,][]{Lafreniere+Marois+Doyon+etal_2007} for data taken in angular differential imaging mode \citep[ADI,][]{Marois+Lafreniere+Doyon+etal_2006}.  The data processing pipeline is described in \cite{Bowler+Liu+Shkolnik+etal_2015}; the distortion corrections were derived by \cite{Yelda+Lu+Ghez+etal_2010} for pre-2015 data and by \cite{Service+Lu+Campbell+etal_2016} for subsequent data, after a pupil realignment.  Detections of two background objects in the field were used to validate the astrometry.  The calibrated Keck/NIRC2 high-contrast imaging gives relative positions of Gl~758A and Gl~758B to an accuracy of 5~mas and marginally detect curvature in the orbit.  

\subsection{HR~7672}

HR~7672A was first monitored as part of the Lick planet search in 1987 \citep{Cumming+Marcy+Butler_1999}.  Since 1994, it has also been monitored by HIRES on Keck \citep{Butler+Vogt+Laughlin+etal_2017}.  HR~7672A displayed a strong linear trend in its radial velocities; the brown dwarf responsible for this trend was first imaged in 2001 \citep{Liu+Fischer+Graham+etal_2002}.  The published HIRES radial velocity time series now extends to 2014, a total baseline of twenty years \citep{Butler+Vogt+Laughlin+etal_2017}.  It comprises 188 measurements with a median uncertainty of 1.45~m\,s$^{-1}$.  Lick spectroscopy adds another 80 measurements with a reported median uncertainty of 1.0~m\,s$^{-1}$, extending the temporal coverage to nearly thirty years.  Curvature is clearly visible in the full time series, which now covers a substantial fraction of the orbit.

HR~7672B has a long history of being imaged with corresponding relative astrometry.  \cite{Liu+Fischer+Graham+etal_2002} obtained the first measurement in 2001 using the Hokupa'a adaptive optics system on Gemini-North \citep{Graves+Northcott+Roddier+etal_1998}; they also obtained two astrometric measurements using natural guide star adaptive optics on Keck \citep{Wizinowich+Acton+Shelton+etal_2000}.  \cite{Liu+Fischer+Graham+etal_2002} obtained a dynamical lower limit on the mass of 48~$M_{\rm Jup}$.  In 2002, \cite{Boccaletti+Chauvin+Lagrange+etal_2003} obtained astrometry and photometry using the PALAO adaptive optics system on Palomar \citep{Troy+Dekany+Brack+etal_2000}; they obtained a mass of 58--72~$M_{\rm Jup}$ from models of substellar evolution.  

In 2007, \cite{Serabyn+Mawet+Bloemhof+etal_2009} detected the brown dwarf using a small, well-corrected aperture on the Palomar-Hale telescope.  The small aperture leads to large astrometric uncertainties.  While we do include the measurement, the error bars are large enough that it does not significantly inform our fit.  Most recently, \cite{Crepp+Johnson+Fischer+etal_2012} obtained precise astrometry using NIRC2 on Keck.  Those authors also reduced NACO observations from 2007.  The NACO images, like the Palomar-Hale measurements, are too noisy to be of much value in our orbit fitting. 

\cite{Crepp+Johnson+Fischer+etal_2012} performed an orbit fit using both radial velocities and companion astrometry, and they obtained a mass of $m=68.7^{+2.4}_{-3.1}$~$M_{\rm Jup}$ using a model-derived host star mass of $M = 1.08\pm0.04$\ \Msun.

\begin{deluxetable}{lccccr}
\tablewidth{0pt}
\tablecaption{Summary of Direct Imaging Astrometry}
\tablehead{
    \colhead{Date} &
    \colhead{$\rho$} &
    \colhead{$\sigma_\rho$} &
    \colhead{PA} &
    \colhead{$\sigma_{\rm PA}$} &
    \colhead{Ref} \\
    \colhead{} &
    \multicolumn{2}{c}{(arcsec)} & 
    \multicolumn{2}{c}{(degrees)} &
    \colhead{}
    }
\startdata
\cutinhead{HD 4747}
2008.69 & 0.608 & 0.011 & 156.4 & 1.3 & P18 \\
2014.78 & 0.6065 & 0.0070 & 180.04 & 0.62 & C16 \\
2015.02 & 0.6066 & 0.0064 & 180.52 & 0.58 & C16 \\
2015.73 & 0.604 & 0.007 & 184.9 & 0.9 & C16 \\
2015.98 & 0.585 & 0.014 & 185.2 & 0.3 & C18 \\
2015.98 & 0.583 & 0.014 & 184.4 & 0.3 & C18 \\
2016.95 & 0.5944 & 0.0051 & 187.2 & 0.3 & P18 \\
2016.95 & 0.5950 & 0.0051 & 187.6 & 0.3 & P18 \\
2017.74 & 0.5812 & 0.0058 & 190.6 & 0.5 & P18 \\
2017.74 & 0.5808 & 0.0063 & 190.6 & 0.7 & P18 \\
\cutinhead{Gl 86}
2000.82 & 1.73  & 0.03  & 119   & 1   & E01 \\
2003.87 & 1.906 & 0.015 & 107.5 & 0.5 & L06 \\
2004.73 & 1.941 & 0.017 & 105.3 & 0.6 & L06 \\
2005.03 & 1.93  & 0.02  & 104.0 & 0.4 & M05 \\
2005.57 & 1.969 & 0.015 & 102.7 & 0.5 & L06 \\
2012.25 & 2.351 & 0.002 & 88.96 & 0.04 & F13 \\
\cutinhead{HD 68017}
2011.15 & 0.5945 & 0.0015 & 248.20 & 0.17 & C12b \\
2012.02 & 0.5746 & 0.0015 & 240.30 & 0.17 & C12b \\
\cutinhead{Gl 758}
2010.33 & 1.848 & 0.005 & 200.6 & 0.3 & B18 \\
2013.50 & 1.743 & 0.005 & 205.7 & 0.2 & B18 \\
2016.49 & 1.626 & 0.005 & 210.3 & 0.4 & B18 \\
2017.77 & 1.588 & 0.005 & 213.5 & 0.3 & B18 \\
\cutinhead{HR 7672}
2001.64 & 0.786 & 0.006 & 157.9 & 0.5 & L02 \\
2001.94 & 0.794 & 0.005 & 157.3 & 0.6 & L02 \\
2002.54 & 0.788 & 0.006 & 156.6 & 0.9 & B03 \\
2006.69 & 0.750 & 0.080 & 155.0 & 5.0 & S09 \\
2007.73 & 0.742 & 0.035 & 151.8 & 2.9 & C12a \\
2011.37 & 0.519 & 0.006 & 147.1 & 0.5 & C12a 
\enddata
\tablenotetext{*}{References abbreviated as: B03 \citep{Boccaletti+Chauvin+Lagrange+etal_2003}; B18 \citep{Bowler+Dupuy+Endl+etal_2018}; C12a \citep{Crepp+Johnson+Fischer+etal_2012}; C12b \citep{Crepp+Johnson+Howard+etal_2012}; 
C16 \citep{Crepp+Gonzales+Bechter+etal_2016};
C18 \citep{Crepp+Principe+Wolff+etal_2018};
E01 \citep{Els+Sterzik+Marchis+etal_2001};
F13 \citep{Farihi+Bond+Dufour+etal_2013}; 
L06 \citep{Lagrange+Beust+Udry+etal_2006};
L02 \citep{Liu+Fischer+Graham+etal_2002}; 
M05 \citep{Mugrauer+Neuhauser_2005};
P18 \citep{Peretti+Segransan+Lavie+etal_2018};
S09 \citep{Serabyn+Mawet+Bloemhof+etal_2009}.}
\label{tab:companion_astrometry}
\end{deluxetable}

\section{Host-Star Astrometry} \label{sec:gaia_astrometry}

This paper adds absolute astrometric measurements of the host stars to existing radial velocity data and direct imaging astrometry.  The host-star absolute astrometry comes from a combination of {\it Hipparcos} and {\it Gaia}, two satellite missions with a $\sim$25-year time baseline between them.   To measure acceleration we use the deviations between three proper motion measurements:
\begin{itemize}
    \item The {\it Hipparcos} proper motions near epoch 1991.25;
    \item The {\it Gaia} DR2 proper motions near epoch 2015.5; and
    \item The ${\it Gaia} - {\it Hipparcos}$ positional difference divided by the $\sim$25 year time baseline (hereinafter referred to as the scaled positional difference).
\end{itemize}
The long baseline between the missions makes them sensitive to companions with periods as long as centuries.  It also renders the scaled positional difference 
our most precise proper motion measurement.  The differences between these proper motion measurements probe the acceleration of the star in an inertial reference frame.  

\cite{Brandt_2018} has performed a cross-calibration of {\it Hipparcos} and {\it Gaia}, placing all three proper motion measurements in the reference frame defined by {\it Gaia} DR2 \citep{Gaia_General_2018, Gaia_Astrometry_2018}.  This reference frame is a close approximation of the International Celestial Reference System \citep[ICRS,][]{Ma+Arias+Eubanks+etal_1998, Fey+Gordon+Jacobs+etal_2015}.  Figure~1 of \cite{Brandt_2018} confirms that such a calibration is necessary: the proper motions taken directly from the catalogs are inconsistent with the standard assumptions of Gaussianity.  There are several components to the final cross-calibration:
\begin{itemize}
    \item Weights for a linear combination of the two {\it Hipparcos} reductions \citep{ESA_1997,vanLeeuwen_2007};
    \item Propagation of positions to their central epochs in each catalog;
    \item Local offsets between the reference frames defined by the {\it Hipparcos} proper motions, the {\it Gaia} DR2 proper motions, and the scaled positional differences; 
    \item A global error inflation term for {\it Hipparcos}, to be added diagonally to the published covariance matrices; and
    \item A local error inflation factor for the {\it Gaia}  DR2 covariance matrices (averaging $\sim$1.7 for the errors, or $\sim$1.7$^2$ for the covariances).
\end{itemize}
The \cite{Brandt_2018} catalog provides the three proper motions all in the reference frame of {\it Gaia} DR2.  Figure~9 of that paper demonstrates that the distribution of residuals do follow the assumed Gaussian distributions after applying these cross-calibrations.  Close binaries in which one star makes a non-negligible contribution to the flux can still be problematic.  Also, Figure 9 of \cite{Brandt_2018} shows that the lowest-precision stars in {\it Gaia} have non-Gaussian tails in their proper motion residuals.

For the stars presented in this paper, the scaled difference in right ascension and declination between {\it Hipparcos} or {\it Gaia} is easily the most precise proper motion measurement ($\mu_{\it HG}$).  We therefore adopt the differences between this and the {\it Hipparcos} or {\it Gaia} proper motion as our astrometric constraints on the host-star orbit, computing two differences from our three proper motions.  We define them here as, e.g.,
\begin{equation}
    \Delta \mu_{\alpha*,\it Hip} = \mu_{\alpha*,\it Hip} - \mu_{\alpha*,\it HG},
    \label{eq:pmdiff}
\end{equation}
where $\mu_{\alpha*,\it Hip}$ is the {\it Hipparcos} proper motion with the $\cos \delta$ factor included, and $\mu_{\alpha*,\it HG}$ is the position difference between {\it Hipparcos} and {\it Gaia} DR2 divided by the time baselines between the measurements.  

The {\it Hipparcos} and {\it Gaia} proper motions are almost entirely independent, apart from a tiny covariance arising from the use of {\it Gaia} parallaxes to improve the other {\it Hipparcos} astrometry \citep{Brandt_2018}.  For this reason, and because the scaled positional difference is so precise, we may neglect the (tiny) covariance between $\Delta \mu_{\alpha*,\it Hip}$ and $\Delta \mu_{\alpha*,\it Gaia}$.  If this covariance were significant, we would have to treat the three proper motions separately and solve for the proper motion of the system's center of mass.

Table \ref{tab:star_astrometry} lists the two proper motion differences and their associated covariance matrices, as computed from the \cite{Brandt_2018} catalog.

\begin{deluxetable*}{lcccccccccccr}
\tablewidth{0pt}
\tablecaption{Summary of {\it Hipparcos} and {\it Gaia} Astrometry}
\tablehead{
    \colhead{Star} &
    \colhead{Data} &
    \colhead{$\Delta \mu_\alpha$\tablenotemark{$\dagger$}} &
    \colhead{$\sigma[\mu_\alpha]$} &
    \colhead{$\Delta \mu_{\delta}$\tablenotemark{$\dagger$}} &
    \colhead{$\sigma[\mu_{\delta}]$} &
    \colhead{Correlation} &
    \colhead{Epoch, $\alpha$} &
    \colhead{Epoch, $\delta$}    \\
    \colhead{} &
    \colhead{Source} &
    \multicolumn{2}{c}{(mas/yr)} & 
    \multicolumn{2}{c}{(mas/yr)} &
    \colhead{Coefficient} &
    \colhead{year} &
    \colhead{year}    
}
\startdata
  HD~4747 & {\it Hip} & $  1.315$ & $0.737$ & $ -5.118$ & $0.588$ & $ 0.067$ & 1991.43 & 1991.67 \\
          & {\it Gaia} &  $  3.067$ & $0.552$ & $ -1.715$ & $0.588$ & $ 0.751$ & 2015.20 & 2015.19 \\
    Gl~86 & {\it Hip} &  $-15.072$ & $0.428$ & $ 12.726$ & $0.463$ & $-0.083$ & 1991.23 & 1991.38 \\
          & {\it Gaia} & $ 17.898$ & $0.134$ & $ -3.528$ & $0.115$ & $-0.072$ & 2015.77 & 2015.75 \\
 HD~68017 & {\it Hip} &  $  9.697$ & $0.925$ & $ -5.444$ & $0.524$ & $-0.170$ & 1991.07 & 1991.34 \\
          & {\it Gaia} & $-13.017$ & $0.159$ & $ -3.951$ & $0.099$ & $-0.219$ & 2015.76 & 2015.71 \\
   Gl~758 & {\it Hip} &  $  0.616$ & $0.475$ & $  1.432$ & $0.447$ & $-0.024$ & 1990.97 & 1991.21 \\
          & {\it Gaia} & $ -0.386$ & $0.061$ & $ -0.933$ & $0.071$ & $-0.036$ & 2015.62 & 2015.67 \\
 HR~7672  & {\it Hip} &  $ -2.430$ & $0.492$ & $  5.176$ & $0.516$ & $ 0.017$ & 1991.32 & 1991.13 \\
          & {\it Gaia} & $  4.561$ & $0.144$ & $ -7.080$ & $0.144$ & $ 0.095$ & 2015.60 & 2015.64 
   
\enddata
\tablenotetext{$\dagger$}{Defined as in Equation \eqref{eq:pmdiff}} 
\label{tab:star_astrometry}
\end{deluxetable*}

\section{Single Epoch Companion Masses} \label{sec:singleepochmasses}

The combination of {\it Hipparcos}, {\it Gaia} DR2, direct imaging astrometry, and radial velocity monitoring makes it possible to determine companion masses to high precision even for very long period systems, and without needing any external information about the host star's mass.  Together, these measurements give the projected separation of the companion relative to the host star $\rho$, the host star's astrometric acceleration $a_{\alpha \delta}$, and the host star's acceleration along the line-of-sight $a_{\rm RV}$ (as measured using radial velocity).  These two accelerations are in the inertial frame defined by the system's center of mass.  Together, $\rho$, $a_{\alpha \delta}$ and $a_{\rm RV}$ uniquely determine the companion mass via the equations
\begin{align}
    a_{\alpha\delta} &= \frac{G M_{\rm B}}{r_{\rm AB}^2} \cos \phi, \\
    a_{\rm RV} &= \frac{G M_{\rm B}}{r_{\rm AB}^2} \sin \phi, \quad {\rm and} \\
    \rho &= r_{\rm AB} \varpi \cos \phi,
\end{align}
where $\phi$ is the angle between the position vector separating the two bodies and the plane of the sky, $r_{\rm AB}$ is the absolute separation of the two bodies, $\varpi$ is the parallax, $M_B$ is the companion mass, and $G$ is the gravitational constant.  Assuming all the measurements can be approximated as representing the same orbital epoch, then combining these three equations gives
\begin{equation}
    M_B = \frac{\rho^2 \left(a_{\alpha\delta}^2 + a_{\rm RV}^2\right)^{3/2}}{\varpi^2 G a_{\alpha\delta}^2} .
    \label{eq:mass}
\end{equation}
If the errors on $\rho$, $a_{\alpha \delta}$, and $a_{\rm RV}$ are small and Gaussian (and the error on parallax is negligible), we may use standard propagation of errors to obtain
\begin{align}
    \frac{\sigma^2[M_B]}{M_B^2} \approx 4 &\frac{\sigma^2[\rho]}{\rho^2} 
    + 9 \frac{a_{\rm RV}^2 \sigma^2 [a_{\rm RV}]}{\big( a_{\rm RV}^2 + a_{\alpha\delta}^2 \big)^2} \nonumber \\
    &+ \left( 3 \frac{a_{\alpha\delta} \sigma [a_{\alpha\delta}]}{a_{\rm RV}^2 + a_{\alpha\delta}^2} - 2 \frac{\sigma[a_{\alpha\delta}]}{a_{\alpha\delta}} \right)^2.
    \label{eq:mass_err}
\end{align}
If the error on parallax is not negligible, it would have to also be propagated through the $a_{\alpha\delta}$ terms, making Equation \eqref{eq:mass_err} slightly more involved.  The parallax is needed to convert $a_{\alpha\delta}$ from angular to linear units; its observational uncertainty is negligible for all systems studied here.

The difference between the {\it Gaia} DR2 proper motion and the {\it Hipparcos}--{\it Gaia} scaled positional difference provides our highest signal-to-noise ratio measurement of acceleration in the plane of the sky.  This is not an instantaneous measurement, but is rather the difference between a mean proper motion over a baseline of $\sim$25 years and a mean proper motion during the {\it Gaia} observing period.  For long-period systems like the ones we study here, we may take the latter proper motion to be effectively instantaneous.  The central epoch of {\it Hipparcos} and {\it Gaia} is near 2003.5, while {\it Gaia} DR2 gives a nearly instantaneous measurement at an epoch close to 2015.5.  In this section, we consider the accelerations in right ascension and declination to be measured at the same time, equal to the average of the two central epochs (we will drop this approximation when fitting orbits in the next section).  Assuming nearly constant acceleration, we have
\begin{equation}
    a_{\alpha\delta}\left[t_a \right] \sim \frac{\Delta \mu_{\it Gaia} }{(t_{\it Gaia} - t_{\it Hip})/2},
\end{equation}
with 
\begin{equation}
    t_a = \frac{3 t_{\it Gaia} + t_{\it Hip}}{4}.
\end{equation}
Table \ref{tab:star_astrometry} lists the components of $\Delta \mu_{\it Gaia}$ and the epochs of the two catalogs for our sample stars.

For $a_{\rm RV}$, we fit a quadratic in ${\rm date} - (t_{\it Gaia} - t_{\it Hip})/2$ to the radial velocity data.  We take the linear term and its standard error for $a_{\rm RV}$ and $\sigma[a_{\rm RV}]$ after adjusting the jitter to obtain a reduced $\chi^2$ of unity.  For Gl~86, we first subtract the best-fit radial velocity signal from the inner planet Gl~86b as determined in the following section.  Finally, for the separation, we fit the functions
\begin{align}
    \Delta\alpha &= b_0 + b_1 t + \frac{1}{2} \gamma t^2 a_{\alpha} \quad {\rm and} \\
    \Delta{\delta} &= c_0 + c_1 t + \frac{1}{2}\gamma t^2 a_{\delta}
\end{align}
where $t$ is the date relative to our desired epoch of $t_a$, $\sqrt{b_0^2 + c_0^2}$ is the separation at our desired epoch, $a_{\alpha}$ and $a_{\delta}$ are the stellar astrometric accelerations in right ascension and declination, and we expect $\gamma = M_{\rm tot}/M_B$.  We account for the errors in $a_\alpha$ and $a_\delta$ by first ignoring them and fitting for the coefficients, and then iteratively updating the covariance matrix with the previous $\gamma$ and redoing the fit.  We then numerically compute the covariance matrix of $b_0$ and $c_0$ about the best $\chi^2$ to determine a standard error on separation at $t_a$.  

Once we have approximate measurements at a single epoch, we apply Equations \eqref{eq:mass} and \eqref{eq:mass_err} to obtain approximate masses with error bars for the objects in our sample with long orbital periods.  Table \ref{tab:approx_masses} shows our results.  We achieve a typical formal precision of $\sim$5\% even for systems with orbital periods of more than a century.  
However, our assumptions in deriving the equations of this section are not fully satisfied for the three systems in Table \ref{tab:approx_masses}, and are not at all satisfied by HD~4747 and HR~7672.  In general, a finite difference measurement of the astrometric acceleration will tend to underestimate the instantaneous acceleration.  This is particularly true if the orbital period is only a few times larger than the {\it Hipparcos}-{\it Gaia} temporal baseline.  As a result, the values in Table \ref{tab:approx_masses} systematically underestimate the true masses.  While this section demonstrates {\it why} we can obtain excellent companion masses, full orbit fits are necessary to obtain accurate constraints.  

\begin{deluxetable}{lcccr}
\tablewidth{0pt}
\tablecaption{Approximate Companion Masses from Short Orbital Arcs}
\tablehead{
    \colhead{Star} &
    \colhead{$\rho$} &
    \colhead{$a_{\alpha\delta}$} &
    \colhead{$a_{\rm RV}$} & 
    \colhead{$M_2$} \\
    \colhead{} &
    \colhead{(mas)} &
    \multicolumn{2}{c}{(m\,s$^{-1}$\,yr$^{-1}$)} &    
    \colhead{($M_{\rm Jup}$)} 
}  
\startdata
Gl\,86 & $2210 \pm 18$ & $76.5 \pm 0.6$\phn & $-61.0 \pm 3.0$\tablenotemark{$\dagger$} & $512 \pm 30$\tablenotemark{$\dagger$}\\
HD\,68017 & $655 \pm 8$ & $113.8 \pm 1.2$\phn\phn & \phs$6.77 \pm 0.35$ & $129 \pm 3$\phn \\
Gl\,758 & $1875 \pm 18$ & $6.10 \pm 0.42$ & $-2.44 \pm 0.16$ & $37 \pm 2$\phn  
\enddata
\tablenotetext{$\dagger$}{The best-fit planet signal from Section \ref{sec:orbitfitting} has been subtracted off.}
\label{tab:approx_masses}
\end{deluxetable}

\section{Orbit Fitting} \label{sec:orbitfitting}

The previous section shows that precise masses are possible with a nearly instantaneous measurement of the three-dimensional acceleration and projected separation.  However, all of the systems presented here trace out a non-negligible fraction of the orbit over the {\it Hipparcos}--{\it Gaia} baseline.  We therefore perform full orbit fitting of our combined astrometric and RV data sets in a way similar to \cite{Bowler+Dupuy+Endl+etal_2018}, with the main difference being that we include the astrometric acceleration between {\it Hipparcos} and {\it Gaia}~DR2.  

We use the parallel-tempering Markov chain Monte Carlo (PT-MCMC) ensemble sampler in \texttt{emcee~v2.1.0} \citep{Foreman-Mackey+Hogg+Lang+etal_2013} that is based on the algorithm described by \citet{Earl+Deem_2005}. Our results are based on the ``coldest'' of 30 chains, where the ``hottest'' chain effectively samples all of the allowed parameter space. We use 100 walkers to sample our 11-parameter model over $2\times10^5$ steps. Two of these parameters are the masses of the host ($M_{\star}$) and companion ($M_{\rm comp}$). Six parameters define the orbit: semimajor axis ($a$), inclination ($i$), PA of the ascending node ($\Omega$), mean longitude at a reference epoch ($t_{\rm ref}$) of 2455197.5~JD ($\lambda_{\rm ref}$; 2010 Jan 1 00:00 UT), and finally eccentricity ($e$) and the argument of periastron ($\omega$) fitted as $\sqrt{e}\sin{\omega}$ and $\sqrt{e}\cos{\omega}$. Two additional orbit parameters are needed for the RV data set to define the zero point of the system velocity (RV$_{\rm zero}$) and the intrinsic jitter ($\sigma_{\rm jit}$). Lastly, we include parallax ($\varpi$) as a fitted parameter in order to impose a Gaussian prior based on the DR2 measurement and its formal error. We assume log-flat priors for $a$, $M_{\star}$, $M_{\rm comp}$, and $\sigma_{\rm jit}$, a prior of $\sin{i}$ for inclination, and uniform priors for all other fitted parameters. For objects with RV data sets from more than one instrument, we use two additional parameters for RV$_{\rm zero}$ and $\sigma_{\rm jit}$ for each extra instrument. The final likelihood used in our MCMC is
\begin{align}
    \ln {\cal L} =  &-0.5 \left( \chi^2_\rho + \chi^2_\theta + \chi^2_{\rm RV} + \chi^2_\varpi + \chi^2_{\it G} + \chi^2_{\it H} \right)\nonumber \\
    &+ \ln\left[\sin[i]\right] - \ln[a] - \ln\left[M_{\star}\right] - \ln\left[M_{\rm comp}\right]
\end{align}
with 
\begin{align}
    \chi^2_\rho &= \sum_{k=1}^{N_{\rm ast}} \frac{\left( \rho_k-\rho\left[t_k\right] \right)^2}{\sigma^2 [\rho_k]} \\
    \chi^2_\theta &= \sum_{k=1}^{N_{\rm ast}} \frac{\left( \arctan\left[\sin\left[\theta_k-\theta\left[t_k\right]\right],\cos\left[\theta_k-\theta\left[t_k\right]\right]\right] \right)^2}{\sigma^2[\theta_k]} \\
    \chi^2_{\rm RV} &= \sum_{j=1}^{N_{\rm inst}}\sum_{k=1}^{N_{\rm RV}} \frac{\left({\rm RV}_{{\rm rel},k}+{\rm RV}_{{\rm zero},j}-{\rm RV}\left[t_k\right] \right)^2}{\sigma^2[{\rm RV}_k] + \sigma^2_{{\rm jit}, j}} \label{eq:chisq_RV} \nonumber \\
    & \qquad + \sum_{j=1}^{N_{\rm inst}}\sum_{k=1}^{N_{\rm RV}} \ln \left[ \sigma^2[{\rm RV}_k] + \sigma^2_{{\rm jit},j} \right] \\
    \chi^2_\varpi &= \frac{\left( \varpi-\varpi_{\rm DR2} \right)^2}{\sigma^2[\varpi_{\rm DR2}]} \\
    \chi^2_{\it G} &= \left(\Delta \mu_{\alpha, {\it G}} - \Delta \mu_{\alpha,{\it G}^\prime} \right)^2 C^{-1}_{\alpha \alpha,\it G} \nonumber \\
    &{\quad} + \left(\Delta \mu_{\delta, {\it G}} - \Delta \mu_{\delta,{\it G}^\prime} \right)^2 C^{-1}_{\delta \delta,\it G} \nonumber \\
    &{\quad} + 2 \left(\Delta \mu_{\alpha, {\it G}} - \Delta \mu_{\alpha,{\it G}^\prime} \right)\left(\Delta \mu_{\delta, {\it G}} - \Delta \mu_{\delta,{\it G}^\prime} \right) C^{-1}_{\alpha \delta,\it G} \label{eq:chisq_gaia_orbitfit}
    \end{align}
    \begin{align}
    \chi^2_{\it H} &= \left(\Delta \mu_{\alpha, {\it H}} - \Delta \mu_{\alpha,{\it H}^\prime} \right)^2 C^{-1}_{\alpha \alpha,\it H} \nonumber \\
    &{\quad} + \left(\Delta \mu_{\delta, {\it H}} - \Delta \mu_{\delta,{\it H}^\prime} \right)^2 C^{-1}_{\delta \delta,\it H} \nonumber \\
    &{\quad} + 2 \left(\Delta \mu_{\alpha, {\it H}} - \Delta \mu_{\alpha,{\it H}^\prime} \right)\left(\Delta \mu_{\delta, {\it H}} - \Delta \mu_{\delta,{\it H}^\prime} \right) C^{-1}_{\alpha \delta,\it H} \label{eq:chisq_hip_orbitfit}
\end{align}
In Equations \eqref{eq:chisq_gaia_orbitfit} and \eqref{eq:chisq_hip_orbitfit}, $\Delta \mu_{\alpha,{\it G}^\prime}$, $\Delta \mu_{\delta,{\it G}^\prime}$, $\Delta \mu_{\alpha,{\it H}^\prime}$ and $\Delta \mu_{\delta,{\it H}^\prime}$ refer to a model orbit's differences between the proper motion at either the {\it Gaia}~DR2 epoch near 2015.5 or the {\it Hipparcos} epoch near 1991.25 and the mean proper motion between the {\it Hipparcos} and {\it Gaia} epochs.  The $C^{-1}$ of Equations \eqref{eq:chisq_gaia_orbitfit} and \eqref{eq:chisq_hip_orbitfit} are the relevant elements of the inverse of the covariance matrix; the covariance matrices themselves are computed from the uncertainties and correlations listed in Table \ref{tab:star_astrometry}.  We omit the $\ln \sigma^2$ factors from all of the equations above apart from Equation \eqref{eq:chisq_RV} for the radial velocities.  The jitter parameters $\sigma^2_{{\rm jit},j}$ are the only uncertainties we vary in our MCMC; the other variances are constants and do not affect the relative likelihood of two sets of parameters.  

When modeling the orbits, we attempt to account for the fact that {\it Hipparcos} and {\it Gaia} do not provide truly instantaneous proper motions.  We calculate the proper motions
$\Delta \mu_{\alpha,{\it G}^\prime}$, $\Delta \mu_{\delta,{\it G}^\prime}$, $\Delta \mu_{\alpha,{\it H}^\prime}$ and $\Delta \mu_{\delta,{\it H}^\prime}$
using a quadratic fit to the star's modeled position, centered on the epoch provided by the HGCA and with a time baseline corresponding to the 3.4-year duration of {\it Hipparcos} or the 22 months of {\it Gaia} DR2, as appropriate.  

For Gl~86, the RV data set includes a signal from the close-in Jupiter-mass planet, so we include five more parameters in our orbit model: orbital period ($P_{\rm pl}$), mean longitude at $t_{\rm ref}$ ($\lambda_{\rm ref, pl}$), eccentricity ($e_{\rm pl}$) and argument of periastron ($\omega_{\rm pl}$) fitted as $\sqrt{e_{\rm pl}}\sin{\omega_{\rm pl}}$ and $\sqrt{e_{\rm pl}}\cos{\omega_{\rm pl}}$, and the RV semiamplitude of the planet's orbit ($K_{1, {\rm pl}}$). We assume log-flat priors on $P_{\rm pl}$ and $K_{1, {\rm pl}}$ and uniform priors on the other three parameters.

In our PT-MCMC analysis, we have experimented with different chain lengths.  We found that after $2\times10^5$, steps our 100-walker chains had stabilized in the mean and rms of the posteriors of each of the model parameters for all objects analyzed. We saved every 50th step of our chains and discarded the first 75\% of the chain as the burn-in portion, leaving $10^5$ PT-MCMC samples in the cold chain.  Tables~\ref{tbl:mcmc-HD4747}--\ref{tbl:mcmc-HR7672} list information on the posterior distributions of our fitted parameters, as well as parameters that can be directly computed from them. The 1- and 2-$\sigma$ confidence intervals are computed as the minimum range in that parameter that contains 68.3\% and 95.4\% of the values, respectively. The best-fit solution quoted is the one with the maximum posterior probability density (likelihood times prior).  

Figure~\ref{fig:orbit-sky} shows the relative orbits of all systems on the sky. Figures~\ref{fig:HD4747-corner}, \ref{fig:GL86-corner}, \ref{fig:HD68017-corner}, \ref{fig:GL758-corner}, and \ref{fig:HR7672-corner} display the companion mass posterior and the most relevant parameter correlations; Figures~\ref{fig:HD4747-hist}, \ref{fig:GL86-hist}, \ref{fig:HD68017-hist}, \ref{fig:GL86-hist}, and \ref{fig:HR7672-hist} display all other marginalized parameter posteriors. Figures~\ref{fig:HD4747-RV}, \ref{fig:GL86-RV}, \ref{fig:HD68017-RV}, \ref{fig:GL758-RV}, \ref{fig:HR7672-RV} show the RV orbits over the full period of the best-fit orbit as well as zoomed in plots of each RV data set.  Figures~\ref{fig:HD4747-ast}, \ref{fig:GL86-ast}, \ref{fig:HD68017-ast}, \ref{fig:GL758-ast}, and \ref{fig:HR7672-ast} show the relative astrometry and astrometric accelerations compared to our orbit fits. In all of the aforementioned figures, the best-fit orbit is shown as a thick, black line, and 100 randomly drawn orbits from the MCMC posterior are plotted as thin lines colored according to the corresponding companion mass from low mass (pink) to high mass (green).

\section{Results and Discussion} \label{sec:results}

In this section we discuss the results of our orbital fits to each star.  For each of the systems except HD~4747, the mass of the secondary has the tightest constraint (for HD~4747 it is the orbital period).  For Gl~758B, HD~68017B, and Gl~86B, these masses represent major improvements on the constraints available in the literature.  

For the three ultracool dwarfs in our sample, we also compare our dynamical masses to the predictions of some commonly-used brown dwarf models.  Our substellar cooling models are Cond \citep{Baraffe+Chabrier+Barman+etal_2003}, the models of \cite{Saumon+Marley_2008} with three different cloud treatments, and the \cite{Burrows+Marley+Hubbard+etal_1997} grid.  The three families of models from \cite{Saumon+Marley_2008} have no clouds (``SM-NC''), a shift in cloud cover at the L/T transition (``SM-Hybrid''), or thick clouds at all temperatures (``SM-f2'').  We use the age posteriors derived from both stellar activity and isochrone fits, the green dot-dashed curves in the lower-middle panels of Figure \ref{fig:age_mass_posteriors}.  Probability distributions of age and bolometric luminosity combine with a given brown dwarf cooling model to provide the model-derived mass distributions.  

\begin{figure*}
    \centering
    \includegraphics[width=0.88\textwidth]{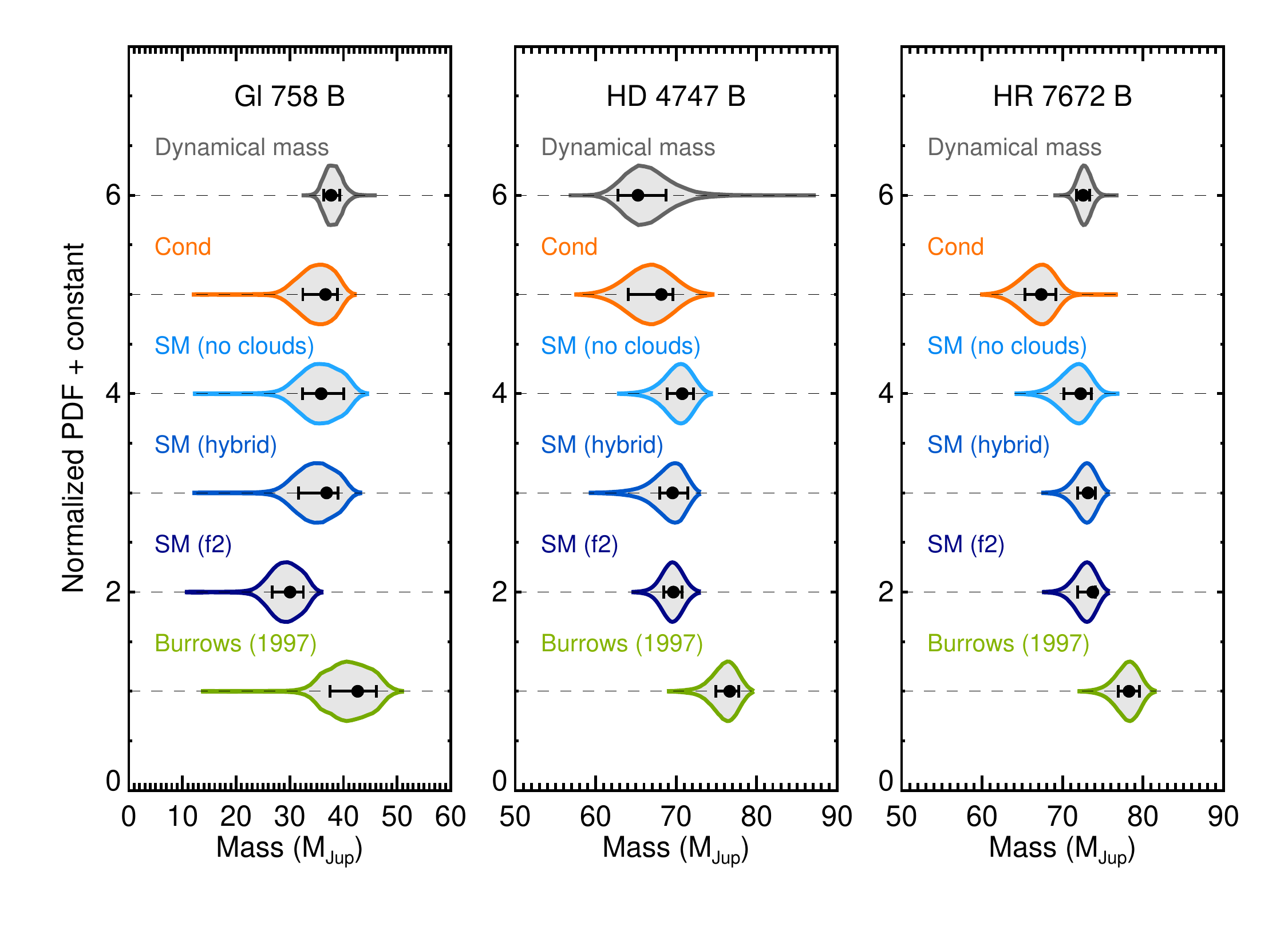} \\
    \vskip -0.18 truein
    \caption{Comparison of the dynamical masses of our ultracool dwarfs with masses inferred from the companion's age and bolometric luminosity using a range of substellar cooling models.  We adopt the activity+isochrone age posteriors, the dot-dashed green curves from the lower-middle panels of Figure~\ref{fig:age_mass_posteriors}.  The brown dwarf evolution models are Cond \citep{Baraffe+Chabrier+Barman+etal_2003}, \cite{Saumon+Marley_2008} with three treatments of clouds, and \cite{Burrows+Marley+Hubbard+etal_1997}.  For each object, at least one substellar cooling model agrees well with our dynamical mass measurement.  This good agreement for Gl~758B is in contrast to some earlier results, and is due to our low measured mass and old inferred age.}
    \label{fig:modelcompare}
\end{figure*}

Figure \ref{fig:modelcompare} shows our results.  We obtain good agreement between dynamical and model-derived masses for all of our systems for at least one model, although some models show better agreement than others.

\subsection{HD~4747}

HD~4747B is an excellent test case for our method because its orbit has been studied extensively in previous work. Our dynamical mass of $m=66.2^{+2.5}_{-3.0}$~$M_{\rm Jup}$ is in reasonable agreement with published values of $m=65.3^{+4.4}_{-3.3}$\,\Mjup\ \citep{Crepp+Principe+Wolff+etal_2018} and $m=70.2\pm 1.6$\,\Mjup\ \citep{Peretti+Segransan+Lavie+etal_2018} that had to assume a mass for the host star. We find a dynamical mass for HD~4747A of $M=0.82^{+0.07}_{-0.08}$\,\Msun\ that agrees well with published isochrone analysis as well as our own. The $\chi^2$ of the best-fit orbit compared to the input astrometric measurements indicates a good fit and accurate measurement errors, with values of 7.39 and 5.89 for separations and PAs (eight measurements each) and 3.09 for the four proper motion differences.

Our orbital parameters are generally in good agreement with the results of \cite{Peretti+Segransan+Lavie+etal_2018} and somewhat discrepant with \cite{Crepp+Principe+Wolff+etal_2018}. This is most likely because we used the same relative astrometry as \cite{Peretti+Segransan+Lavie+etal_2018}, and they noted the same differences in orbital parameters compared to \cite{Crepp+Principe+Wolff+etal_2018} that we find. Despite this fact, our companion mass $m=66.2^{+2.5}_{-3.0}$~$M_{\rm Jup}$ agrees better with the $m=65.3^{+4.4}_{-3.3}$\,\Mjup\ found by \cite{Crepp+Principe+Wolff+etal_2018} than the $m = 70.2\pm 1.6$\,\Mjup\ derived by \cite{Peretti+Segransan+Lavie+etal_2018}.  Our best-fit $
sin i$ differs by 5\% from that found by \cite{Peretti+Segransan+Lavie+etal_2018}, accounting for most of the discrepancy in mass and illustrating
the importance of including the {\it Hipparcos}--{\it Gaia} acceleration. Our model-independent dynamical mass has somewhat larger errors (4\%) compared to the 2.3\% mass of \cite{Peretti+Segransan+Lavie+etal_2018}.  This may be a consequence of fitting for the host star mass directly rather than using an isochronal mass.  

HD~4747AB has the shortest orbital period of any of the systems in our sample ($34.0^{+0.8}_{-1.0}$\,yr), but its period is still much longer than either the {\it Hipparcos} or {\it Gaia} mission duration. The period is comparable to the temporal baseline between the missions, resulting in a large astrometric signal. This star also has low precision in {\it Gaia}, perhaps because it heavily saturated the {\it Gaia} CCDs.  Despite this, our results for HD~4747 demonstrate the ability of {\it Hipparcos} and {\it Gaia} to measure dynamical masses of systems with periods of only a few decades. HD~4747 will be an excellent test case for future {\it Gaia} data releases, which will both measure a larger fraction of the orbit and improve the processing for very bright stars.

Figure \ref{fig:modelcompare} compares the dynamical mass of HD~4747B with mass predictions from theoretical models based on the objects age and luminosity.  We adopt the age posterior derived using both stellar activity and isochrone fitting.  Our bolometric luminosity of  $\log [L_{\rm bol}/L_\odot] = -4.54\pm 0.06$~dex is derived using the $K_s$-band absolute magnitude of \cite{Crepp+Gonzales+Bechter+etal_2016} together with the empirical magnitude--luminosity of \cite{Dupuy+Liu_2017} based on the luminosities of \cite{2015ApJ...810..158F}.  All of the brown dwarf cooling models except for those of \cite{Burrows+Marley+Hubbard+etal_1997} provide consistent masses within the errors.  

\subsection{Gl~86}

The significance of proper motion acceleration in the difference between {\it Gaia} proper motion for Gl~86 and the scaled {\it Hipparcos}--{\it Gaia} positional difference is over 130\,$\sigma$.  Even at such a high significance, the formal goodness-of-fit is excellent for relative and absolute astrometry, making Gl~86 a robust demonstration of the power of using {\it Hipparcos} and {\it Gaia} astrometry to determine orbits.

Thanks to the $>$100\,$\sigma$ significance of acceleration in {\it Hipparcos} and {\it Gaia}, we obtain a mass on Gl~86B of $0.595 \pm 0.010$~$M_\odot$, better than 2\% precision.  This model-independent mass is broadly consistent with the dynamical analysis of \cite{Lagrange+Beust+Udry+etal_2006}, who found a possible range in mass of 0.48--0.62\,\Msun, and our orbital eccentricity of $0.53_{-0.03}^{+0.04}$ is consistent with their conclusion that $e>0.4$. Our mass is also consistent with the value of $0.59\pm0.01$\,\Msun\ derived in the photometric and spectroscopic analysis of Gl~86B by \cite{Farihi+Bond+Dufour+etal_2013}.  According to their analysis, this mass implies a progenitor mass of $1.9\pm0.1$\,\Msun, main sequence lifetime of $1.4\pm0.2$\,Gyr, and cooling age of $1.25\pm0.05$\,Gyr.  A total system age of 2.7\,Gyr agrees well with our stellar age for Gl~86A, and \cite{Farihi+Bond+Dufour+etal_2013} also noted that it agrees with the age--activity relationship of \cite{Mamajek+Hillenbrand_2008}. In contrast, \cite{Fuhrmann+Chini+Buda+etal_2014} noted the tension between such a young age and the chemistry and kinematics of Gl~86A, which they determined was intermediate between the thin and thick disks implying an age of $\sim$10\,Gyr. Such an old age would require a lower-mass progenitor, and thus a lower mass for the white dwarf of $0.49 \pm 0.02$~$M_\odot$. Our model-independent mass for the white dwarf seems to be inconsistent with the scenario proposed by \cite{Fuhrmann+Chini+Buda+etal_2014}, but the chemical and kinematic peculiarity of Gl~86A calls for further study.

Some caution is still warranted in using dynamical masses that rely on relative astrometry from different instruments that have been calibrated to varying levels of accuracy (HST/WFC3, VLT/NACO, and ADONIS/SHARPII here). For example,  \cite{Bowler+Dupuy+Endl+etal_2018} found that even instruments thought to be well-calibrated sometimes delivered astrometry in disagreement with each other, but that was only evident when there were many degrees of freedom (many measurements per instrument). This is not the case for Gl~86, where each literature source provides only one or a few measurements. Given that Gl~86 has the highest {\it precision} of all our masses, it is worth noting that it may not have comparable {\it accuracy} due to the heterogeneous relative astrometry used in our orbit analysis.

\subsection{HD~68017}

HD~68017 represents the second highest-mass companion and the second highest signal-to-noise ratio in the {\it Hipparcos}-{\it Gaia} catalog of the objects we study here.  We obtain a dynamical mass of $0.147 \pm 0.003$~$M_\odot$ for the secondary.  Our $\chi^2$ for the absolute astrometry provided by {\it Hipparcos} and {\it Gaia} is somewhat high at 9.6 for four measurements.  It is difficult to assign a number of degrees of freedom to the absolute astrometry, as the parameters are jointly constrained by absolute and relative astrometry and by radial velocities.  For four degrees of freedom, a $\chi^2$ value of 9.6 occurs about 5\% of the time.  

Our adopted separation uncertainties are inflated over the values published by \cite{Crepp+Johnson+Howard+etal_2012}, but are still much lower than the uncertainties of 5~mas ultimately adopted by \cite{Bowler+Dupuy+Endl+etal_2018} for their orbital fit of Gl~758B.  Those authors added 4.3~mas in quadrature with their uncertainties to achieve a reduced $\chi^2$ of unity on the relative astrometry.  Adding a similar amount of uncertainty to our separation would allow for a satisfactory fit to the {\it Hipparcos} and {\it Gaia} astrometry.  Because the system is accelerating through the {\it Hipparcos} and {\it Gaia} measurements, the exact measured proper motions will also depend somewhat on the distribution of astrometric epochs.

\cite{Crepp+Johnson+Howard+etal_2012} originally estimated the mass of HD~68017B from isochrones \citep[$0.16\pm0.02$\,\Msun,][]{2008ApJS..178...89D} and an empirical mass relation \citep[$0.15\pm0.01$\,\Msun,][]{Delfosse+Forveille+Segransan+etal_2000}. Our dynamical mass is in excellent agreement with these values and will help refine the mass--metallicity--magnitude relation in the future given that it is a companion to a metal-poor G~dwarf \citep[${[\rm Fe/H]}= -0.44\pm0.03$\,dex,][]{Crepp+Johnson+Howard+etal_2012}.

\subsection{Gl~758}

Our mass of $m = 38.1^{+1.7}_{-1.5}$~$M_{\rm Jup}$ for Gl~758B, obtained with the aid of the cross-calibrated astrometric catalog of \cite{Brandt_2018}, significantly improves previous constraints.  \cite{Bowler+Dupuy+Endl+etal_2018} obtained a mass of $m = 42^{+19}_{-7}$~$M_{\rm Jup}$ from the same companion astrometry and radial velocity measurements that we use but assuming a host star mass prior of $0.97\pm0.02$\,\Msun.  \cite{Calissendorff+Janson_2018} refined this constraint to $m = 42.4^{+5.6}_{-5.0}$~$M_{\rm Jup}$ by using the {\it Hipparcos} and {\it Gaia} DR2 proper motions and \cite{Bowler+Dupuy+Endl+etal_2018} posteriors.  They did not, however, use the scaled positional difference (the most precise proper motion measurement), nor did they cross-calibrate the {\it Hipparcos} and {\it Gaia} catalogs.  

Our mass is firmly on the low end of that determined by \cite{Bowler+Dupuy+Endl+etal_2018}, consistent with their lower mass limit, as well as \cite{Calissendorff+Janson_2018}.  
In Section \ref{sec:ages_masses}, we show that the activity and rotation of Gl~758A favor an age $\gtrsim$6~Gyr, while isochrone fitting, depending on the band(s) used, is consistent with any age.  We favor both an older system age and a lower companion mass than previous work, eliminating much of the tension with evolutionary models of brown dwarfs.  Figure \ref{fig:modelcompare} demonstrates that Cond, the \cite{Burrows+Marley+Hubbard+etal_1997} models, and the \cite{Saumon+Marley_2008} models (apart from those with thick clouds at all temperatures) all make predictions consistent with our dynamical mass and age posteriors.

Gl~758B is the lowest luminosity brown dwarf ($\log[\Lbol/\Lsun] = -6.07\pm0.03$\,dex, \citealt{Bowler+Dupuy+Endl+etal_2018}) for which a precise dynamical mass has been measured, as the sample of \cite{Dupuy+Liu_2017} extends to $-5.0$\,dex, and the components of $\epsilon$~Ind~B \citep{2018arXiv180709880D} have luminosities of $-4.71$\,dex and $-5.35$\,dex \citep{2010A&A...510A..99K}.

\subsection{HR~7672}

\begin{figure*}
\centerline{
\includegraphics[width=0.35\linewidth]{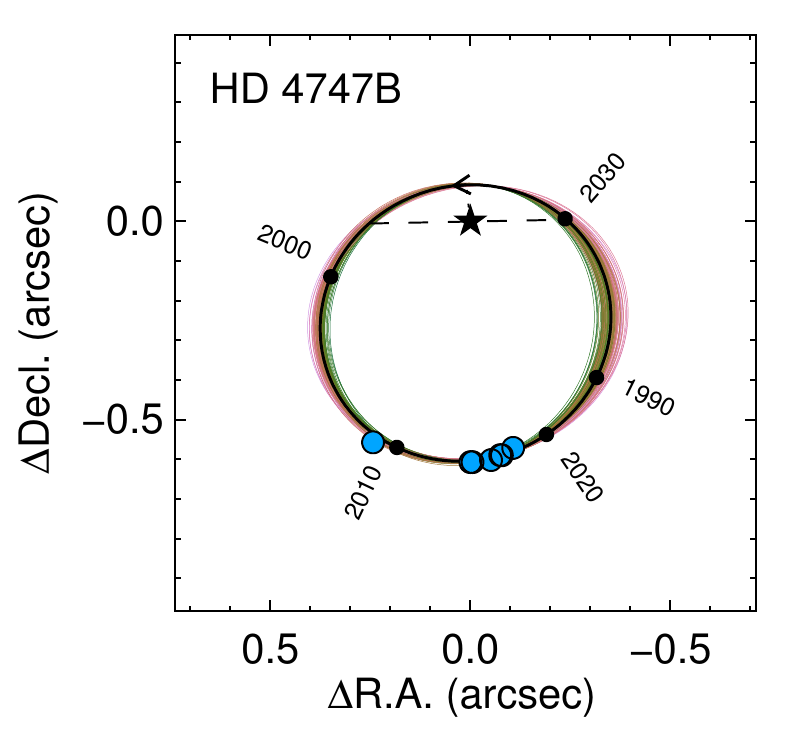}
\hskip -1.9in
\includegraphics[width=0.35\linewidth]{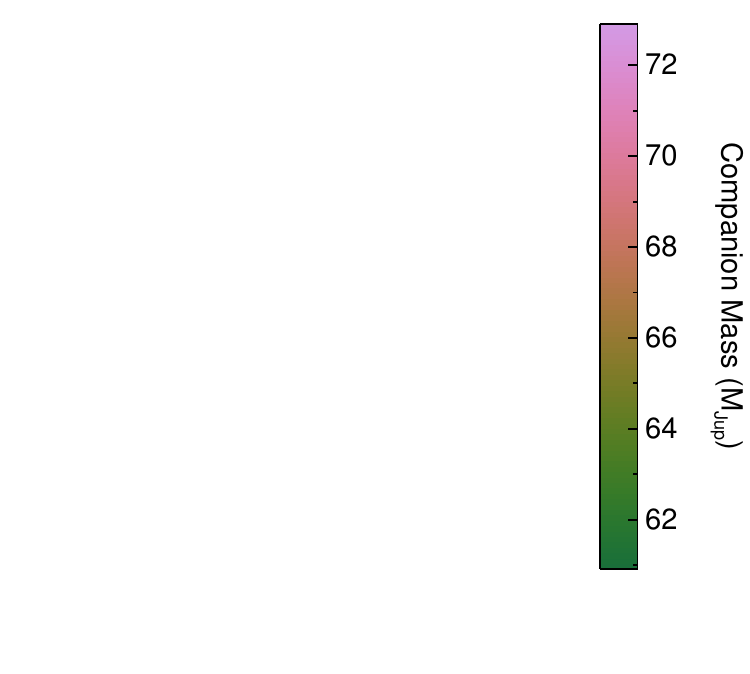}
\hskip 0.5in
\includegraphics[width=0.35\linewidth]{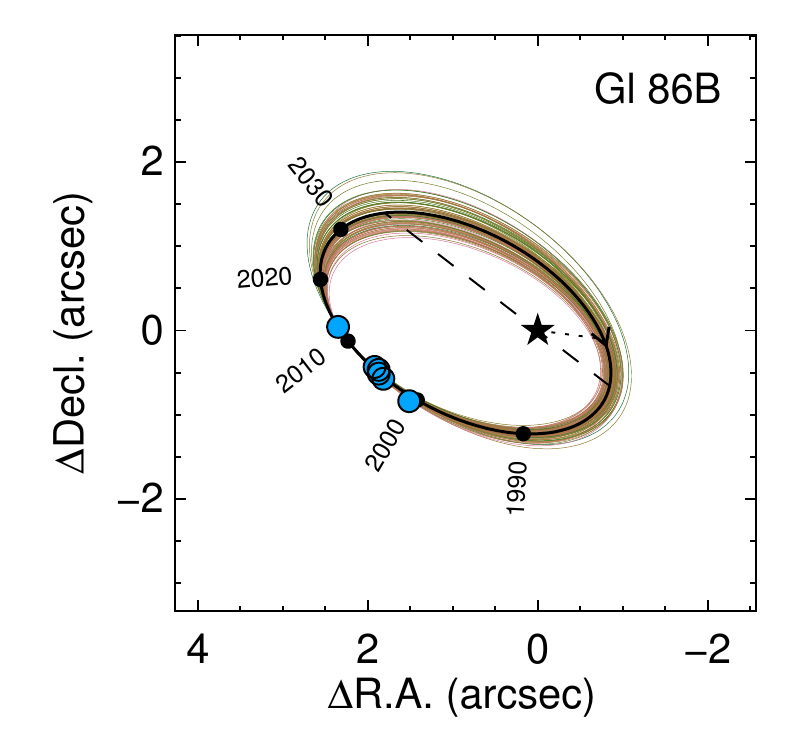}
\hskip -1.9in
\includegraphics[width=0.35\linewidth]{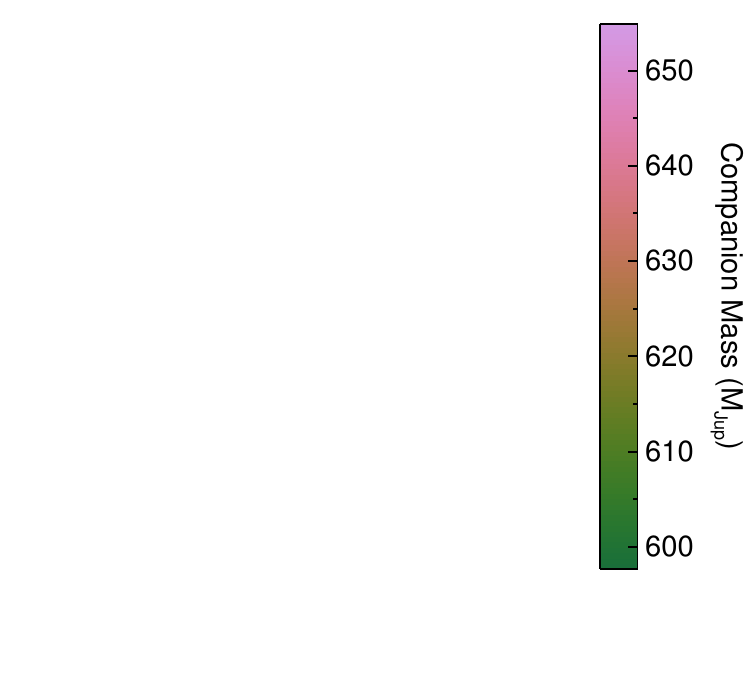}
}
\centerline{
\includegraphics[width=0.35\linewidth]{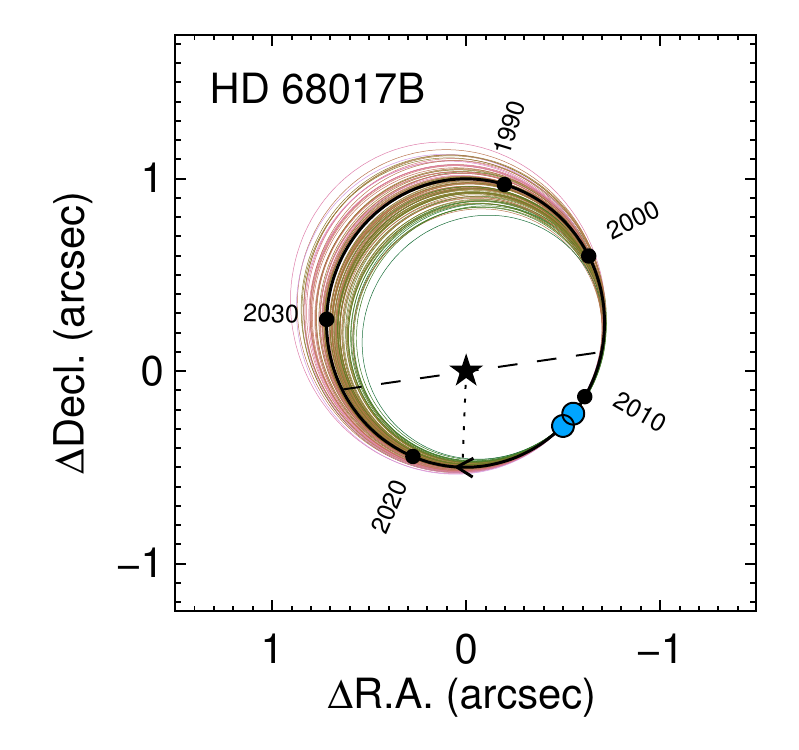}
\hskip -1.9in
\includegraphics[width=0.35\linewidth]{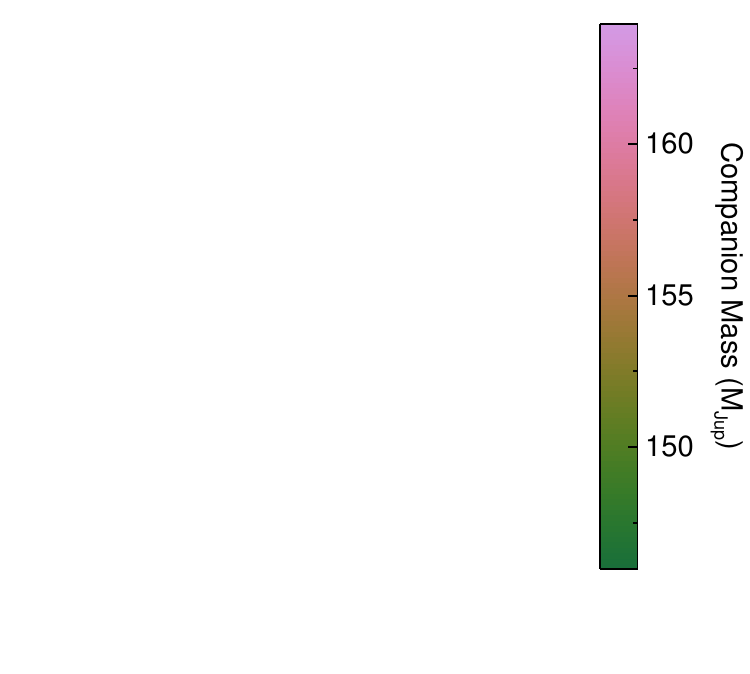}
\hskip 0.5in
\includegraphics[width=0.35\linewidth]{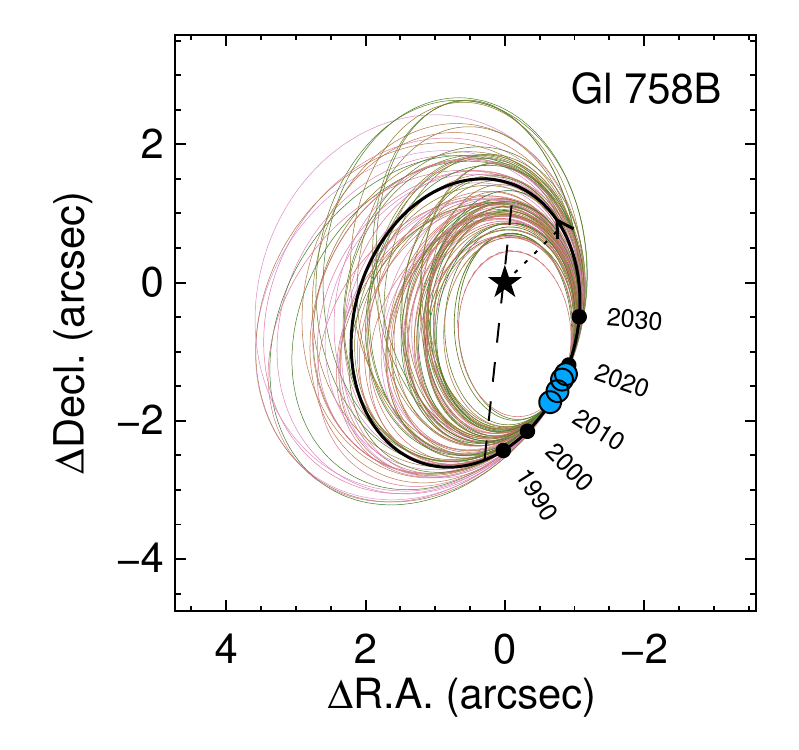}
\hskip -1.9in
\includegraphics[width=0.35\linewidth]{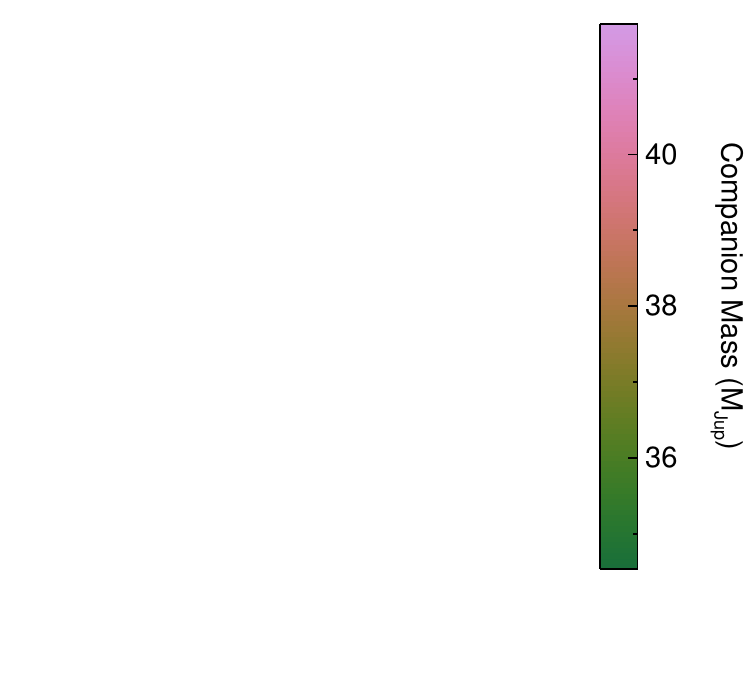}
}
\centerline{
\includegraphics[width=0.35\linewidth]{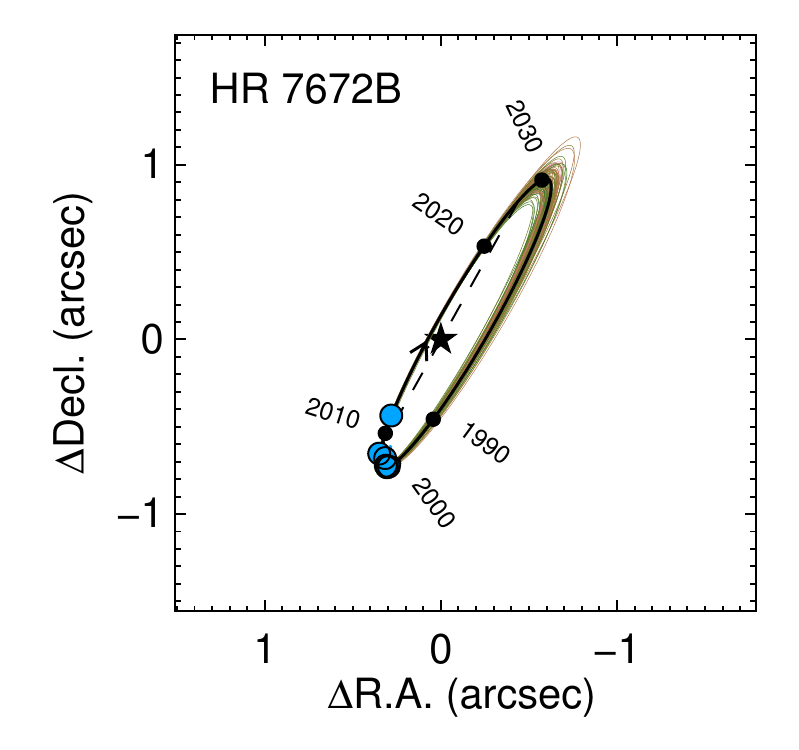}
\hskip -1.9in
\includegraphics[width=0.35\linewidth]{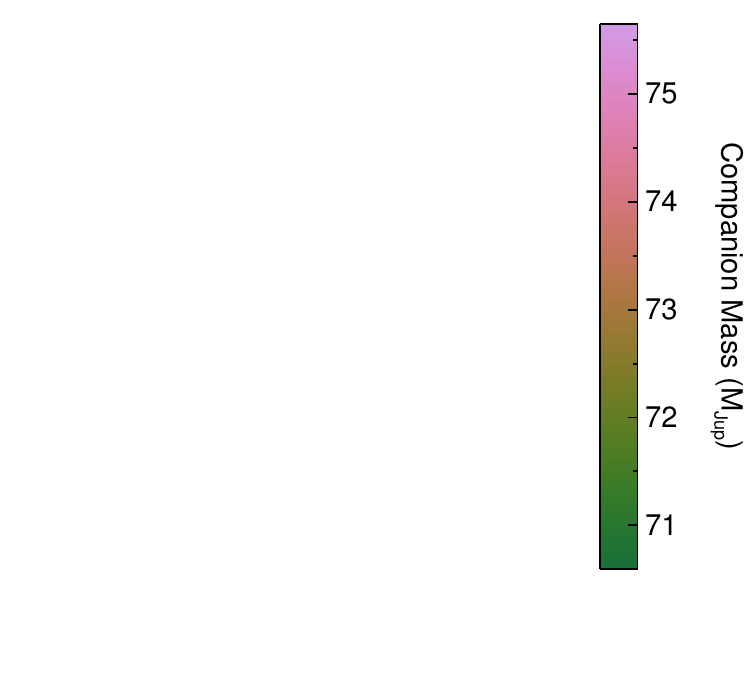}
}
\vskip 0.0 truein
\caption{Relative astrometric orbits of companions. The thick black line is the highest likelihood orbit and thin lines are 100 orbits drawn randomly from our posterior distribution colored according to companion mass from pink (low mass) to green (high mass). Dotted lines connect the host star to the periastron passage. Arrows plotted at periastron passage indicate orbital direction. Dashed lines indicate the line of nodes, that is the intersection of orbital and sky planes. Past and future points along the orbit are indicated by small black dots. Relative astrometry measurements are plotted as larger filled circles, where measurement errors are typically smaller than the plotted symbols.}
\label{fig:orbit-sky}
\end{figure*}

\begin{figure*}
\includegraphics[width=1.0\linewidth]{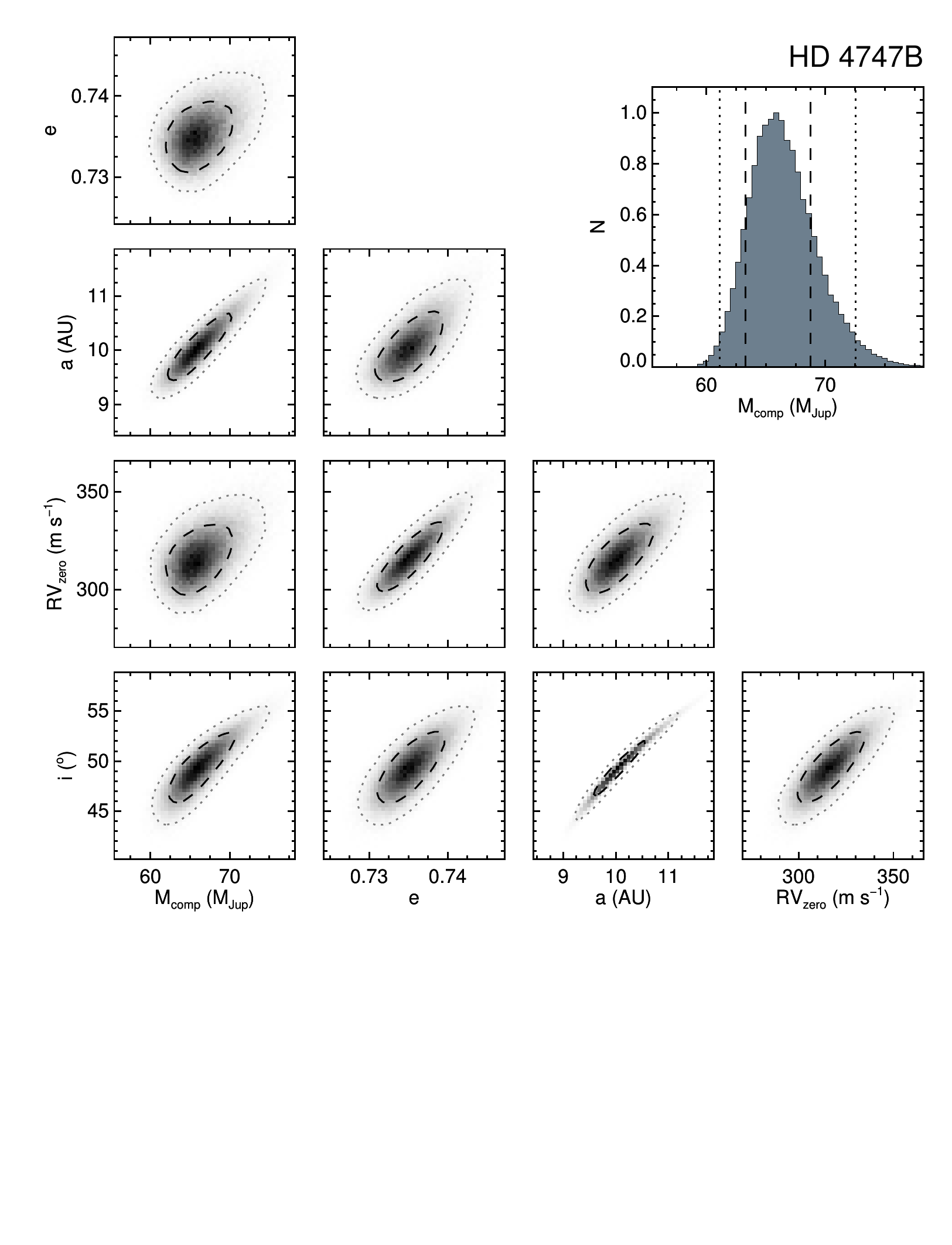}
\vskip -2.25 truein
\caption{Joint posterior distributions for selected orbital parameters. Dark dashed contours indicate 1$\sigma$ ranges and lighter dotted contours indicate 2$\sigma$ ranges. The top right panel shows the posterior distribution of the companion's mass.}
\label{fig:HD4747-corner}
\end{figure*}

\begin{figure*}
\includegraphics[width=1.0\linewidth]{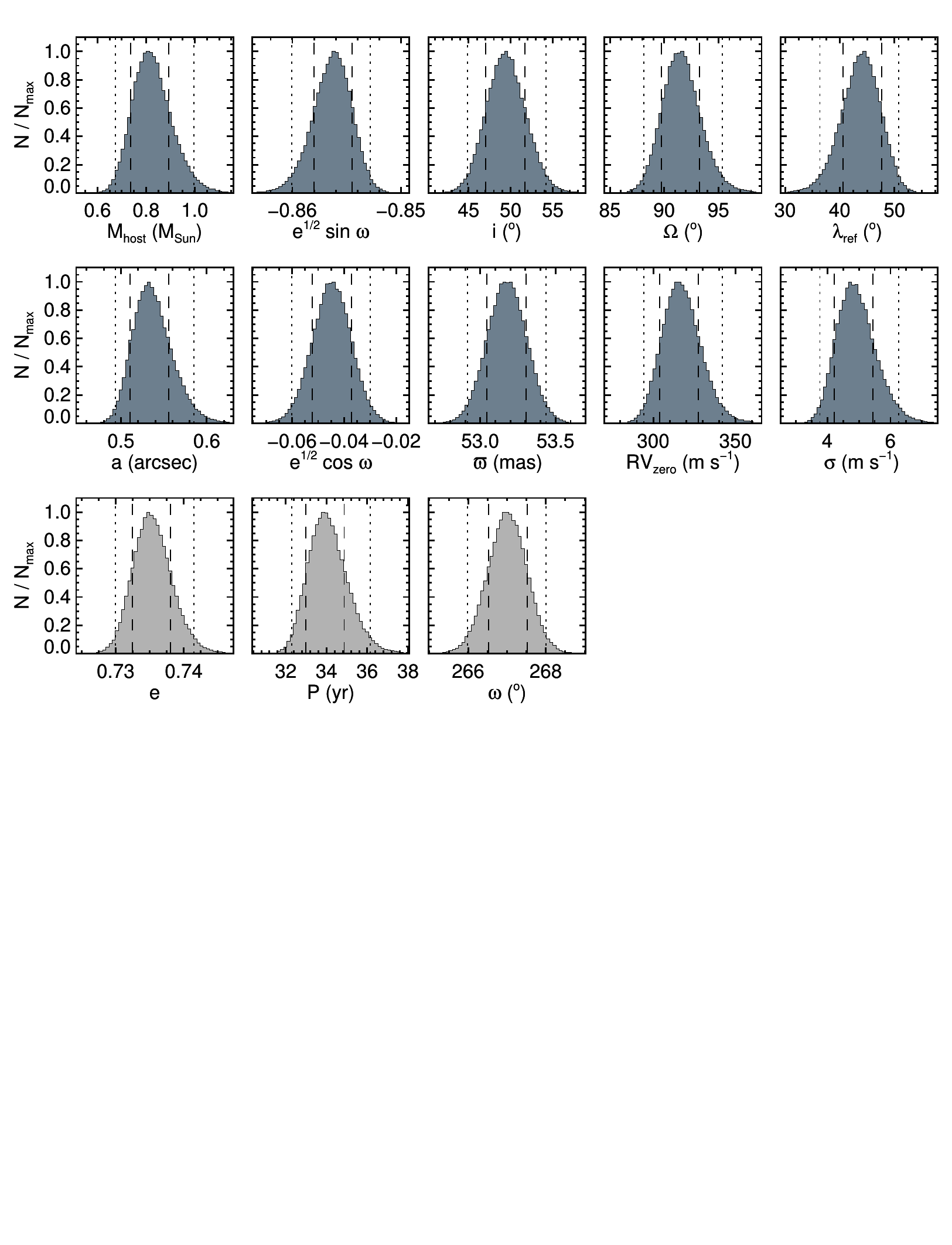}
\vskip -4.0 truein
\caption{Marginalized posterior distributions for all directly fitted orbital parameters (dark gray histograms), aside from companion mass which is shown in Figure~\ref{fig:HD4747-corner}. Posteriors for properties computed from the directly fitted parameters are shown in light gray histograms.}
\label{fig:HD4747-hist}
\end{figure*}

\begin{figure*}
\centerline{
\includegraphics[width=0.4\linewidth]{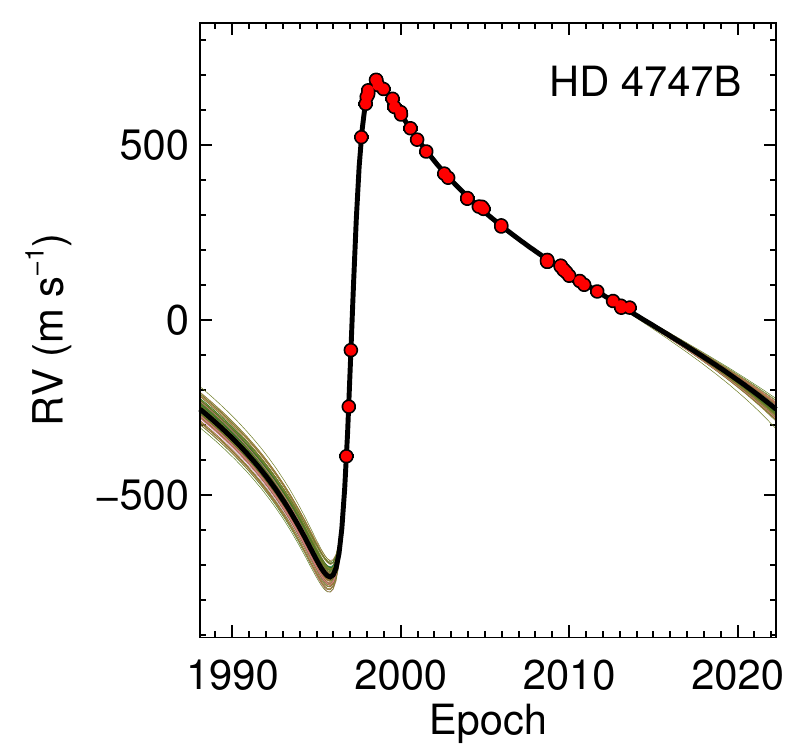}
\includegraphics[width=0.4\linewidth]{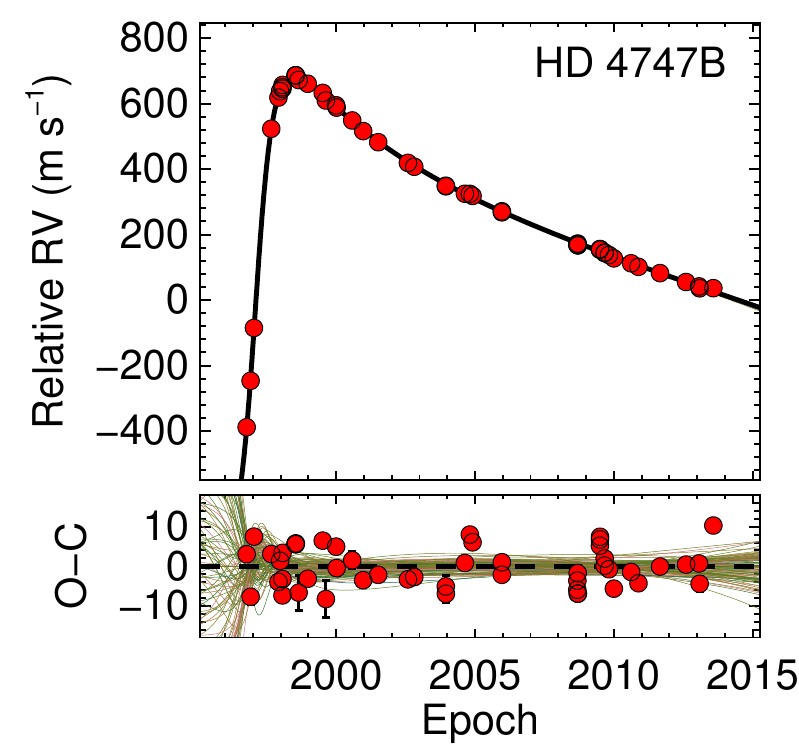}
\hskip -2.2in
\includegraphics[width=0.4\linewidth]{HD4747-colorbar.pdf}
}
\vskip 0.0 truein
\caption{Left: The RV orbit induced by the directly imaged companion over a full orbital period, where the thick black line is the highest likelihood orbit and the thin lines are 100 orbits drawn randomly from our posterior distribution. These orbits are colored according to companion mass from pink (low mass) to green (high mass). RV measurements of the host star with the best-fit RV zero point added in are overplotted. RV of zero on this panel is the system's barycentric velocity. Right: RV orbit induced by the directly imaged companion over the time range sampled by RV measurements. The zero point is arbitrary here as we plot relative RV measurements. The bottom panel shows residuals after subtracting the RV orbit from the measurements. Error bars are too small to be visible except for some points on the plot of residuals.}
\label{fig:HD4747-RV}
\end{figure*}

\begin{figure*}
\centerline{
\includegraphics[width=0.23\linewidth]{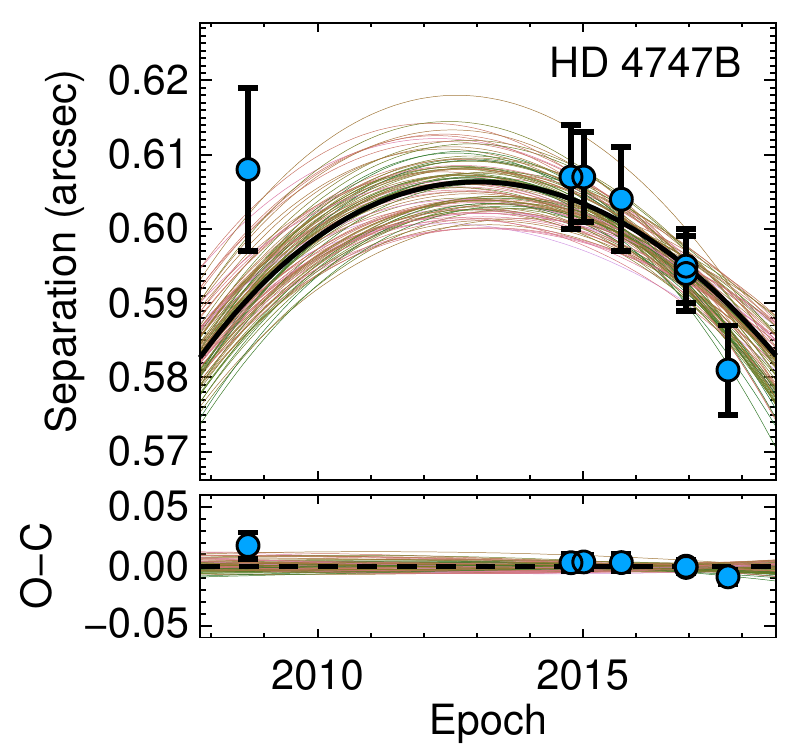}
\includegraphics[width=0.23\linewidth]{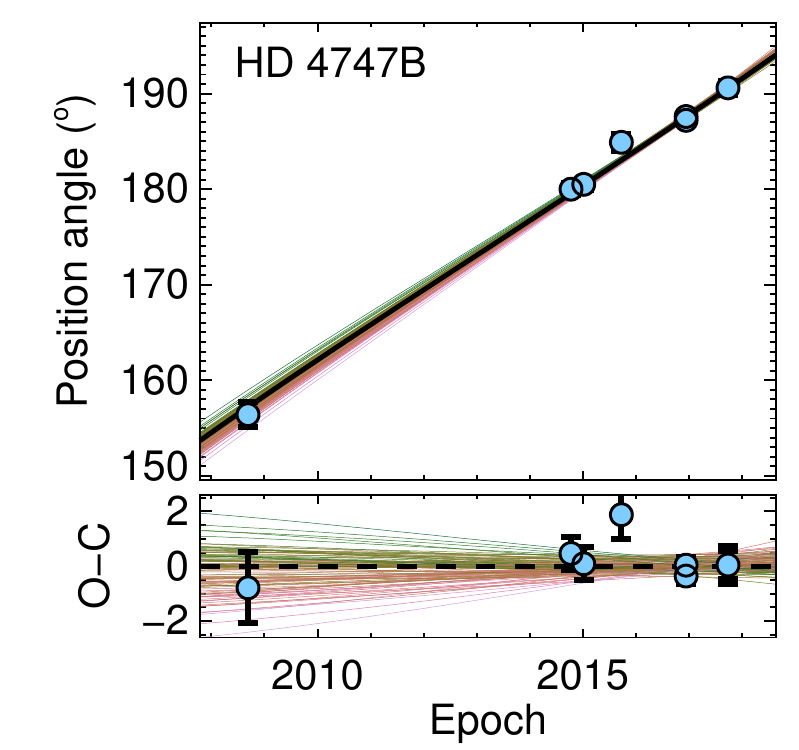}
\includegraphics[width=0.23\linewidth]{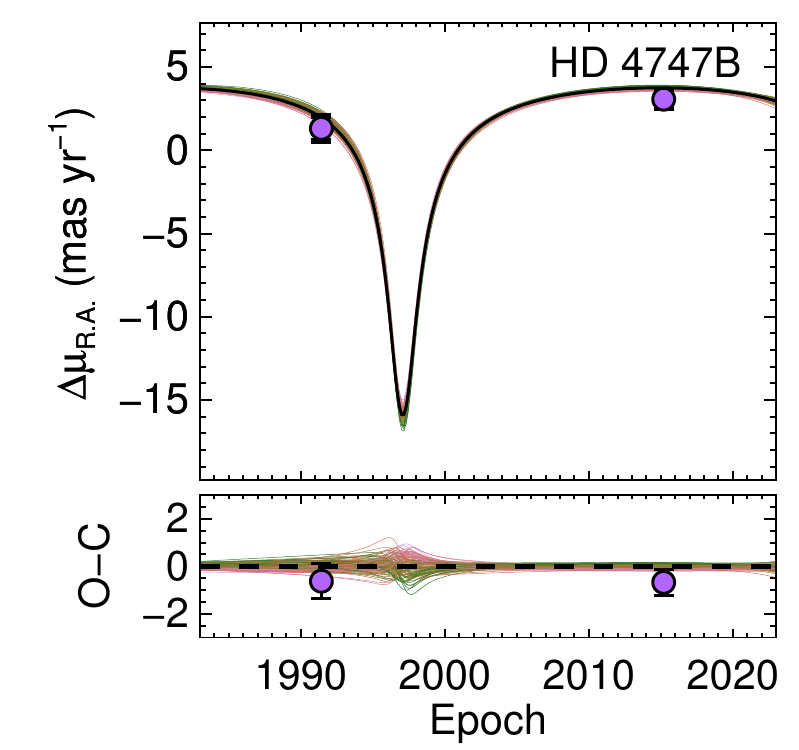}
\includegraphics[width=0.23\linewidth]{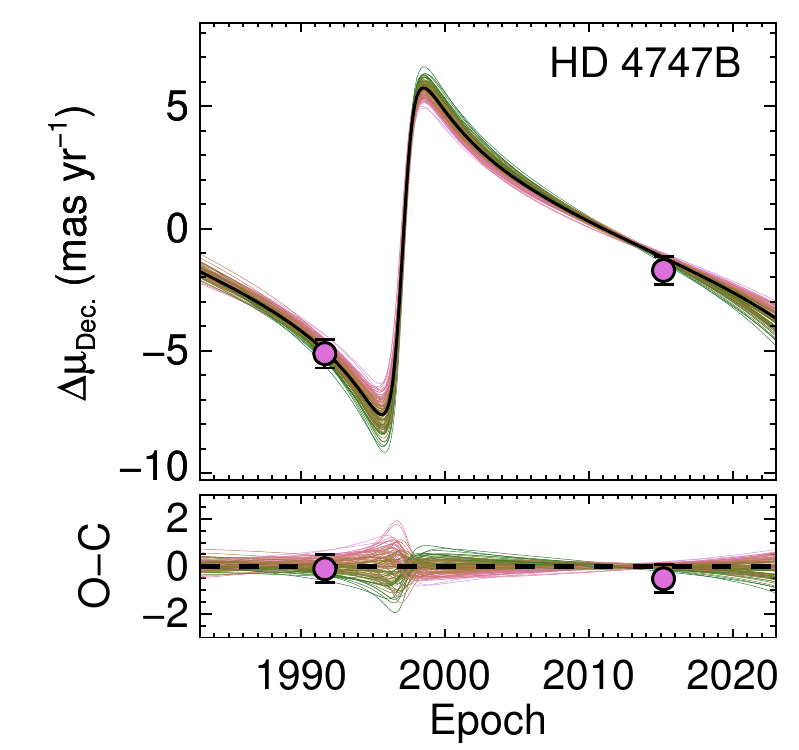}
\hskip -1.2in
\includegraphics[width=0.23\linewidth]{HD4747-colorbar.pdf}
}
\vskip 0.0 truein
\caption{Top: astrometry of the directly imaged companion relative to the host star. Bottom: acceleration induced by the companion on the host star as measured from absolute astrometry from Hipparcos and Gaia. This acceleration is plotted in the same way as it is used in our orbital analysis, that is the difference between the instantaneous linear motion of the host star at a given epoch (Hipparcos and Gaia DR2) as compared to the proper motion computed from the position differences between those two epochs. In every panel, the thick black line is the highest likelihood orbit and the thin lines are 100 orbits drawn randomly from our posterior distribution. These orbits are colored according to companion mass from pink (low mass) to green (high mass).}
\label{fig:HD4747-ast}
\end{figure*}

\begin{figure*}
\includegraphics[width=1.0\linewidth]{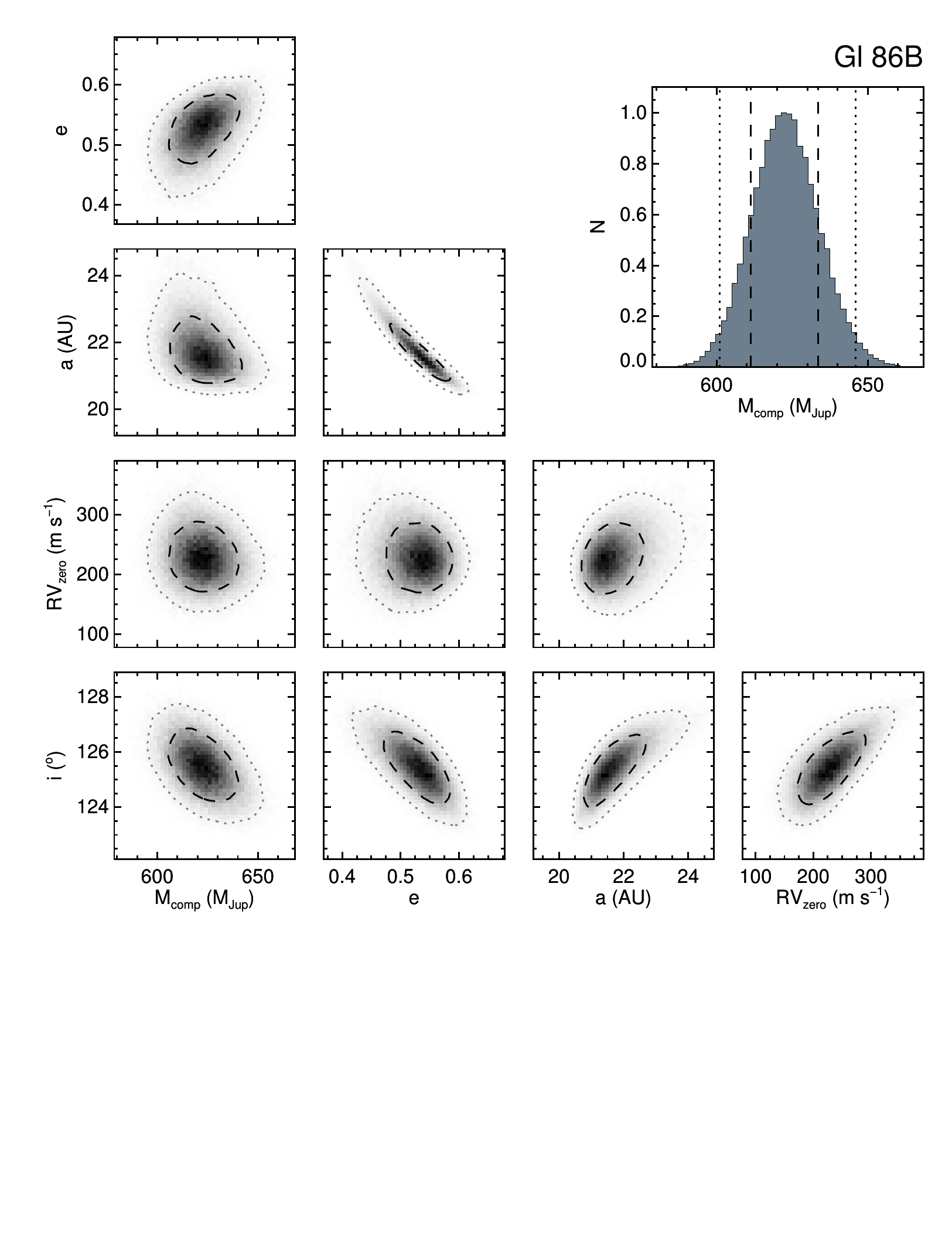}
\vskip -2.25 truein
\caption{Same as Figure~\ref{fig:HD4747-corner}.}
\label{fig:GL86-corner}
\end{figure*}

\begin{figure*}
\includegraphics[width=1.0\linewidth]{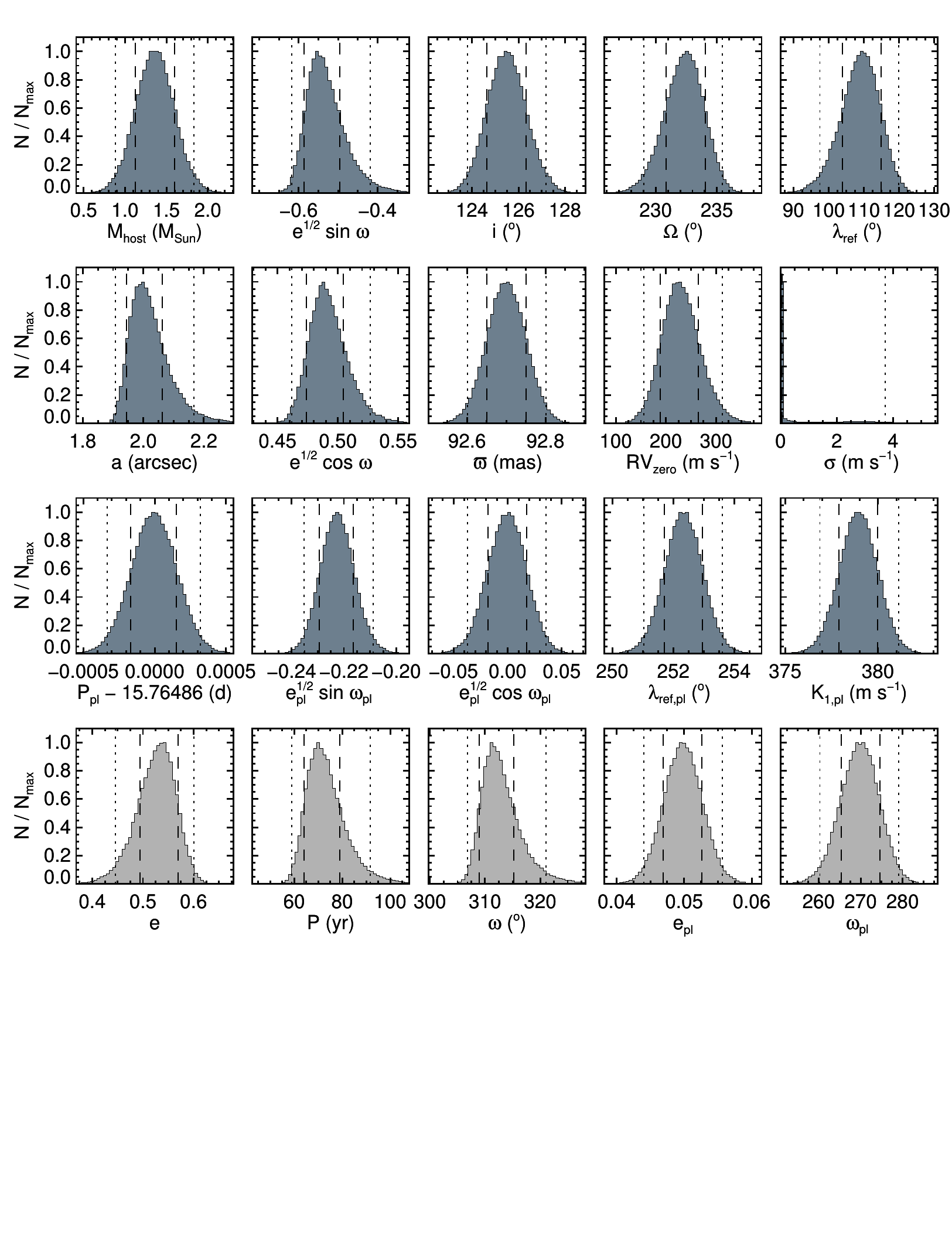}
\vskip -2.25 truein
\caption{Same as Figure~\ref{fig:HD4747-hist}.}
\label{fig:GL86-hist}
\end{figure*}

\begin{figure*}
\centerline{
\includegraphics[width=0.4\linewidth]{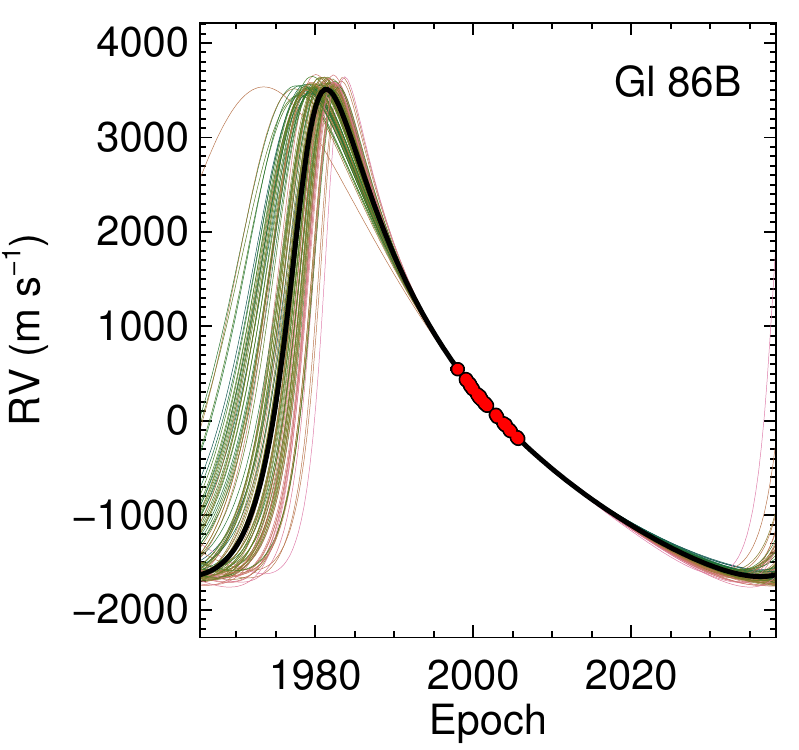}
\includegraphics[width=0.4\linewidth]{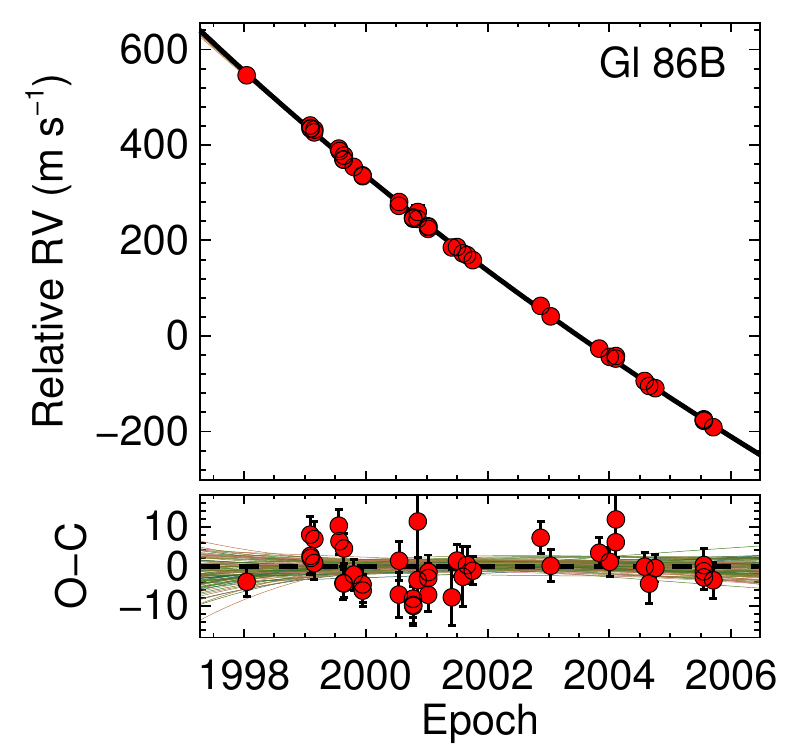}
\hskip -2.2in
\includegraphics[width=0.4\linewidth]{GL86-colorbar.pdf}
}
\vskip 0.0 truein
\caption{Same as Figure~\ref{fig:HD4747-RV}, and here the orbit of the inner planet Gl~86~b has been subtracted off of the measured RVs.}
\label{fig:GL86-RV}
\end{figure*}

\begin{figure*}
\centerline{
\includegraphics[width=0.23\linewidth]{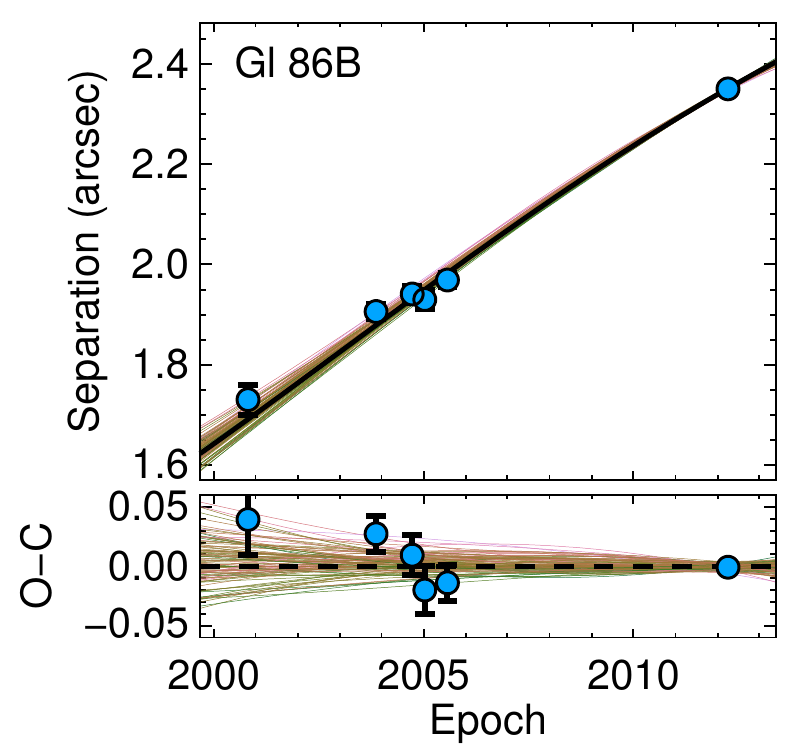}
\includegraphics[width=0.23\linewidth]{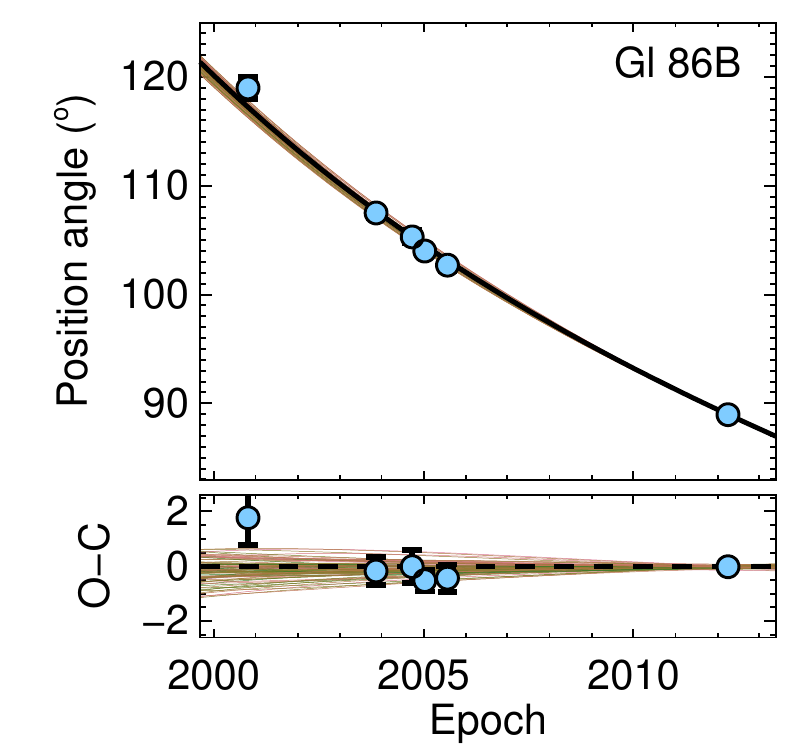}
\includegraphics[width=0.23\linewidth]{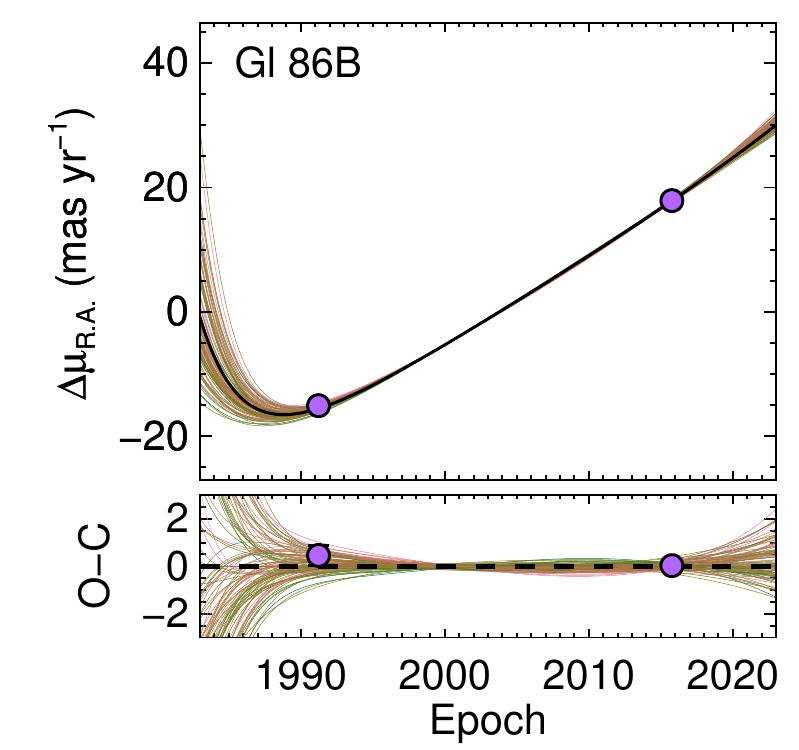}
\includegraphics[width=0.23\linewidth]{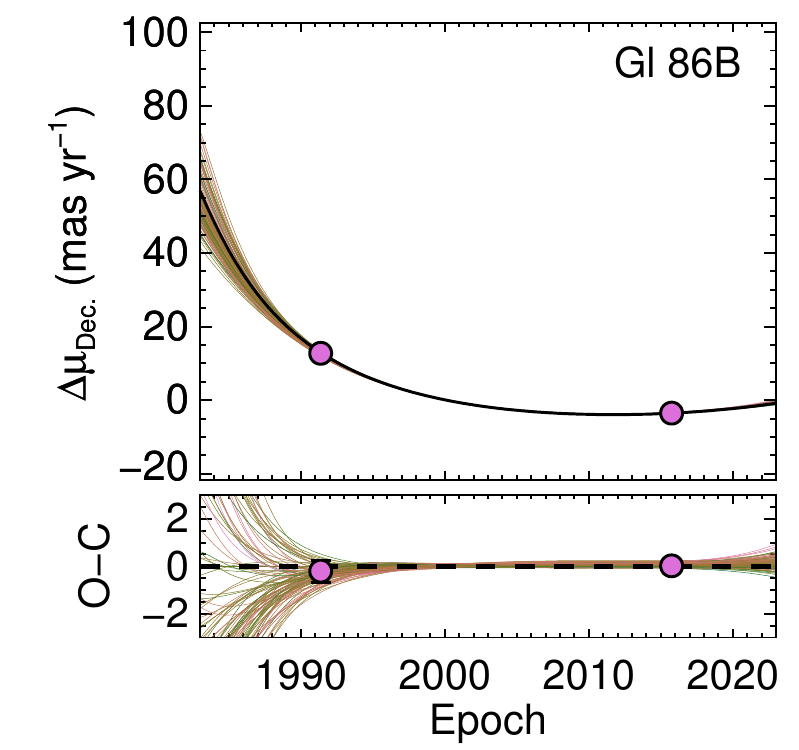}
\hskip -1.2in
\includegraphics[width=0.23\linewidth]{GL86-colorbar.pdf}
}
\vskip 0.0 truein
\caption{Same as Figure~\ref{fig:HD4747-ast}.}
\label{fig:GL86-ast}
\end{figure*}

\begin{figure*}
\includegraphics[width=1.0\linewidth]{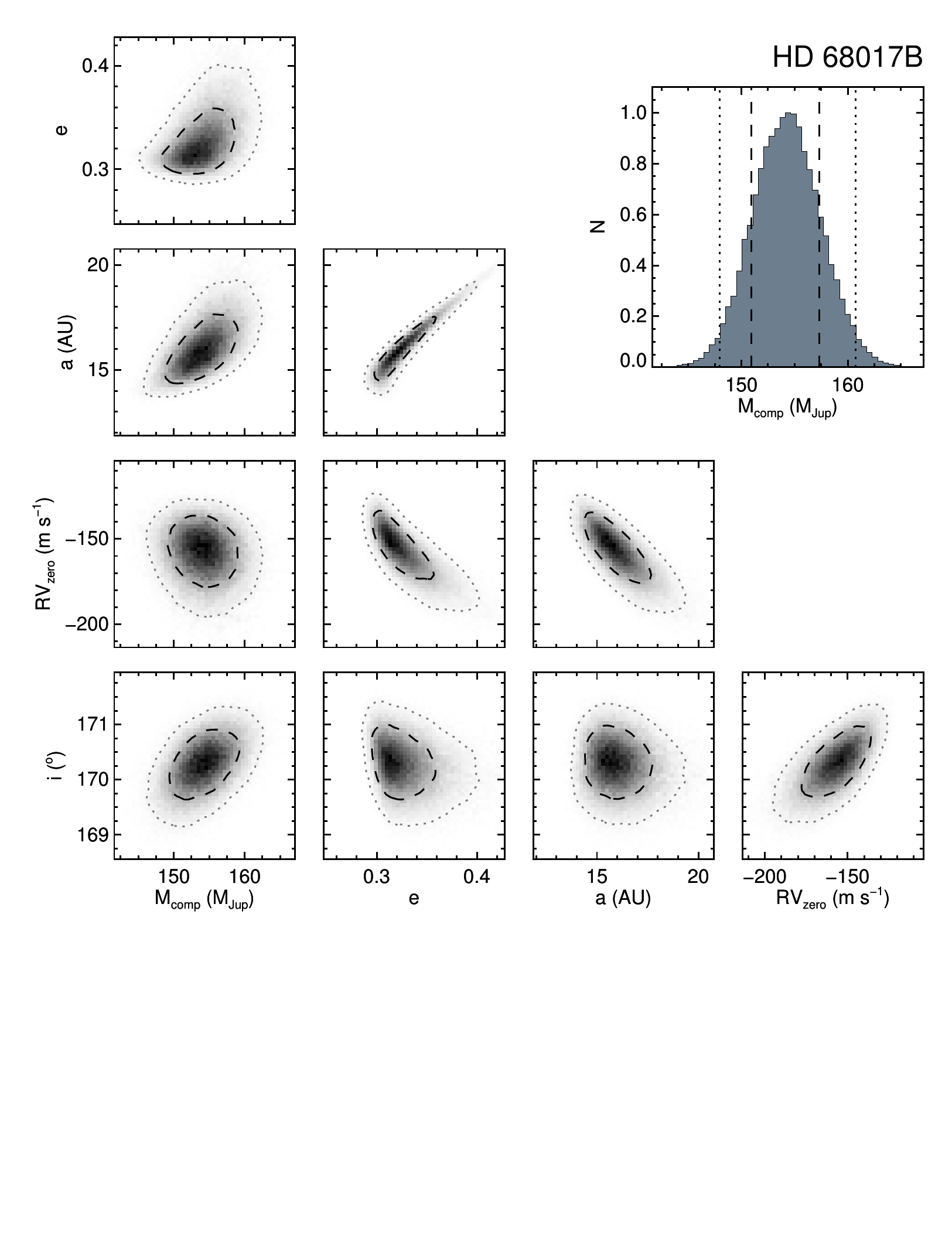}
\vskip -2.25 truein
\caption{Same as Figure~\ref{fig:HD4747-corner}.}
\label{fig:HD68017-corner}
\end{figure*}

\begin{figure*}
\includegraphics[width=1.0\linewidth]{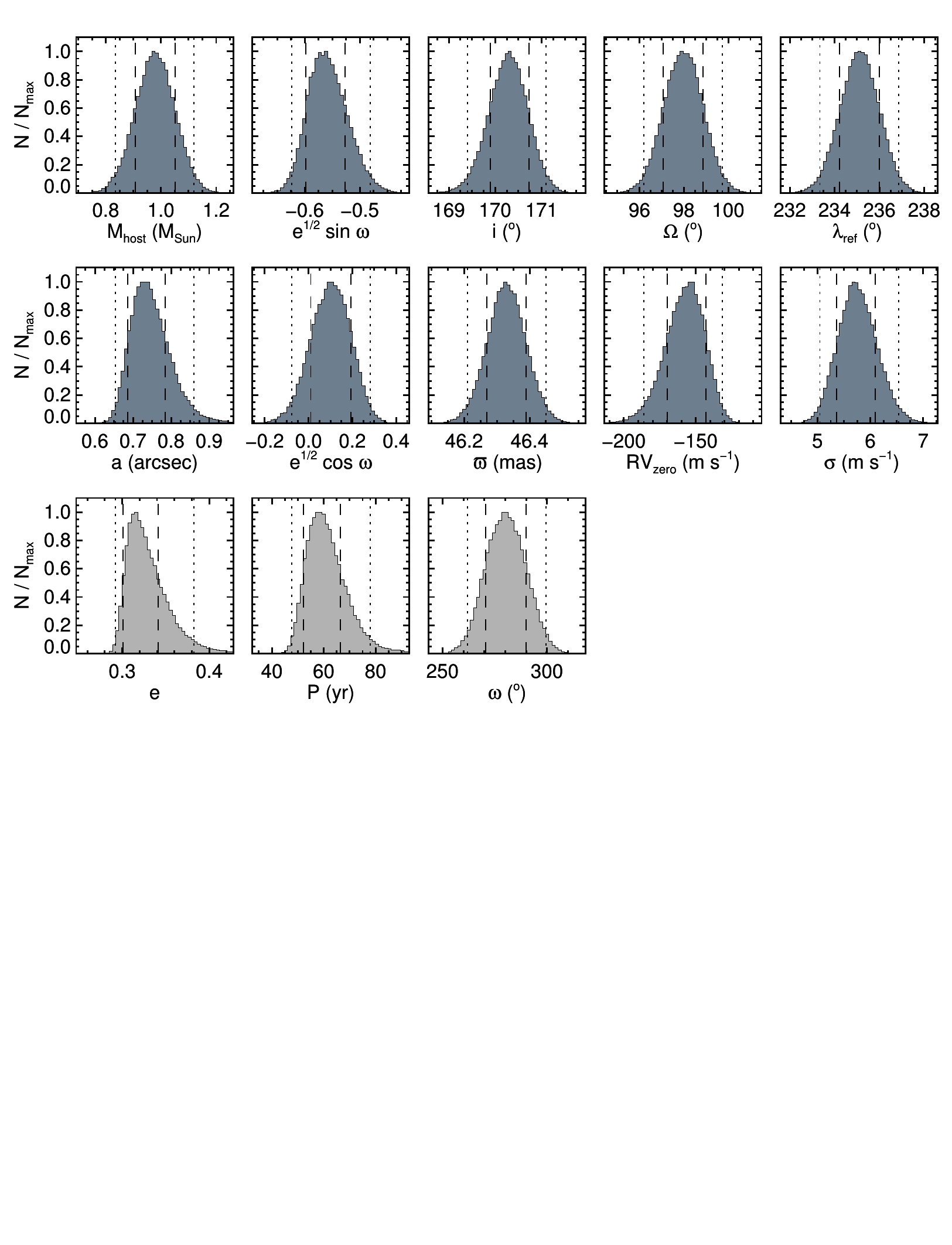}
\vskip -4.0 truein
\caption{Same as Figure~\ref{fig:HD4747-hist}.}
\label{fig:HD68017-hist}
\end{figure*}

\begin{figure*}
\centerline{
\includegraphics[width=0.4\linewidth]{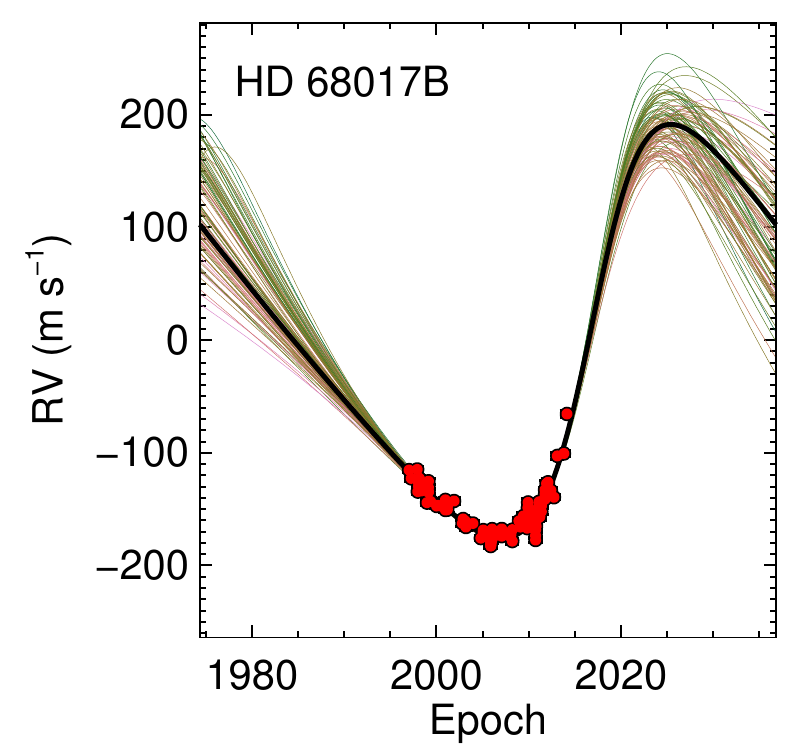}
\includegraphics[width=0.4\linewidth]{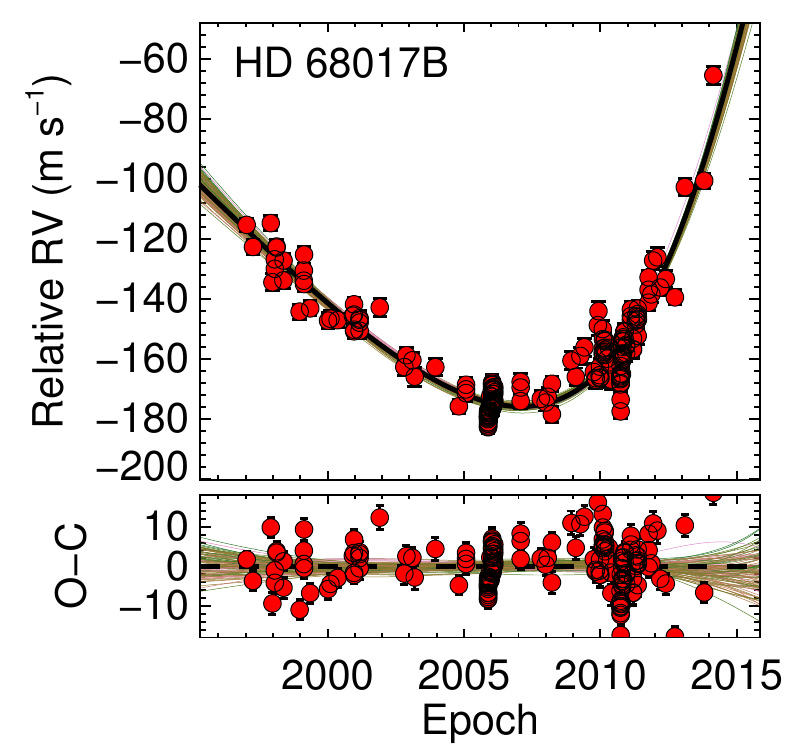}
\hskip -2.2in
\includegraphics[width=0.4\linewidth]{HD68017-colorbar.pdf}
}
\vskip 0.0 truein
\caption{Same as Figure~\ref{fig:HD4747-RV}.}
\label{fig:HD68017-RV}
\end{figure*}

\begin{figure*}
\centerline{
\includegraphics[width=0.23\linewidth]{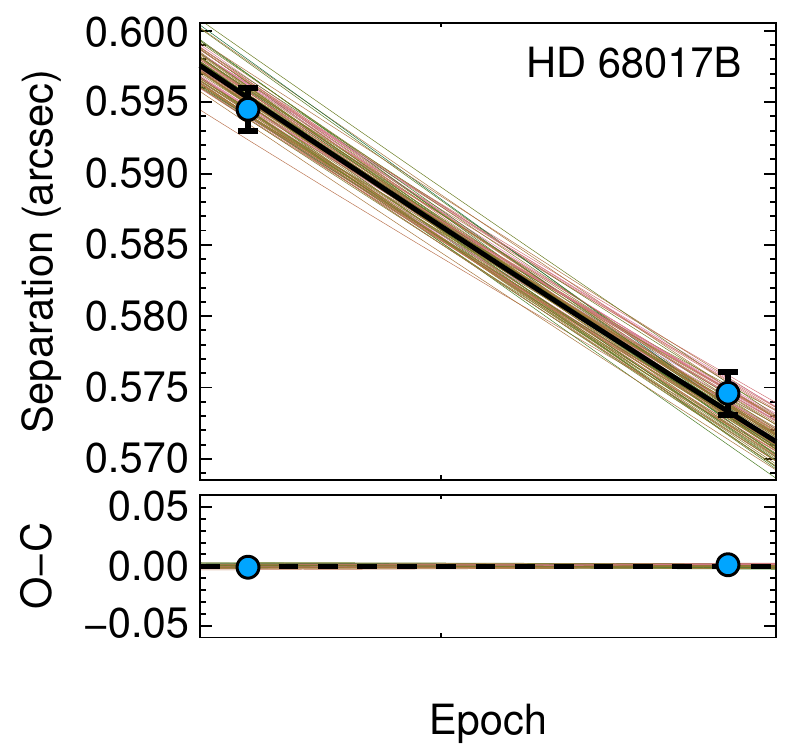}
\includegraphics[width=0.23\linewidth]{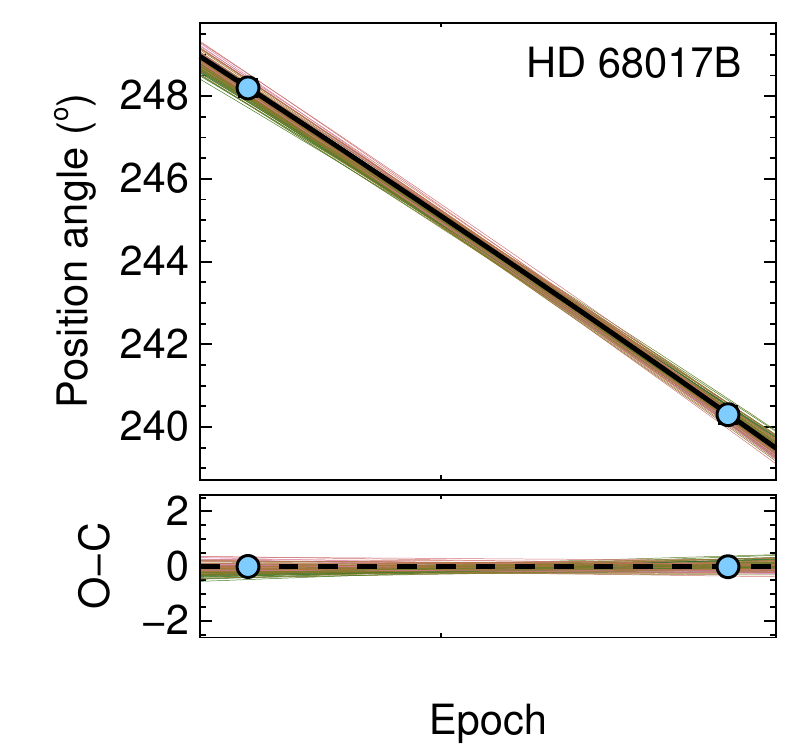}
\includegraphics[width=0.23\linewidth]{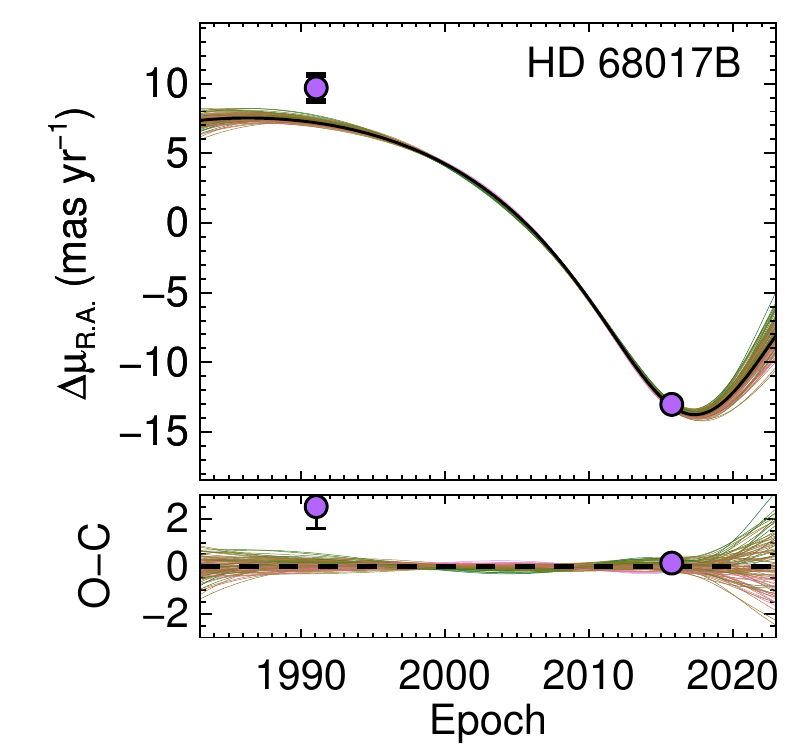}
\includegraphics[width=0.23\linewidth]{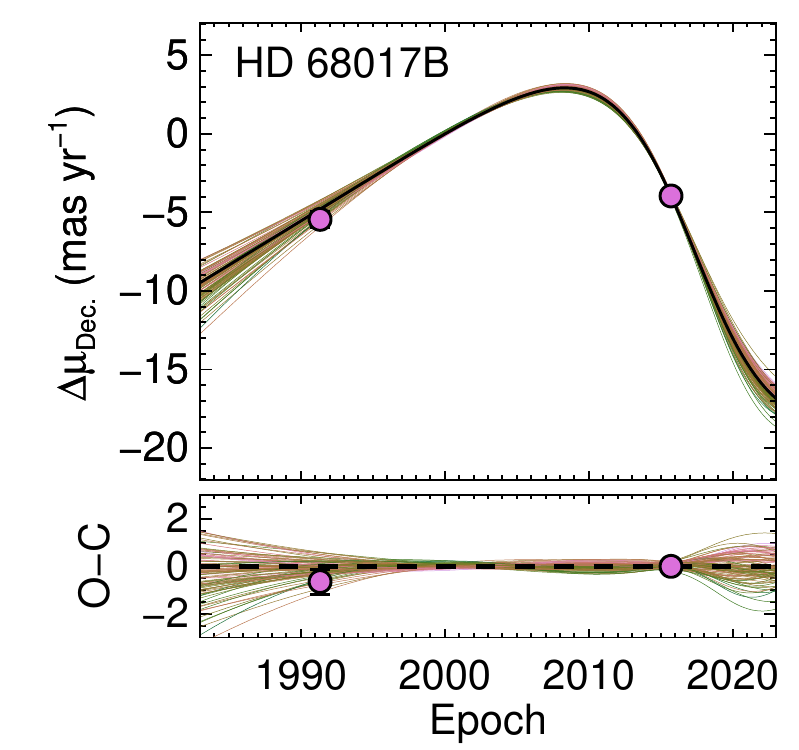}
\hskip -1.2in
\includegraphics[width=0.23\linewidth]{HD68017-colorbar.pdf}
}
\vskip 0.0 truein
\caption{Same as Figure~\ref{fig:HD4747-ast}.}
\label{fig:HD68017-ast}
\end{figure*}

\begin{figure*}
\includegraphics[width=1.0\linewidth]{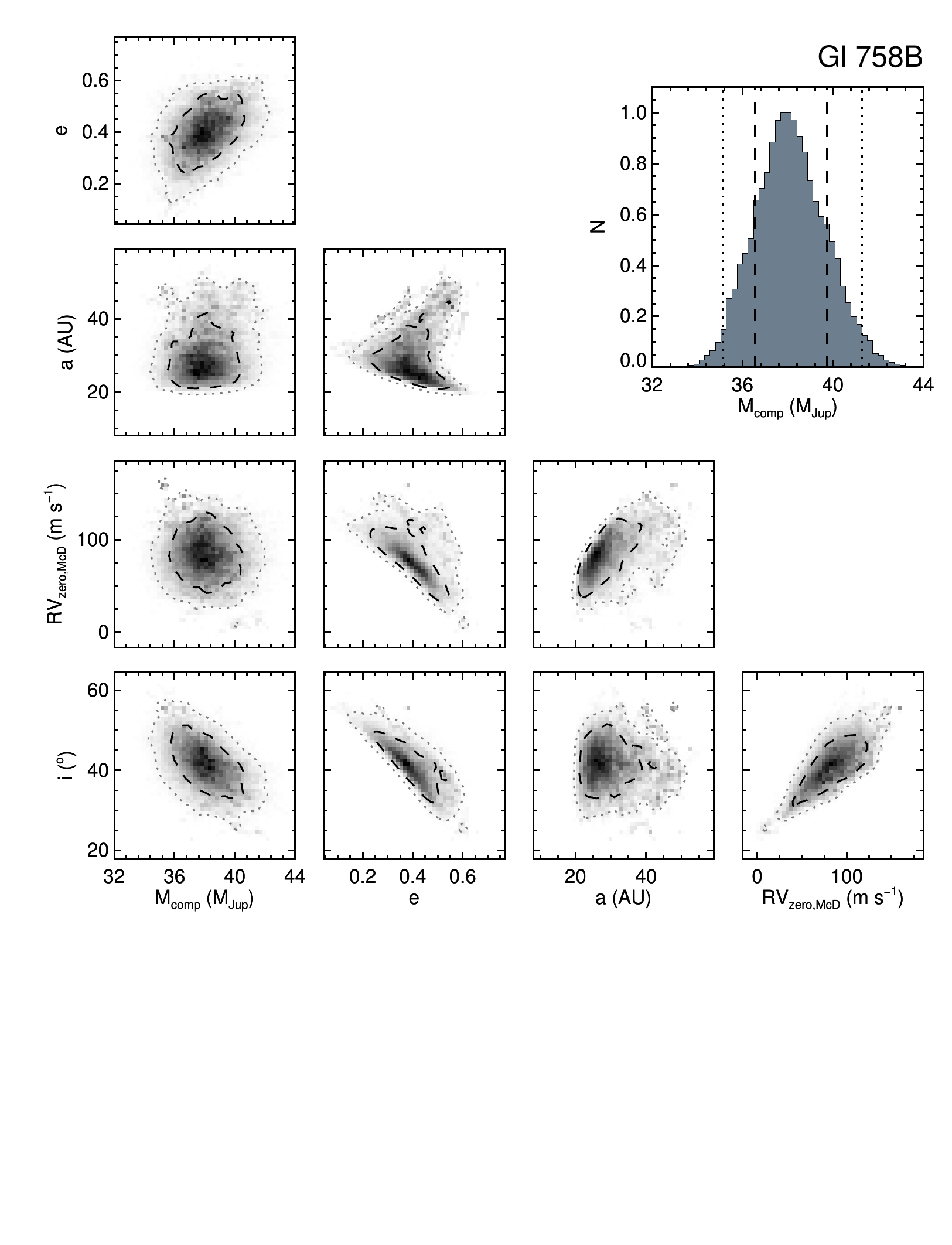}
\vskip -2.25 truein
\caption{Same as Figure~\ref{fig:HD4747-corner}.}
\label{fig:GL758-corner}
\end{figure*}

HR~7672B has a previous dynamical mass of $m = 68.7^{+2.4}_{-3.1}$\,\Mjup\ measured from radial velocities and relative astrometry \citep{Crepp+Johnson+Fischer+etal_2012}.  As such, it presents another excellent test case of our method, like HD~4747B.  HR~7672B passed close to its host star in projection around 2015 (Figure~\ref{fig:orbit-sky}), as predicted by \cite{Crepp+Johnson+Fischer+etal_2012}; this results in proper motion changing sign within the time frame of {\it Gaia} observations (see Figure~\ref{fig:HR7672-ast}). We obtain a dynamical mass of $m = 72.7 \pm 0.8$~$M_{\rm Jup}$ for HR~7672B, 1.6\,$\sigma$ higher than \cite{Crepp+Johnson+Fischer+etal_2012}. We also find somewhat higher eccentricity and longer orbital period for the orbit. HR~7672A has the most precise dynamical mass among the host stars in our sample, $0.96^{+0.04}_{-0.05}$\,\Msun, which is somewhat lower than the value of $1.08\pm0.04$\,\Msun\ inferred from the star's color and luminosity and used in the \cite{Crepp+Johnson+Fischer+etal_2012} dynamical analysis. There is not significant correlation between companion mass and other parameters like eccentricity and period in our MCMC posteriors (Figure~\ref{fig:HR7672-corner}), so it is not apparent why our mass for the companion is systematically higher than that of \cite{Crepp+Johnson+Fischer+etal_2012}.  

Our best-fit orbit for HR~7672 is suspiciously good in separation and position angle, suggesting that errors in relative astrometry may have been overestimated.  The $\chi^2$ in position angle, for example, is just 0.55 for six measurements.  The best-fit orbit is poorer for the host-star astrometry, with $\chi^2$ of 11.4 for our four proper motion measurements.  Such a discrepancy is expected 2.2\% of the time in a $\chi^2$ distribution with four degrees of freedom.  If we attribute the discrepancy to systematics in the catalog or to the heavy tails in the low-precision {\it Gaia} proper motions, and inflate uncertainties beyond the calibrations of the \cite{Brandt_2018} catalog, it may increase the uncertainty on our dynamical mass.  Underestimated uncertainties or unaccounted systematics in the \cite{Liu+Fischer+Graham+etal_2002} astrometry could also explain the relatively poor fit to the calibrated {\it Hipparcos} and {\it Gaia} astrometry.

Our best-fit mass for HR~7672B is very close to the hydrogen burning limit, placing this object near the stellar/substellar boundary. We derive a bolometric luminosity of $\log[\Lbol/\Lsun] = -4.14\pm0.06$\,dex by combining the $K_s$-band photometry of \cite{Boccaletti+Chauvin+Lagrange+etal_2003} with the calibrations of \cite{Dupuy+Liu_2017}, or $\log[\Lbol/\Lsun] = -4.23\pm0.05$\,dex using $H$-band photometry.  Figure \ref{fig:modelcompare} adopts the weighted average of these two values, $\log[\Lbol/\Lsun] = -4.19\pm0.04$\,dex.  This bolometric luminosity falls within the range that \cite{Dupuy+Liu_2017} identified as the end of the main sequence ($10^{-4.3}$--$10^{-3.9}$\,\Lsun). Evolutionary models from \cite{2008ApJ...689.1327S} predict that an object with the luminosity and mass of HR~7672B is still cooling but that it will eventually stabilize on the main sequence once it reaches an age of several Gyr.  Our best-fit mass is marginally consistent with the Cond \citep{Baraffe+Chabrier+Barman+etal_2003} and \cite{Burrows+Marley+Hubbard+etal_1997} models; it agrees well with the cooling curves of \cite{Saumon+Marley_2008} regardless of the assumptions about cloud cover.

\section{Conclusions} \label{sec:conclusions}

In this paper, we have combined the astrometric catalog of \cite{Brandt_2018} with radial velocities and relative astrometry to determine absolute orbits and thereby measure model-independent dynamical masses.  The astrometric catalog represents a cross-calibration of {\it Hipparcos} and {\it Gaia}~DR2, including an assessment of systematic errors needed to bring the two catalogs in agreement, constructed with the goal of joint orbit fitting in mind.  We choose an initial sample of five objects: three ultracool dwarfs, one white dwarf, and one low-mass star.  We also perform a uniform age analysis on all five of the host stars.

For our two more massive ultracool dwarfs, HD~4747B and HR~7672B, we determine more precise dynamical masses that improve on and are consistent with previous results, notably using no external constraints on the host star masses.  Both of these brown dwarfs lie near or below the hydrogen burning limit, and both are intermediate in age ($\sim$2--4~Gyr based on activity-age relations and isochrone fitting).  For Gl~758B, the lowest-luminosity imaged brown dwarf with a dynamical mass, we significantly improve on previously determined masses \citep{Bowler+Dupuy+Endl+etal_2018,Calissendorff+Janson_2018}.  The latter authors used {\it Hipparcos} and {\it Gaia} astrometry but without performing a cross-calibration between the two catalogs and without using the scaled position difference (i.e., the proper motion between {\it Hipparcos} and {\it Gaia} epochs computed from the reported catalog RA and Dec positions).  Our mass of $38.1^{+1.7}_{-1.5}$~$M_\odot$ for Gl~758B is firmly on the low end of previous determinations and on the high end of the original estimate from \cite{Thalmann+Carson+Janson+etal_2009}.  Our mass, combined with an older host-star age ($\gtrsim$6~Gyr), resolves previous apparent discrepancies and is consistent with predictions from brown dwarf cooling models.

For the white dwarf Gl~86B, we find a dynamical mass of $0.595\pm0.010$\,\Msun\ that sheds light on the progenitor star's history. Our mass is consistent with the spectroscopic and photometric analysis of this white dwarf companion by \cite{Farihi+Bond+Dufour+etal_2013}, who found that the progenitor star was massive ($1.9\pm0.1$\,\Msun) with a relatively short main sequence lifetime ($1.4\pm0.2$\,Gyr) and subsequent white dwarf cooling time ($1.25\pm0.05$\,Gyr). This scenario is also consistent with our activity-based age of the host star. However, our mass is inconsistent with the scenario proposed by \cite{Fuhrmann+Chini+Buda+etal_2014}, who noted that the chemistry and kinematics of Gl~86A imply a much older age that would require a lower mass progenitor and lower mass white dwarf ($0.49\pm0.02$\,\Msun). The present-day orbit we determine will also inform the formation history of this system where the lower-mass star in the system (Gl~86A) formed a massive planet (Gl~86b).

Finally, we measure a precise mass for the low-mass star HD~68017B of $0.147\pm0.003$\,\Msun\ in good agreement with previous estimates based on its absolute magnitude. As a rare example of an M~dwarf with a low-metallicity (${[\rm Fe/H]}= -0.44\pm0.03$\,dex) and precisely measured dynamical mass, HD~68017B will help constrain the mass--magnitude--metallicity relation in the future.

Our analysis here can serve as a prototype for further dynamical masses derived using the \cite{Brandt_2018} catalog. By combining the absolute astrometry from this catalog with data for well-characterized companions, we have quantitatively vetted the errors reported in the catalog for stars with significant acceleration detections. Thousands of stars show astrometric acceleration, and many of the companions responsible for these accelerations will be amenable to direct imaging and dynamical mass measurements.  Model-independent masses of stars, brown dwarfs, and white dwarfs will provide new anchors to theoretical models of their formation and evolution.

\acknowledgements{The authors thank the anonymous referee for a thorough and helpful report.  Some of the data presented herein were obtained at the W. M. Keck Observatory, which is operated as a scientific partnership among the California Institute of Technology, the University of California and the National Aeronautics and Space Administration. The Observatory was made possible by the generous financial support of the W. M. Keck Foundation.  The authors wish to recognize and acknowledge the very significant cultural role and reverence that the summit of Maunakea has always had within the indigenous Hawaiian community.  We are most fortunate to have the opportunity to conduct observations from this mountain.  This work has made use of data from the European Space Agency (ESA) mission {\it Gaia} (https://www.cosmos.esa.int/gaia), processed by the Gaia Data Processing and Analysis Consortium (DPAC, https://www.cosmos.esa.int/web/gaia/dpac/consortium). Funding for the DPAC has been provided by national institutions, in particular the institutions participating in the {\it Gaia} Multilateral Agreement.  T.D.B.~gratefully acknowledges support from the Heising-Simons foundation and from NASA under grant \#80NSSC18K0439. T.J.D.~acknowledges research support from Gemini Observatory.}



\begin{figure*}
\includegraphics[width=1.0\linewidth]{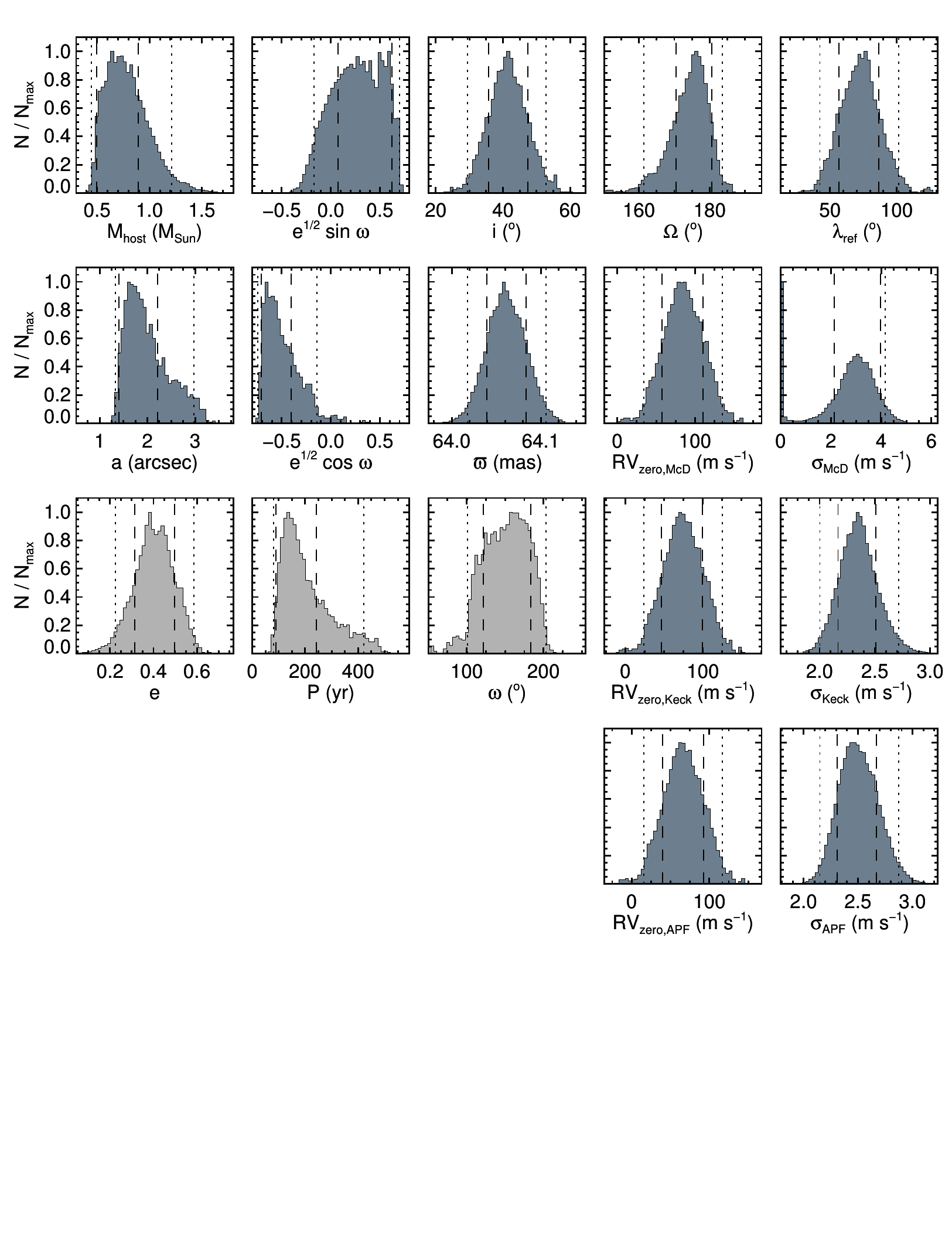}
\vskip -2.25 truein
\caption{Same as Figure~\ref{fig:HD4747-hist}.}
\label{fig:GL758-hist}
\end{figure*}

\begin{figure*}
\centerline{
\includegraphics[width=0.23\linewidth]{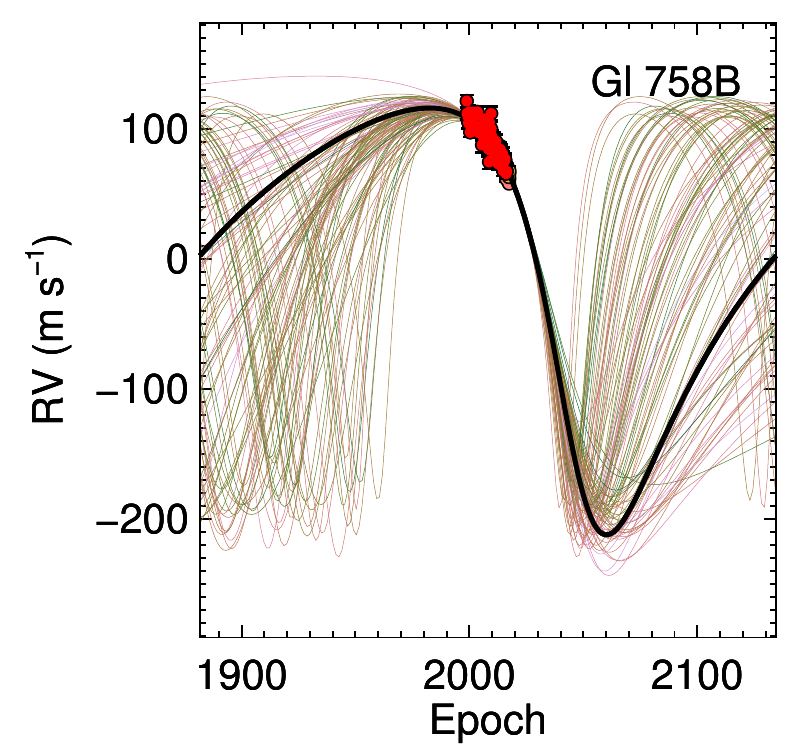}
\includegraphics[width=0.23\linewidth]{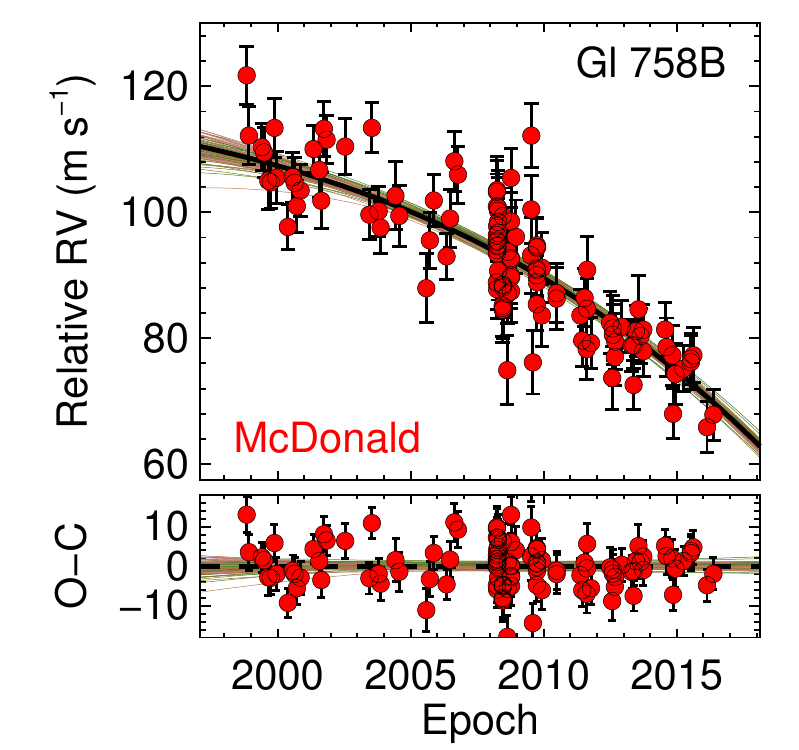}
\includegraphics[width=0.23\linewidth]{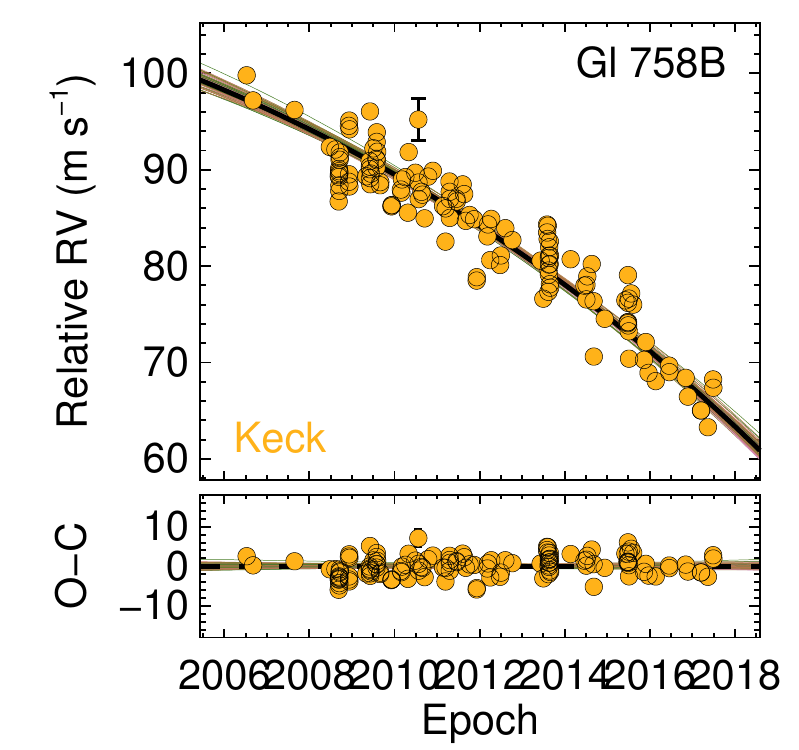}
\includegraphics[width=0.23\linewidth]{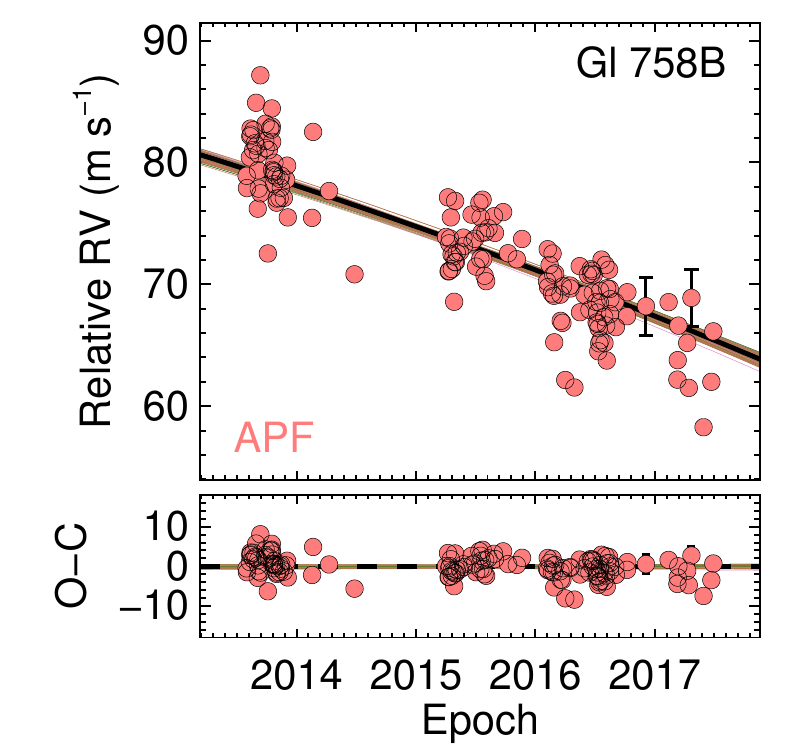}
\hskip -1.2in
\includegraphics[width=0.23\linewidth]{GL758-colorbar.pdf}
}
\vskip 0.0 truein
\caption{Same as Figure~\ref{fig:HD4747-RV}.}
\label{fig:GL758-RV}
\end{figure*}

\begin{figure*}
\centerline{
\includegraphics[width=0.23\linewidth]{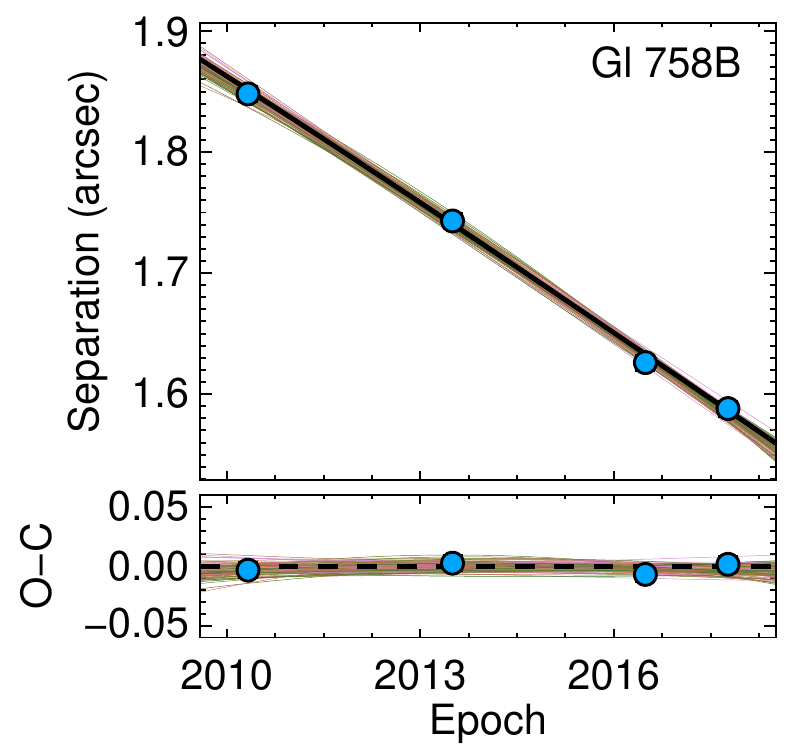}
\includegraphics[width=0.23\linewidth]{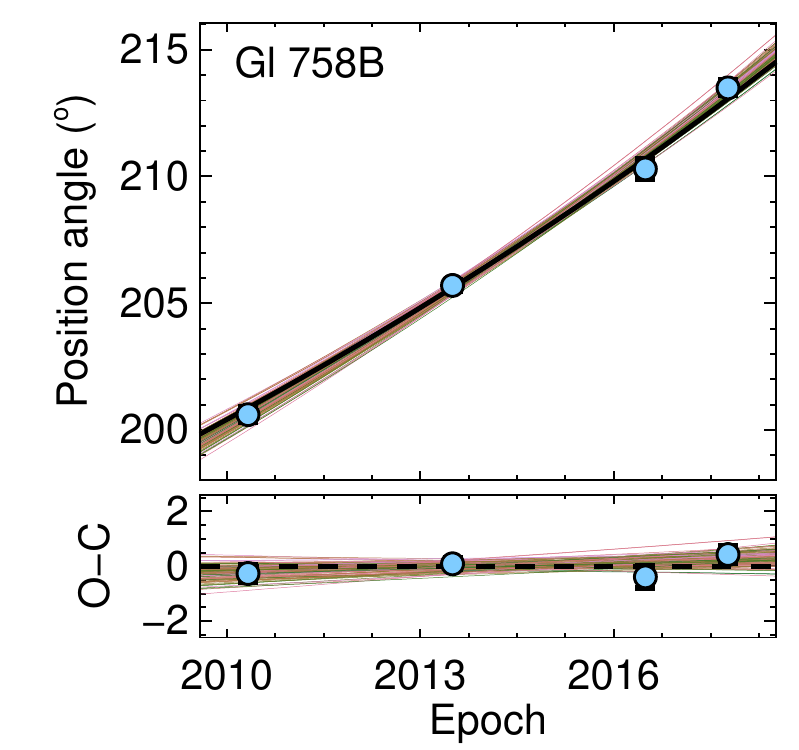}
\includegraphics[width=0.23\linewidth]{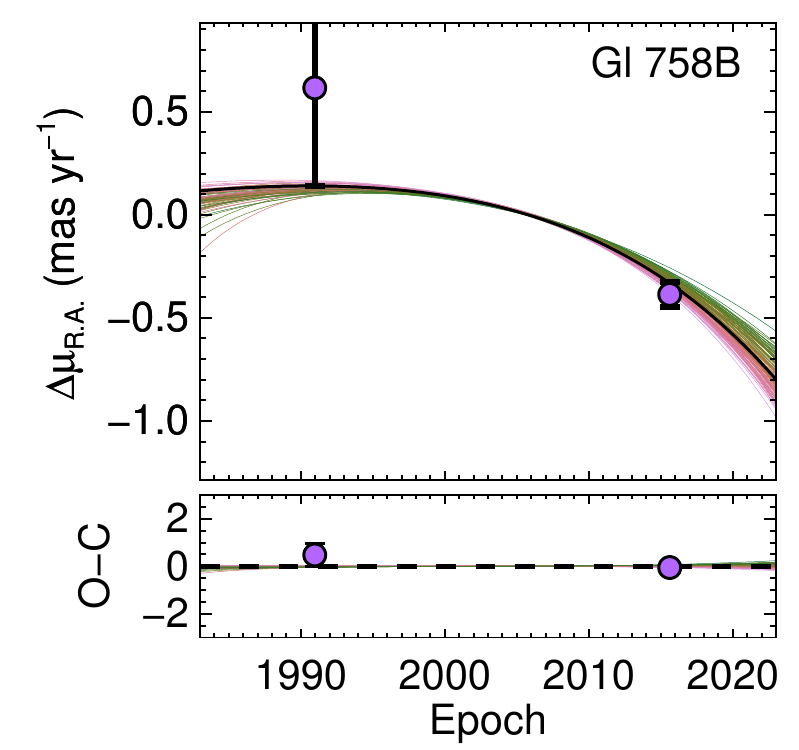}
\includegraphics[width=0.23\linewidth]{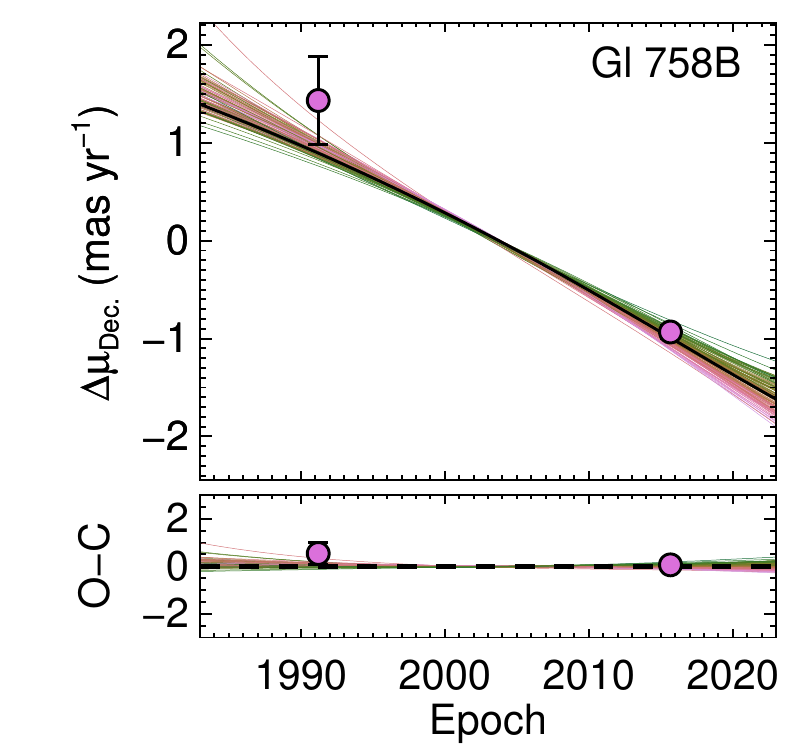}
\hskip -1.2in
\includegraphics[width=0.23\linewidth]{GL758-colorbar.pdf}
}
\vskip 0.0 truein
\caption{Same as Figure~\ref{fig:HD4747-ast}.}
\label{fig:GL758-ast}
\end{figure*}
\begin{figure*}
\includegraphics[width=1.0\linewidth]{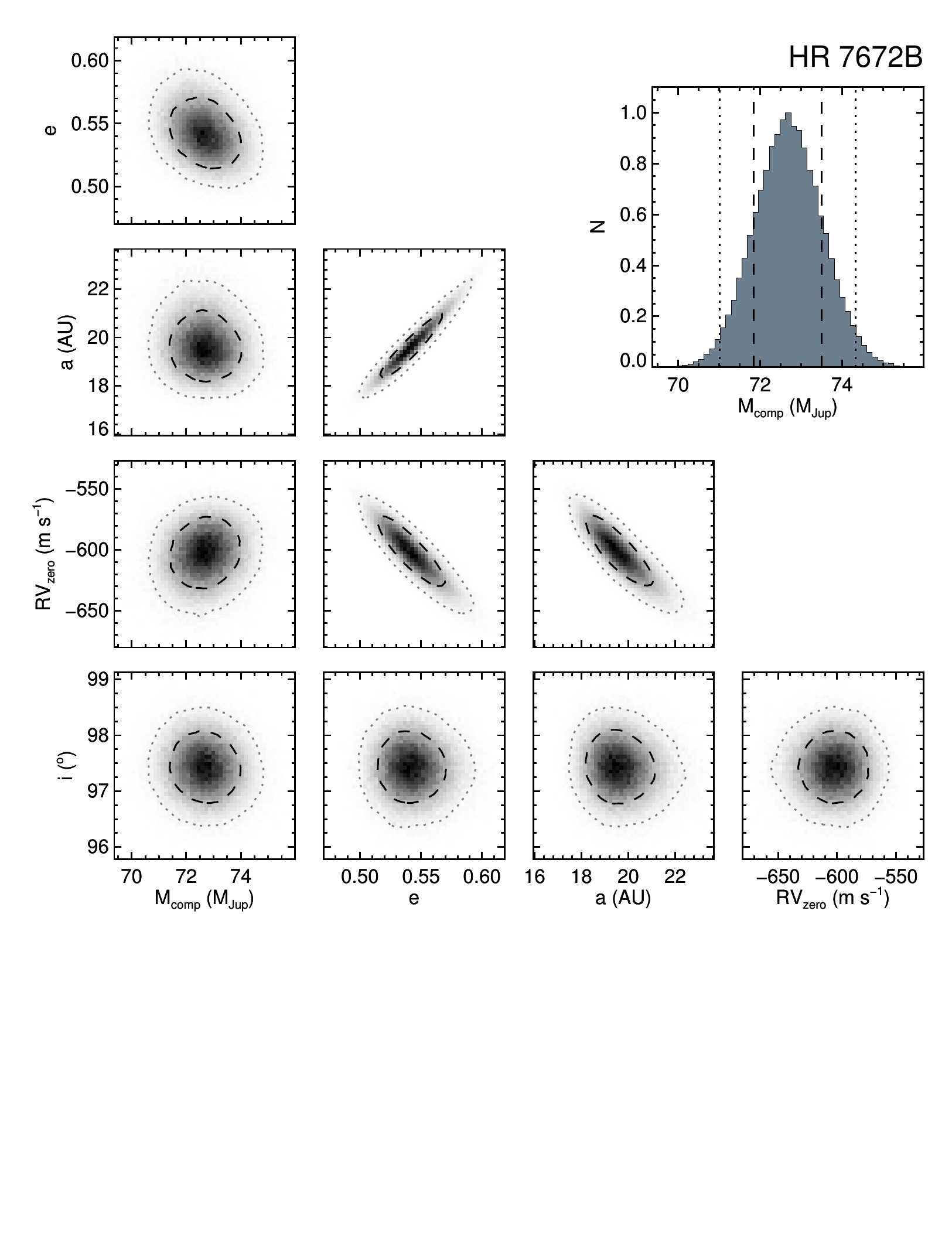}
\vskip -2.25 truein
\caption{Same as Figure~\ref{fig:HD4747-corner}.}
\label{fig:HR7672-corner}
\end{figure*}

\begin{figure*}
\includegraphics[width=1.0\linewidth]{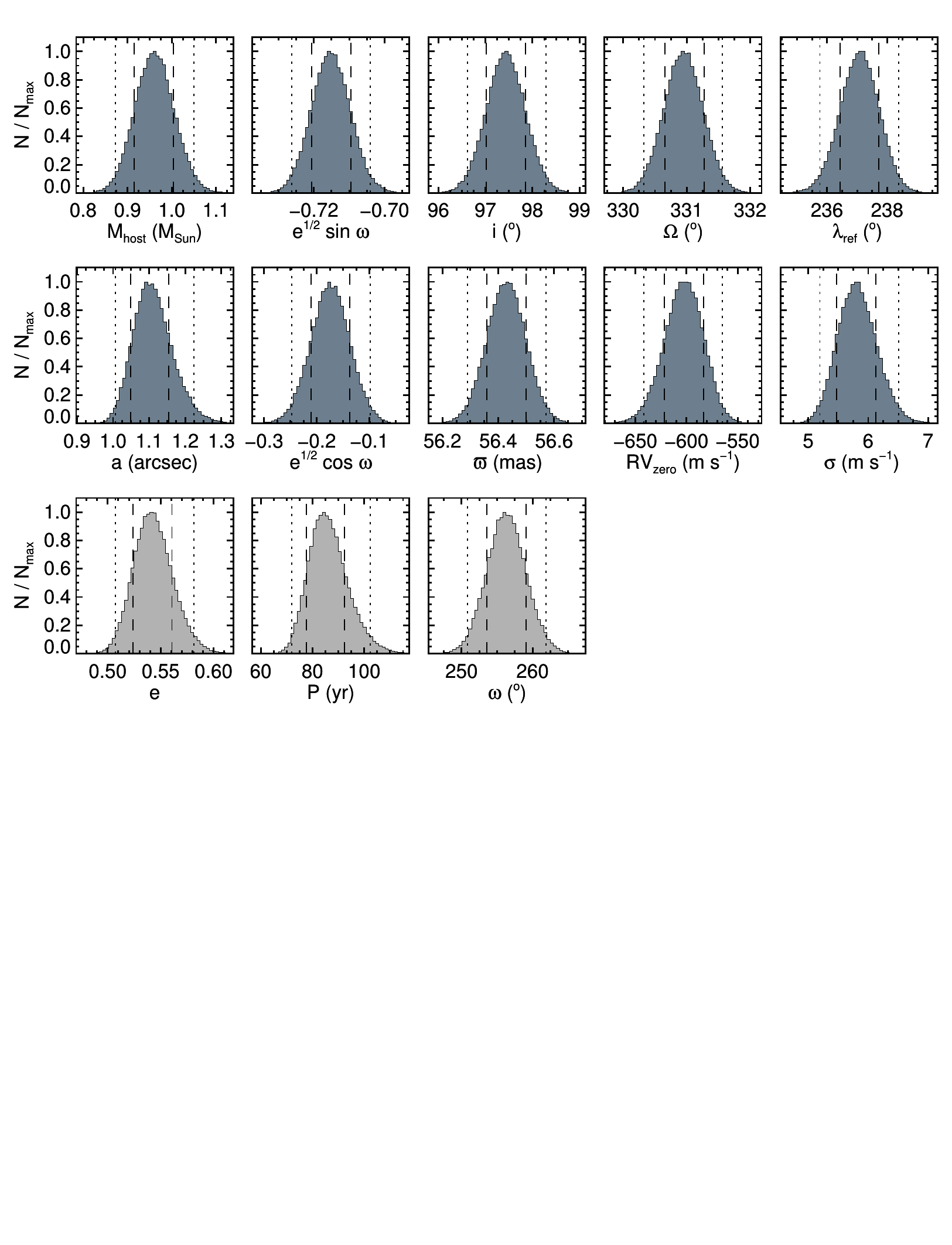}
\vskip -4.0 truein
\caption{Same as Figure~\ref{fig:HD4747-hist}.}
\label{fig:HR7672-hist}
\end{figure*}

\begin{figure*}
\centerline{
\includegraphics[width=0.4\linewidth]{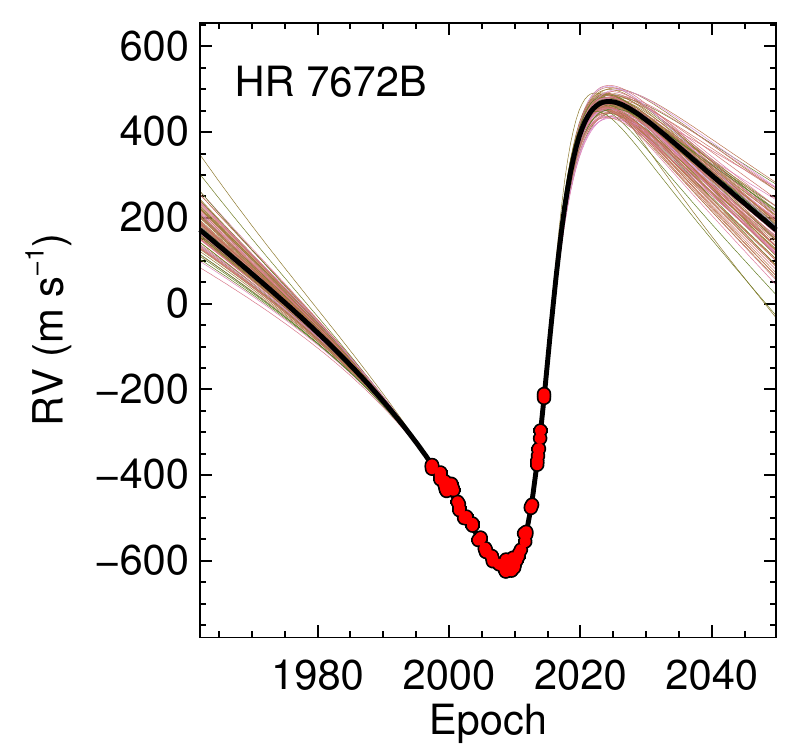}
\includegraphics[width=0.4\linewidth]{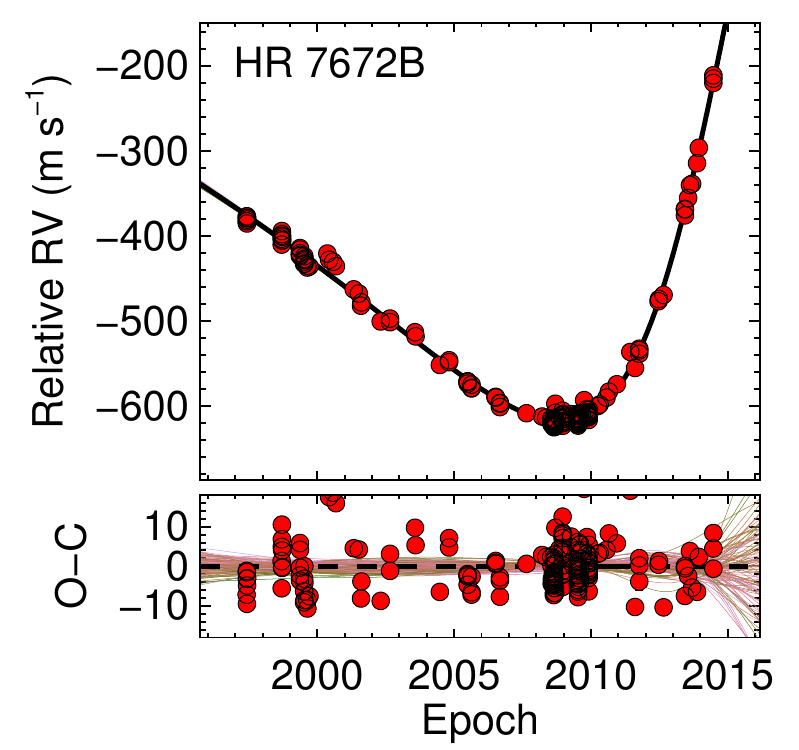}
\hskip -2.2in
\includegraphics[width=0.4\linewidth]{HR7672-colorbar.pdf}
}
\vskip 0.0 truein
\caption{Same as Figure~\ref{fig:HD4747-RV}.}
\label{fig:HR7672-RV}
\end{figure*}

\begin{figure*}
\centerline{
\includegraphics[width=0.23\linewidth]{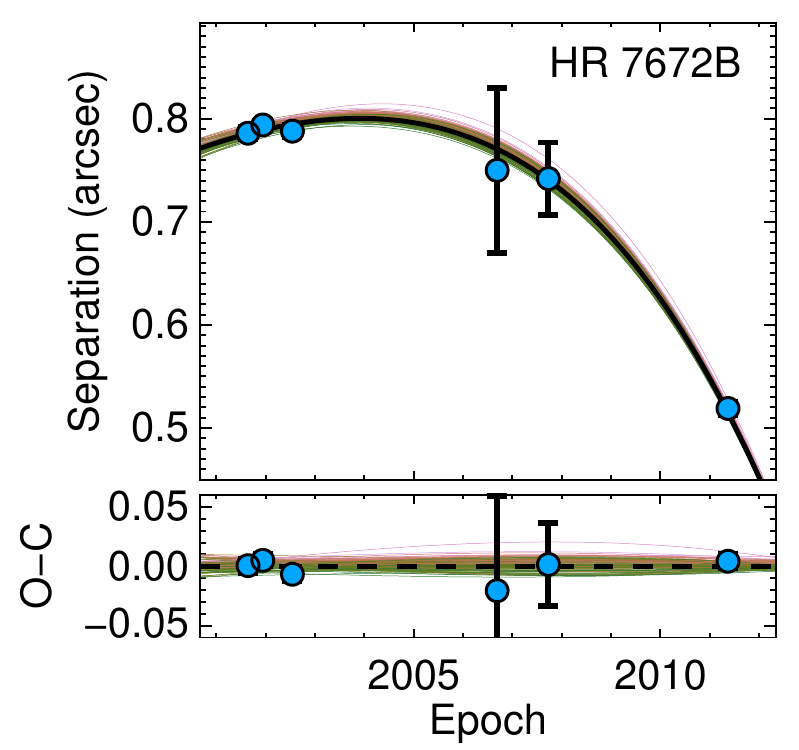}
\includegraphics[width=0.23\linewidth]{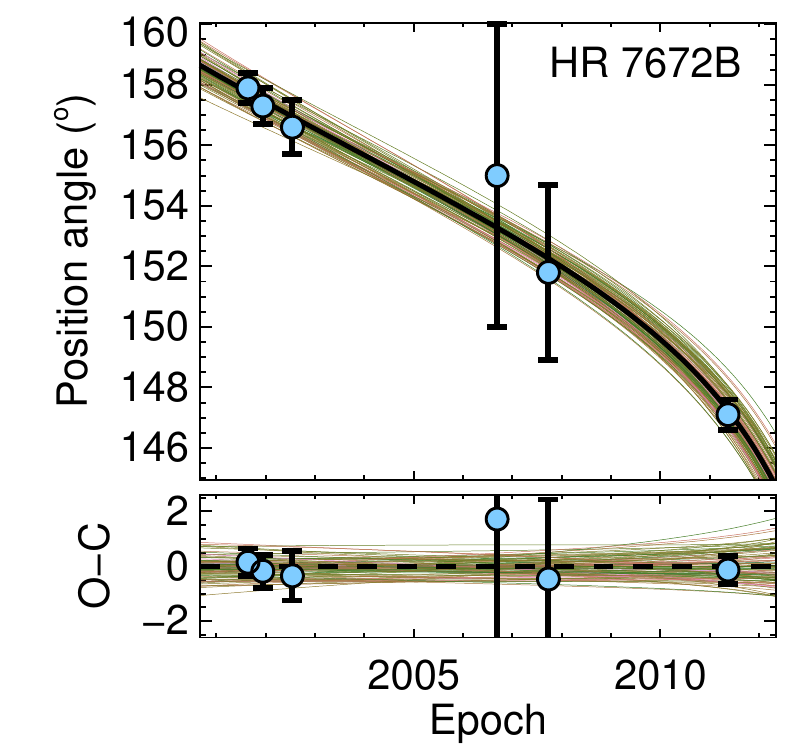}
\includegraphics[width=0.23\linewidth]{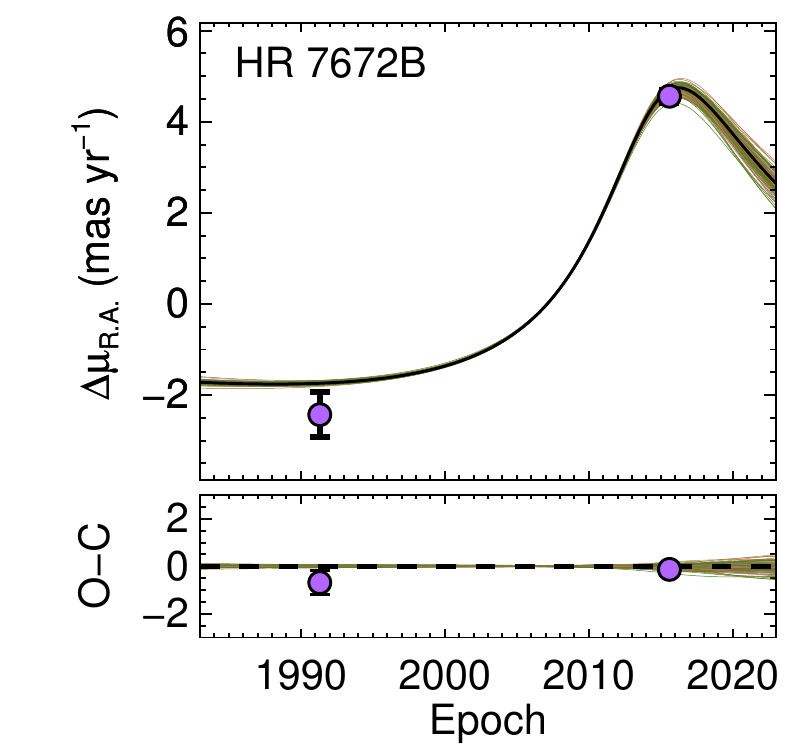}
\includegraphics[width=0.23\linewidth]{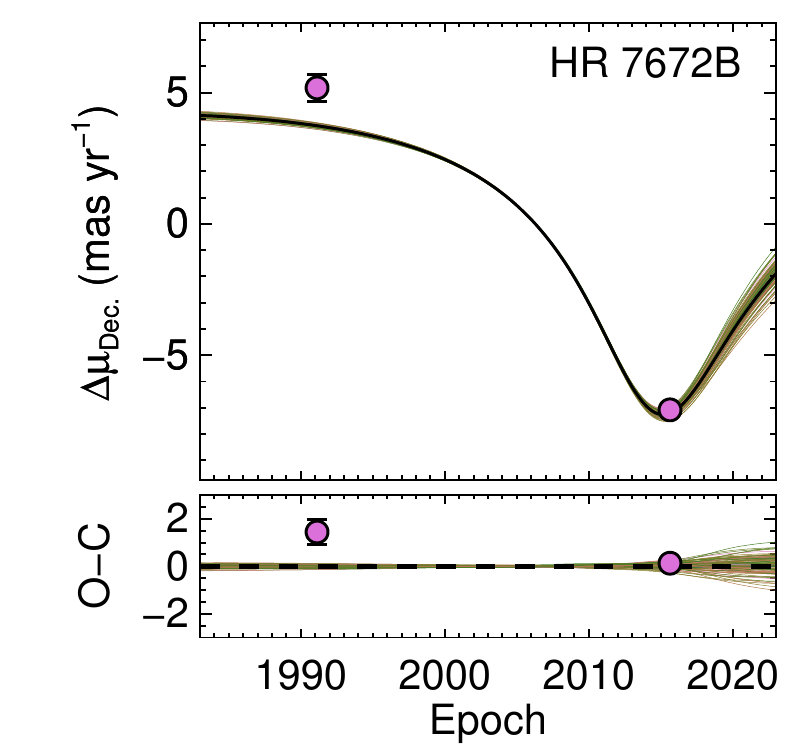}
\hskip -1.2in
\includegraphics[width=0.23\linewidth]{HR7672-colorbar.pdf}
}
\vskip 0.0 truein
\caption{Same as Figure~\ref{fig:HD4747-ast}.}
\label{fig:HR7672-ast}
\end{figure*}

\begin{deluxetable*}{lccc}
\tablecaption{MCMC Orbital Posteriors for HD 4747B \label{tbl:mcmc-HD4747}}
\setlength{\tabcolsep}{0.10in}
\tabletypesize{\tiny}
\tablewidth{0pt}
\tablehead{
\colhead{Property}              &
\colhead{Median $\pm$1$\sigma$} &
\colhead{95.4\% c.i.}           &
\colhead{Prior}                 }
\startdata
\multicolumn{4}{c}{Fitted parameters} \\[1pt]
\cline{1-4}
\multicolumn{4}{c}{} \\[-5pt]
Companion mass $M_{\rm comp}$ (\Mjup)                                       & $66.2_{-3.0}^{+2.5}$             &         61.1, 72.5         & $1/M$ (log-flat)                                                   \\[3pt]
Host-star mass $M_{\rm host}$ (\Msun)                                       & $0.82_{-0.08}^{+0.07}$           &         0.67, 1.00         & $1/M$ (log-flat)                                                   \\[3pt]
Parallax (mas)                                                              & $53.18\pm0.13$                   &        52.92, 53.44        & $\exp[-0.5((\varpi-\varpi_{\rm DR2})/\sigma[\varpi_{\rm DR2}])^2]$ \\[3pt]
Semimajor axis $a$ (AU)                                                     & $10.1_{-0.5}^{+0.4}$             &          9.3, 11.0         & $1/a$ (log-flat)                                                   \\[3pt]
Inclination $i$ (\degree)                                                   & $49.4_{-2.4}^{+2.3}$             &         44.9, 54.2         & $\sin(i)$, $0\degree < i < 180\degree$                             \\[3pt]
$\sqrt{e}\sin{\omega}$                                                      & $-0.8563_{-0.0017}^{+0.0018}$    &    $-$0.8601, $-$0.8528    & uniform                                                            \\[3pt]
$\sqrt{e}\cos{\omega}$                                                      & $-0.045_{-0.007}^{+0.008}$       &     $-$0.060, $-$0.030     & uniform                                                            \\[3pt]
Mean longitude at $t_{\rm ref}=2455197.5$~JD, $\lambda_{\rm ref}$ (\degree) & $44_{-3}^{+4}$                   &           36, 51           & uniform                                                            \\[3pt]
PA of the ascending node $\Omega$ (\degree)                                 & $91.5_{-1.8}^{+1.7}$             &         88.1, 95.4         & uniform                                                            \\[3pt]
RV zero point (m\,s$^{-1}$)                                      & $317_{-13}^{+11}$                &          294, 342          & uniform                                                            \\[3pt]
RV jitter $\sigma$ (m\,s$^{-1}$)                                 & $4.9_{-0.7}^{+0.6}$              &          3.8, 6.3          & $1/\sigma$ (log-flat)                                              \\[3pt]
\cline{1-4}
\multicolumn{4}{c}{} \\[-5pt]
\multicolumn{4}{c}{Computed properties} \\[1pt]
\cline{1-4}
\multicolumn{4}{c}{} \\[-5pt]
Orbital period $P$ (yr)                                                     & $34.0_{-1.0}^{+0.8}$             &         32.3, 36.1         & \nodata                                                            \\[3pt]
Semimajor axis (mas)                                                        & $535_{-25}^{+20}$                &          494, 585          & \nodata                                                            \\[3pt]
Eccentricity $e$                                                            & $0.7353_{-0.0029}^{+0.0027}$     &       0.7300, 0.7415       & \nodata                                                            \\[3pt]
Argument of periastron $\omega$ (\degree)                                   & $267.0\pm0.5$                    &        266.0, 268.0        & \nodata                                                            \\[3pt]
Time of periastron $T_0=t_{\rm ref}-P\frac{\lambda-\omega}{360\degree}$ (JD)& $2450471\pm5$                    &      2450460, 2450482      & \nodata                                                            \\[3pt]
Mass ratio $q = M_{\rm comp}/M_{\rm host}$                                  & $0.077_{-0.005}^{+0.004}$        &        0.069, 0.086        & \nodata                                                            \\[3pt]
\enddata
\tablecomments{The $\chi^2$ of relative astrometry is 7.69 for separations and 6.82 for PAs, with 8 measurements for each. The $\chi^2$ of the Hipparcos and Gaia proper motion differences is 3.09 for four measurements.}
\end{deluxetable*}
\begin{deluxetable*}{lccc}
\tablecaption{MCMC Orbital Posteriors for Gl~86B and Gl~86~b \label{tbl:mcmc-GL86}}
\setlength{\tabcolsep}{0.10in}
\tabletypesize{\tiny}
\tablewidth{0pt}
\tablehead{
\colhead{Property}              &
\colhead{Median $\pm$1$\sigma$} &
\colhead{95.4\% c.i.}           &
\colhead{Prior}                 }
\startdata
\multicolumn{4}{c}{Fitted parameters} \\[1pt]
\cline{1-4}
\multicolumn{4}{c}{} \\[-5pt]
Companion mass $M_{\rm comp}$ (\Mjup)                                       & $623\pm11$                       &          601, 646          & $1/M$ (log-flat)                                                   \\[3pt]
Host-star mass $M_{\rm host}$ (\Msun)                                       & $1.36\pm0.23$                    &         0.88, 1.83         & $1/M$ (log-flat)                                                   \\[3pt]
Parallax (mas)                                                              & $92.70\pm0.05$                   &        92.60, 92.80        & $\exp[-0.5((\varpi-\varpi_{\rm DR2})/\sigma[\varpi_{\rm DR2}])^2]$ \\[3pt]
Semimajor axis $a$ (AU)                                                     & $21.7_{-0.7}^{+0.5}$             &         20.6, 23.4         & $1/a$ (log-flat)                                                   \\[3pt]
Inclination $i$ (\degree)                                                   & $125.5_{-0.9}^{+0.8}$            &        123.8, 127.2        & $\sin(i)$, $0\degree < i < 180\degree$                             \\[3pt]
$\sqrt{e}\sin{\omega}$                                                      & $-0.54_{-0.05}^{+0.04}$          &      $-$0.62, $-$0.42      & uniform                                                            \\[3pt]
$\sqrt{e}\cos{\omega}$                                                      & $0.491_{-0.017}^{+0.014}$        &        0.462, 0.527        & uniform                                                            \\[3pt]
Mean longitude at $t_{\rm ref}=2455197.5$~JD, $\lambda_{\rm ref}$ (\degree) & $109_{-5}^{+6}$                  &           98, 120          & uniform                                                            \\[3pt]
PA of the ascending node $\Omega$ (\degree)                                 & $232.4_{-1.5}^{+1.7}$            &        229.0, 235.4        & uniform                                                            \\[3pt]
Semiamplitude of Gl~86~b (m\,s$^{-1}$)                                      & $378.9\pm1.0$                    &        376.9, 381.1        & $1/K_1$ (log-flat)                                                 \\[3pt]
Orbital period of Gl~86~b $P_{\rm pl}$ (d)                                  & $15.76486_{-0.00017}^{+0.00016}$ &     15.76453, 15.76518     & $1/P_{\rm pl}$ (log-flat)                                          \\[3pt]
Mean longitude of Gl~86~b at $t_{\rm ref}$ $\lambda_{\rm ref,plx}$ (\degree)& $252.3\pm0.6$                    &        251.0, 253.6        & uniform                                                            \\[3pt]
$\sqrt{e_{\rm pl}}\sin{\omega_{\rm pl}}$                                    & $-0.223_{-0.007}^{+0.006}$       &     $-$0.235, $-$0.209     & uniform                                                            \\[3pt]
$\sqrt{e_{\rm pl}}\cos{\omega_{\rm pl}}$                                    & $-0.001\pm0.018$                 &     $-$0.037, 0.036        & uniform                                                            \\[3pt]
RV zero point (m\,s$^{-1}$)                                      & $230\pm40$                       &          160, 310          & uniform                                                            \\[3pt]
RV jitter $\sigma$ (m\,s$^{-1}$)                                 & $0.00028_{-0.00028}^{+0.05988}$  &      0.00000, 3.72509      & $1/\sigma$ (log-flat)                                              \\[3pt]
\cline{1-4}
\multicolumn{4}{c}{} \\[-5pt]
\multicolumn{4}{c}{Computed properties} \\[1pt]
\cline{1-4}
\multicolumn{4}{c}{} \\[-5pt]
Orbital period $P$ (yr)                                                     & $72_{-8}^{+7}$                   &           59, 92           & \nodata                                                            \\[3pt]
Semimajor axis (mas)                                                        & $2010_{-70}^{+50}$               &         1910, 2170         & \nodata                                                            \\[3pt]
Eccentricity $e$                                                            & $0.53_{-0.03}^{+0.04}$           &         0.45, 0.60         & \nodata                                                            \\[3pt]
Argument of periastron $\omega$ (\degree)                                   & $312.5_{-3.5}^{+2.8}$            &        306.8, 321.1        & \nodata                                                            \\[3pt]
Time of periastron $T_0=t_{\rm ref}-P\frac{\lambda-\omega}{360\degree}$ (JD)& $2443700_{-600}^{+700}$          &      2442300, 2444800      & \nodata                                                            \\[3pt]
Mass ratio $q = M_{\rm comp}/M_{\rm host}$                                  & $0.44_{-0.08}^{+0.06}$           &         0.31, 0.62         & \nodata                                                            \\[3pt]
Eccentricity of Gl~86~b $e_{\rm pl}$                                        & $0.0498_{-0.0029}^{+0.0028}$     &       0.0440, 0.0557       & \nodata                                                            \\[3pt]
Argument of periastron of Gl~86~b $\omega_{\rm pl}$ (\degree)               & $270\pm5$                        &          260, 279          & \nodata                                                            \\[3pt]
Time of periastron of Gl~86~b $T_{0,{\rm pl}}$ (JD)                         & $2455198.25_{-0.21}^{+0.22}$     &   2455197.75, 2455198.75   & \nodata                                                            \\[3pt]
\enddata
\tablecomments{The $\chi^2$ of relative astrometry is 7.39 for separations and 5.89 for PAs, with 6 measurements for each. The $\chi^2$ of the Hipparcos and Gaia proper motion differences is 1.39 for four measurements.}
\end{deluxetable*}
\begin{deluxetable*}{lccc}
\tablecaption{MCMC Orbital Posteriors for HD 68017B \label{tbl:mcmc-HD68017}}
\setlength{\tabcolsep}{0.10in}
\tabletypesize{\tiny}
\tablewidth{0pt}
\tablehead{
\colhead{Property}              &
\colhead{Median $\pm$1$\sigma$} &
\colhead{95.4\% c.i.}           &
\colhead{Prior}                 }
\startdata
\multicolumn{4}{c}{Fitted parameters} \\[1pt]
\cline{1-4}
\multicolumn{4}{c}{} \\[-5pt]
Companion mass $M_{\rm comp}$ (\Mjup)                                       & $154\pm3$                        &          148, 161          & $1/M$ (log-flat)                                                   \\[3pt]
Host-star mass $M_{\rm host}$ (\Msun)                                       & $0.98\pm0.07$                    &         0.83, 1.12         & $1/M$ (log-flat)                                                   \\[3pt]
Parallax (mas)                                                              & $46.33\pm0.06$                   &        46.21, 46.45        & $\exp[-0.5((\varpi-\varpi_{\rm DR2})/\sigma[\varpi_{\rm DR2}])^2]$ \\[3pt]
Semimajor axis $a$ (AU)                                                     & $16.0_{-1.2}^{+1.0}$             &         14.1, 18.5         & $1/a$ (log-flat)                                                   \\[3pt]
Inclination $i$ (\degree)                                                   & $170.3\pm0.4$                    &        169.4, 171.1        & $\sin(i)$, $0\degree < i < 180\degree$                             \\[3pt]
$\sqrt{e}\sin{\omega}$                                                      & $-0.56_{-0.04}^{+0.03}$          &      $-$0.62, $-$0.48      & uniform                                                            \\[3pt]
$\sqrt{e}\cos{\omega}$                                                      & $0.10\pm0.09$                    &      $-$0.08, 0.28         & uniform                                                            \\[3pt]
Mean longitude at $t_{\rm ref}=2455197.5$~JD, $\lambda_{\rm ref}$ (\degree) & $235.1\pm0.9$                    &        233.3, 236.9        & uniform                                                            \\[3pt]
PA of the ascending node $\Omega$ (\degree)                                 & $98.0\pm0.9$                     &         96.2, 99.7         & uniform                                                            \\[3pt]
RV zero point (m\,s$^{-1}$)                                      & $-157_{-13}^{+14}$               &       $-$186, $-$132       & uniform                                                            \\[3pt]
RV jitter $\sigma$ (m\,s$^{-1}$)                                 & $5.7_{-0.4}^{+0.3}$              &          5.0, 6.5          & $1/\sigma$ (log-flat)                                              \\[3pt]
\cline{1-4}
\multicolumn{4}{c}{} \\[-5pt]
\multicolumn{4}{c}{Computed properties} \\[1pt]
\cline{1-4}
\multicolumn{4}{c}{} \\[-5pt]
Orbital period $P$ (yr)                                                     & $60_{-8}^{+6}$                   &           48, 78           & \nodata                                                            \\[3pt]
Semimajor axis (mas)                                                        & $740_{-60}^{+40}$                &          650, 860          & \nodata                                                            \\[3pt]
Eccentricity $e$                                                            & $0.325_{-0.024}^{+0.017}$        &        0.292, 0.382        & \nodata                                                            \\[3pt]
Argument of periastron $\omega$ (\degree)                                   & $280\pm10$                       &          262, 300          & \nodata                                                            \\[3pt]
Time of periastron $T_0=t_{\rm ref}-P\frac{\lambda-\omega}{360\degree}$ (JD)& $2435900_{-2400}^{+3300}$        &      2429000, 2440900      & \nodata                                                            \\[3pt]
Mass ratio $q = M_{\rm comp}/M_{\rm host}$                                  & $0.150_{-0.012}^{+0.010}$        &        0.130, 0.174        & \nodata                                                            \\[3pt]
\enddata
\tablecomments{The $\chi^2$ of relative astrometry is 1.01 for separations and 0.0121 for PAs, with 2 measurements for each. The $\chi^2$ of the Hipparcos and Gaia proper motion differences is 9.60 for four measurements.}
\end{deluxetable*}
\begin{deluxetable*}{lccc}
\tablecaption{MCMC Orbital Posteriors for Gl~758B \label{tbl:mcmc-GL758}}
\setlength{\tabcolsep}{0.10in}
\tabletypesize{\tiny}
\tablewidth{0pt}
\tablehead{
\colhead{Property}              &
\colhead{Median $\pm$1$\sigma$} &
\colhead{95.4\% c.i.}           &
\colhead{Prior}                 }
\startdata
\multicolumn{4}{c}{Fitted parameters} \\[1pt]
\cline{1-4}
\multicolumn{4}{c}{} \\[-5pt]
Companion mass $M_{\rm comp}$ (\Mjup)                                       & $38.1_{-1.5}^{+1.7}$             &         35.1, 41.3         & $1/M$ (log-flat)                                                   \\[3pt]
Host-star mass $M_{\rm host}$ (\Msun)                                       & $0.76_{-0.27}^{+0.13}$           &         0.45, 1.21         & $1/M$ (log-flat)                                                   \\[3pt]
Parallax (mas)                                                              & $64.061\pm0.022$                 &       64.018, 64.106       & $\exp[-0.5((\varpi-\varpi_{\rm DR2})/\sigma[\varpi_{\rm DR2}])^2]$ \\[3pt]
Semimajor axis $a$ (AU)                                                     & $30_{-8}^{+5}$                   &           21, 46           & $1/a$ (log-flat)                                                   \\[3pt]
Inclination $i$ (\degree)                                                   & $41\pm6$                         &           29, 53           & $\sin(i)$, $0\degree < i < 180\degree$                             \\[3pt]
$\sqrt{e}\sin{\omega}$                                                      & $0.27_{-0.20}^{+0.35}$           &      $-$0.17, 0.70         & uniform                                                            \\[3pt]
$\sqrt{e}\cos{\omega}$                                                      & $-0.53_{-0.18}^{+0.12}$          &      $-$0.74, $-$0.14      & uniform                                                            \\[3pt]
Mean longitude at $t_{\rm ref}=2455197.5$~JD, $\lambda_{\rm ref}$ (\degree) & $73_{-16}^{+14}$                 &           42, 101          & uniform                                                            \\[3pt]
PA of the ascending node $\Omega$ (\degree)                                 & $175_{-4}^{+6}$                  &          161, 184          & uniform                                                            \\[3pt]
McDonald RV zero point (m\,s$^{-1}$)                                      & $85_{-27}^{+26}$                 &           34, 135          & uniform                                                            \\[3pt]
McDonald RV jitter $\sigma$ (m\,s$^{-1}$)                                 & $2.9_{-0.8}^{+1.1}$              &          0.0, 4.2          & $1/\sigma$ (log-flat)                                              \\[3pt]
Keck RV zero point (m\,s$^{-1}$)                                      & $74_{-28}^{+25}$                 &           24, 125          & uniform                                                            \\[3pt]
Keck RV jitter $\sigma$ (m\,s$^{-1}$)                                 & $2.35_{-0.18}^{+0.16}$           &         2.00, 2.71         & $1/\sigma$ (log-flat)                                              \\[3pt]
APF RV zero point (m\,s$^{-1}$)                                      & $67_{-27}^{+26}$                 &           16, 117          & uniform                                                            \\[3pt]
APF RV jitter $\sigma$ (m\,s$^{-1}$)                                 & $2.50_{-0.19}^{+0.17}$           &         2.15, 2.87         & $1/\sigma$ (log-flat)                                              \\[3pt]
\cline{1-4}
\multicolumn{4}{c}{} \\[-5pt]
\multicolumn{4}{c}{Computed properties} \\[1pt]
\cline{1-4}
\multicolumn{4}{c}{} \\[-5pt]
Orbital period $P$ (yr)                                                     & $180_{-90}^{+60}$                &           80, 420          & \nodata                                                            \\[3pt]
Semimajor axis (mas)                                                        & $1910_{-510}^{+300}$             &         1330, 2980         & \nodata                                                            \\[3pt]
Eccentricity $e$                                                            & $0.40\pm0.09$                    &         0.22, 0.59         & \nodata                                                            \\[3pt]
Argument of periastron $\omega$ (\degree)                                   & $150\pm30$                       &          100, 200          & \nodata                                                            \\[3pt]
Time of periastron $T_0=t_{\rm ref}-P\frac{\lambda-\omega}{360\degree}$ (JD)& $2468700_{-1100}^{+1300}$        &      2465800, 2472000      & \nodata                                                            \\[3pt]
Mass ratio $q = M_{\rm comp}/M_{\rm host}$                                  & $0.048_{-0.015}^{+0.011}$        &        0.026, 0.077        & \nodata                                                            \\[3pt]
\enddata
\tablecomments{The $\chi^2$ of relative astrometry is 2.70 for separations and 3.94 for PAs, with 4 measurements for each. The $\chi^2$ of the Hipparcos and Gaia proper motion differences is 3.95 for four measurements.}
\end{deluxetable*}
\begin{deluxetable*}{lccc}
\tablecaption{MCMC Orbital Posteriors for HR 7672B \label{tbl:mcmc-HR7672}}
\setlength{\tabcolsep}{0.10in}
\tabletypesize{\tiny}
\tablewidth{0pt}
\tablehead{
\colhead{Property}              &
\colhead{Median $\pm$1$\sigma$} &
\colhead{95.4\% c.i.}           &
\colhead{Prior}                 }
\startdata
\multicolumn{4}{c}{Fitted parameters} \\[1pt]
\cline{1-4}
\multicolumn{4}{c}{} \\[-5pt]
Companion mass $M_{\rm comp}$ (\Mjup)                                       & $72.7\pm0.8$                     &         71.0, 74.3         & $1/M$ (log-flat)                                                   \\[3pt]
Host-star mass $M_{\rm host}$ (\Msun)                                       & $0.96_{-0.05}^{+0.04}$           &         0.87, 1.05         & $1/M$ (log-flat)                                                   \\[3pt]
Parallax (mas)                                                              & $56.43\pm0.07$                   &        56.29, 56.57        & $\exp[-0.5((\varpi-\varpi_{\rm DR2})/\sigma[\varpi_{\rm DR2}])^2]$ \\[3pt]
Semimajor axis $a$ (AU)                                                     & $19.6_{-1.0}^{+0.8}$             &         17.9, 21.7         & $1/a$ (log-flat)                                                   \\[3pt]
Inclination $i$ (\degree)                                                   & $97.4\pm0.4$                     &         96.6, 98.3         & $\sin(i)$, $0\degree < i < 180\degree$                             \\[3pt]
$\sqrt{e}\sin{\omega}$                                                      & $-0.715_{-0.005}^{+0.006}$       &     $-$0.726, $-$0.704     & uniform                                                            \\[3pt]
$\sqrt{e}\cos{\omega}$                                                      & $-0.17\pm0.04$                   &      $-$0.25, $-$0.10      & uniform                                                            \\[3pt]
Mean longitude at $t_{\rm ref}=2455197.5$~JD, $\lambda_{\rm ref}$ (\degree) & $237.1_{-0.7}^{+0.6}$            &        235.8, 238.4        & uniform                                                            \\[3pt]
PA of the ascending node $\Omega$ (\degree)                                 & $330.95_{-0.30}^{+0.32}$         &       330.32, 331.56       & uniform                                                            \\[3pt]
RV zero point (m\,s$^{-1}$)                                      & $-602\pm19$                      &       $-$642, $-$565       & uniform                                                            \\[3pt]
RV jitter $\sigma$ (m\,s$^{-1}$)                                 & $5.8\pm0.3$                      &          5.2, 6.5          & $1/\sigma$ (log-flat)                                              \\[3pt]
\cline{1-4}
\multicolumn{4}{c}{} \\[-5pt]
\multicolumn{4}{c}{Computed properties} \\[1pt]
\cline{1-4}
\multicolumn{4}{c}{} \\[-5pt]
Orbital period $P$ (yr)                                                     & $86_{-8}^{+7}$                   &           72, 102          & \nodata                                                            \\[3pt]
Semimajor axis (mas)                                                        & $1110_{-60}^{+50}$               &         1010, 1220         & \nodata                                                            \\[3pt]
Eccentricity $e$                                                            & $0.542\pm0.018$                  &        0.507, 0.581        & \nodata                                                            \\[3pt]
Argument of periastron $\omega$ (\degree)                                   & $256.3_{-2.7}^{+2.8}$            &        250.9, 261.9        & \nodata                                                            \\[3pt]
Time of periastron $T_0=t_{\rm ref}-P\frac{\lambda-\omega}{360\degree}$ (JD)& $2425600_{-2600}^{+2900}$        &      2419400, 2430700      & \nodata                                                            \\[3pt]
Mass ratio $q = M_{\rm comp}/M_{\rm host}$                                  & $0.0722\pm0.0030$                &       0.0664, 0.0784       & \nodata                                                            \\[3pt]
\enddata
\tablecomments{The $\chi^2$ of relative astrometry is 2.83 for separations and 0.553 for PAs, with 6 measurements for each. The $\chi^2$ of the Hipparcos and Gaia proper motion differences is 11.4 for four measurements.}
\end{deluxetable*}

\clearpage

\bibliographystyle{apj_eprint}
\bibliography{refs.bib}

\begin{thebibliography}{}

\bibitem[\protect\citeauthoryear{{Baliunas} et~al.}{{Baliunas}
  et~al.}{1995}]{Baliulnas+Donahue+Soon+etal_1995}
{Baliunas}, S.~L., {Donahue}, R.~A., {Soon}, W.~H., et~al. 1995, \apj, 438, 269

\bibitem[\protect\citeauthoryear{{Baraffe} et~al.}{{Baraffe}
  et~al.}{2003}]{Baraffe+Chabrier+Barman+etal_2003}
{Baraffe}, I., {Chabrier}, G., {Barman}, T.~S., {Allard}, F.,  \& {Hauschildt},
  P.~H. 2003, \aap, 402, 701

\bibitem[\protect\citeauthoryear{{Barnes}}{{Barnes}}{2003}]{Barnes_2003}
{Barnes}, S.~A. 2003, \apj, 586, 464

\bibitem[\protect\citeauthoryear{{Benedict} et~al.}{{Benedict}
  et~al.}{2002}]{2002ApJ...581L.115B}
{Benedict}, G.~F., {McArthur}, B.~E., {Forveille}, T., et~al. 2002, \apjl, 581,
  L115

\bibitem[\protect\citeauthoryear{{Beuzit} et~al.}{{Beuzit}
  et~al.}{1997}]{1997ExA.....7..285B}
{Beuzit}, J.-L., {Demailly}, L., {Gendron}, E., et~al. 1997, Experimental
  Astronomy, 7, 285

\bibitem[\protect\citeauthoryear{{Boccaletti} et~al.}{{Boccaletti}
  et~al.}{2003}]{Boccaletti+Chauvin+Lagrange+etal_2003}
{Boccaletti}, A., {Chauvin}, G., {Lagrange}, A.-M.,  \& {Marchis}, F. 2003,
  \aap, 410, 283

\bibitem[\protect\citeauthoryear{{Bond}, {Bergeron}, \& {B{\'e}dard}}{{Bond}
  et~al.}{2017}]{Bond+Bergeron+Bedard_2017}
{Bond}, H.~E., {Bergeron}, P.,  \& {B{\'e}dard}, A. 2017, \apj, 848, 16

\bibitem[\protect\citeauthoryear{{Bowler} et~al.}{{Bowler}
  et~al.}{2018}]{Bowler+Dupuy+Endl+etal_2018}
{Bowler}, B.~P., {Dupuy}, T.~J., {Endl}, M., et~al. 2018, \aj, 155, 159

\bibitem[\protect\citeauthoryear{{Bowler} et~al.}{{Bowler}
  et~al.}{2015}]{Bowler+Liu+Shkolnik+etal_2015}
{Bowler}, B.~P., {Liu}, M.~C., {Shkolnik}, E.~L.,  \& {Tamura}, M. 2015, \apjs,
  216, 7

\bibitem[\protect\citeauthoryear{{Brandt}}{{Brandt}}{2018}]{Brandt_2018}
{Brandt}, T.~D. 2018, submitted

\bibitem[\protect\citeauthoryear{{Brandt} et~al.}{{Brandt}
  et~al.}{2014}]{Brandt+Kuzuhara+McElwain+etal_2014}
{Brandt}, T.~D., {Kuzuhara}, M., {McElwain}, M.~W., et~al. 2014, \apj, 786, 1

\bibitem[\protect\citeauthoryear{{Bressan} et~al.}{{Bressan}
  et~al.}{2012}]{Bressan+Marigo+Girardi+etal_2012}
{Bressan}, A., {Marigo}, P., {Girardi}, L., et~al. 2012, \mnras, 427, 127

\bibitem[\protect\citeauthoryear{{Brewer} et~al.}{{Brewer}
  et~al.}{2016}]{Brewer+Fischer+Valenti+etal_2016}
{Brewer}, J.~M., {Fischer}, D.~A., {Valenti}, J.~A.,  \& {Piskunov}, N. 2016,
  The Astrophysical Journal Supplement Series, 225

\bibitem[\protect\citeauthoryear{{Burrows} et~al.}{{Burrows}
  et~al.}{1997}]{Burrows+Marley+Hubbard+etal_1997}
{Burrows}, A., {Marley}, M., {Hubbard}, W.~B., et~al. 1997, \apj, 491, 856

\bibitem[\protect\citeauthoryear{{Butler} et~al.}{{Butler}
  et~al.}{2017}]{Butler+Vogt+Laughlin+etal_2017}
{Butler}, R.~P., {Vogt}, S.~S., {Laughlin}, G., et~al. 2017, \aj, 153, 208

\bibitem[\protect\citeauthoryear{{Butler} et~al.}{{Butler}
  et~al.}{2006}]{Butler+Wright+Marcy+etal_2006}
{Butler}, R.~P., {Wright}, J.~T., {Marcy}, G.~W., et~al. 2006, \apj, 646, 505

\bibitem[\protect\citeauthoryear{{Calissendorff} \& {Janson}}{{Calissendorff}
  \& {Janson}}{2018}]{Calissendorff+Janson_2018}
{Calissendorff}, P.,  \& {Janson}, M. 2018, ArXiv e-prints, 1806.07899

\bibitem[\protect\citeauthoryear{{Cheetham} et~al.}{{Cheetham}
  et~al.}{2018}]{Cheetham+Segransan+Peretti+etal_2018}
{Cheetham}, A., {S{\'e}gransan}, D., {Peretti}, S., et~al. 2018, \aap, 614, A16

\bibitem[\protect\citeauthoryear{{Cochran} et~al.}{{Cochran}
  et~al.}{1997}]{Cochran+Hatzes+Butler+etal_1997}
{Cochran}, W.~D., {Hatzes}, A.~P., {Butler}, R.~P.,  \& {Marcy}, G.~W. 1997,
  \apj, 483, 457

\bibitem[\protect\citeauthoryear{{Crepp} et~al.}{{Crepp}
  et~al.}{2016}]{Crepp+Gonzales+Bechter+etal_2016}
{Crepp}, J.~R., {Gonzales}, E.~J., {Bechter}, E.~B., et~al. 2016, \apj, 831,
  136

\bibitem[\protect\citeauthoryear{{Crepp} et~al.}{{Crepp}
  et~al.}{2012a}]{Crepp+Johnson+Fischer+etal_2012}
{Crepp}, J.~R., {Johnson}, J.~A., {Fischer}, D.~A., et~al. 2012a, \apj, 751, 97

\bibitem[\protect\citeauthoryear{{Crepp} et~al.}{{Crepp}
  et~al.}{2012b}]{Crepp+Johnson+Howard+etal_2012}
{Crepp}, J.~R., {Johnson}, J.~A., {Howard}, A.~W., et~al. 2012b, \apj, 761, 39

\bibitem[\protect\citeauthoryear{{Crepp} et~al.}{{Crepp}
  et~al.}{2018}]{Crepp+Principe+Wolff+etal_2018}
{Crepp}, J.~R., {Principe}, D.~A., {Wolff}, S., et~al. 2018, \apj, 853, 192

\bibitem[\protect\citeauthoryear{{Crepp} et~al.}{{Crepp}
  et~al.}{2015}]{Crepp+Rice+Veicht+etal_2015}
{Crepp}, J.~R., {Rice}, E.~L., {Veicht}, A., et~al. 2015, \apjl, 798, L43

\bibitem[\protect\citeauthoryear{{Cumming}, {Marcy}, \& {Butler}}{{Cumming}
  et~al.}{1999}]{Cumming+Marcy+Butler_1999}
{Cumming}, A., {Marcy}, G.~W.,  \& {Butler}, R.~P. 1999, \apj, 526, 890

\bibitem[\protect\citeauthoryear{{Cutri} et~al.}{{Cutri}
  et~al.}{2003}]{Cutri+Skrutskie+vanDyk+etal_2003}
{Cutri}, R.~M., {Skrutskie}, M.~F., {van Dyk}, S., et~al. 2003, VizieR Online
  Data Catalog, 2246

\bibitem[\protect\citeauthoryear{{Delfosse} et~al.}{{Delfosse}
  et~al.}{2000}]{Delfosse+Forveille+Segransan+etal_2000}
{Delfosse}, X., {Forveille}, T., {S{\'e}gransan}, D., et~al. 2000, \aap, 364,
  217

\bibitem[\protect\citeauthoryear{{Demarque} et~al.}{{Demarque}
  et~al.}{2004}]{Demarque+Woo+Kim+etal_2004}
{Demarque}, P., {Woo}, J.-H., {Kim}, Y.-C.,  \& {Yi}, S.~K. 2004, \apjs, 155,
  667

\bibitem[\protect\citeauthoryear{{Diego} et~al.}{{Diego}
  et~al.}{1990}]{Diego+Charalambous+Fish+etal_1990}
{Diego}, F., {Charalambous}, A., {Fish}, A.~C.,  \& {Walker}, D.~D. 1990, in
  \procspie, Vol. 1235, Instrumentation in Astronomy VII, ed. D.~L. {Crawford},
  562

\bibitem[\protect\citeauthoryear{{Dieterich} et~al.}{{Dieterich}
  et~al.}{2018}]{2018arXiv180709880D}
{Dieterich}, S.~B., {Weinberger}, A.~J., {Boss}, A.~P., et~al. 2018, ArXiv
  e-prints, arXiv:1807.09880

\bibitem[\protect\citeauthoryear{{Dotter} et~al.}{{Dotter}
  et~al.}{2008}]{2008ApJS..178...89D}
{Dotter}, A., {Chaboyer}, B., {Jevremovi{\'c}}, D., et~al. 2008, The
  Astrophysical Journal Supplement Series, 178, 89

\bibitem[\protect\citeauthoryear{{Ducati}}{{Ducati}}{2002}]{Ducati_2002}
{Ducati}, J.~R. 2002, VizieR Online Data Catalog

\bibitem[\protect\citeauthoryear{{Dupuy} \& {Liu}}{{Dupuy} \&
  {Liu}}{2017}]{Dupuy+Liu_2017}
{Dupuy}, T.~J.,  \& {Liu}, M.~C. 2017, \apjs, 231, 15

\bibitem[\protect\citeauthoryear{{Dupuy}, {Liu}, \& {Ireland}}{{Dupuy}
  et~al.}{2009a}]{Dupuy+Liu+Ireland_2009a}
{Dupuy}, T.~J., {Liu}, M.~C.,  \& {Ireland}, M.~J. 2009a, \apj, 692, 729

\bibitem[\protect\citeauthoryear{{Dupuy}, {Liu}, \& {Ireland}}{{Dupuy}
  et~al.}{2009b}]{Dupuy+Liu+Ireland_2009b}
{Dupuy}, T.~J., {Liu}, M.~C.,  \& {Ireland}, M.~J. 2009b, \apj, 699, 168

\bibitem[\protect\citeauthoryear{{Earl} \& {Deem}}{{Earl} \&
  {Deem}}{2005}]{Earl+Deem_2005}
{Earl}, D.~J.,  \& {Deem}, M.~W. 2005, Physical Chemistry Chemical Physics
  (Incorporating Faraday Transactions), 7, 3910

\bibitem[\protect\citeauthoryear{{Els} et~al.}{{Els}
  et~al.}{2001}]{Els+Sterzik+Marchis+etal_2001}
{Els}, S.~G., {Sterzik}, M.~F., {Marchis}, F., et~al. 2001, \aap, 370, L1

\bibitem[\protect\citeauthoryear{{Endl} et~al.}{{Endl}
  et~al.}{2016}]{Endl+Brugamver+Cochran+etal_2016}
{Endl}, M., {Brugamyer}, E.~J., {Cochran}, W.~D., et~al. 2016, \apj, 818, 34

\bibitem[\protect\citeauthoryear{{ESA}}{{ESA}}{1997}]{ESA_1997}
{ESA}, ed. 1997, ESA Special Publication, Vol. 1200, {The HIPPARCOS and TYCHO
  catalogues. Astrometric and photometric star catalogues derived from the ESA
  HIPPARCOS Space Astrometry Mission}

\bibitem[\protect\citeauthoryear{{Farihi} et~al.}{{Farihi}
  et~al.}{2013}]{Farihi+Bond+Dufour+etal_2013}
{Farihi}, J., {Bond}, H.~E., {Dufour}, P., et~al. 2013, \mnras, 430, 652

\bibitem[\protect\citeauthoryear{{Fey} et~al.}{{Fey}
  et~al.}{2015}]{Fey+Gordon+Jacobs+etal_2015}
{Fey}, A.~L., {Gordon}, D., {Jacobs}, C.~S., et~al. 2015, \aj, 150, 58

\bibitem[\protect\citeauthoryear{Filippazzo et~al.}{Filippazzo
  et~al.}{2015}]{2015ApJ...810..158F}
Filippazzo, J.~C., {Rice}, E.~L., {Faherty}, J., et~al. 2015, \apj, 810, 158

\bibitem[\protect\citeauthoryear{{Fontaine}, {Brassard}, \&
  {Bergeron}}{{Fontaine} et~al.}{2001}]{Fontaine+Brassard+Bergeron_2001}
{Fontaine}, G., {Brassard}, P.,  \& {Bergeron}, P. 2001, \pasp, 113, 409

\bibitem[\protect\citeauthoryear{{Foreman-Mackey} et~al.}{{Foreman-Mackey}
  et~al.}{2013}]{Foreman-Mackey+Hogg+Lang+etal_2013}
{Foreman-Mackey}, D., {Hogg}, D.~W., {Lang}, D.,  \& {Goodman}, J. 2013, \pasp,
  125, 306

\bibitem[\protect\citeauthoryear{{Fortney} et~al.}{{Fortney}
  et~al.}{2008}]{Fortney+Marley+Saumon+etal_2008}
{Fortney}, J.~J., {Marley}, M.~S., {Saumon}, D.,  \& {Lodders}, K. 2008, \apj,
  683, 1104

\bibitem[\protect\citeauthoryear{{Fuhrmann}}{{Fuhrmann}}{2004}]{Fuhrmann_2004}
{Fuhrmann}, K. 2004, Astronomische Nachrichten, 325, 3

\bibitem[\protect\citeauthoryear{{Fuhrmann} et~al.}{{Fuhrmann}
  et~al.}{2014}]{Fuhrmann+Chini+Buda+etal_2014}
{Fuhrmann}, K., {Chini}, R., {Buda}, L.-S.,  \& {Pozo Nu{\~n}ez}, F. 2014,
  \apj, 785, 68

\bibitem[\protect\citeauthoryear{{Fulton} et~al.}{{Fulton}
  et~al.}{2015}]{Fulton+Weiss+Sinukoff+etal_2015}
{Fulton}, B.~J., {Weiss}, L.~M., {Sinukoff}, E., et~al. 2015, \apj, 805, 175

\bibitem[\protect\citeauthoryear{{Gaia Collaboration} et~al.}{{Gaia
  Collaboration} et~al.}{2018}]{Gaia_General_2018}
{Gaia Collaboration}, {Brown}, A.~G.~A., {Vallenari}, A., et~al. 2018, \aap,
  616, A1

\bibitem[\protect\citeauthoryear{{Gaia Collaboration} et~al.}{{Gaia
  Collaboration} et~al.}{2016}]{Gaia_General_2016}
{Gaia Collaboration}, {Prusti}, T., {de Bruijne}, J.~H.~J., et~al. 2016, \aap,
  595, A1

\bibitem[\protect\citeauthoryear{{Graves} et~al.}{{Graves}
  et~al.}{1998}]{Graves+Northcott+Roddier+etal_1998}
{Graves}, J.~E., {Northcott}, M.~J., {Roddier}, F.~J., {Roddier}, C.~A.,  \&
  {Close}, L.~M. 1998, in \procspie, Vol. 3353, Adaptive Optical System
  Technologies, ed. D.~{Bonaccini} \& R.~K. {Tyson}, 34

\bibitem[\protect\citeauthoryear{{Gray} et~al.}{{Gray}
  et~al.}{2006}]{Gray+Corbally+Garrison+etal_2006}
{Gray}, R.~O., {Corbally}, C.~J., {Garrison}, R.~F., et~al. 2006, \aj, 132, 161

\bibitem[\protect\citeauthoryear{{Gray} et~al.}{{Gray}
  et~al.}{2003}]{Gray+Corbally+Garrison+etal_2003}
{Gray}, R.~O., {Corbally}, C.~J., {Garrison}, R.~F., {McFadden}, M.~T.,  \&
  {Robinson}, P.~E. 2003, \aj, 126, 2048

\bibitem[\protect\citeauthoryear{{Han}, {Black}, \& {Gatewood}}{{Han}
  et~al.}{2001}]{Han+Black+Gatewood_2001}
{Han}, I., {Black}, D.~C.,  \& {Gatewood}, G. 2001, \apjl, 548, L57

\bibitem[\protect\citeauthoryear{{Harris} et~al.}{{Harris}
  et~al.}{2013}]{2013ApJ...779...21H}
{Harris}, H.~C., {Dahn}, C.~C., {Dupuy}, T.~J., et~al. 2013, \apj, 779, 21

\bibitem[\protect\citeauthoryear{{Hillenbrand} \& {White}}{{Hillenbrand} \&
  {White}}{2004}]{Hillenbrand+White_2004}
{Hillenbrand}, L.~A.,  \& {White}, R.~J. 2004, \apj, 604, 741

\bibitem[\protect\citeauthoryear{{H{\o}g} et~al.}{{H{\o}g}
  et~al.}{2000}]{Hog+Fabricius+Makarov+etal_2000}
{H{\o}g}, E., {Fabricius}, C., {Makarov}, V.~V., et~al. 2000, \aap, 355, L27

\bibitem[\protect\citeauthoryear{{Houk} \& {Smith-Moore}}{{Houk} \&
  {Smith-Moore}}{1988}]{Houk+Smith-Moore_1988}
{Houk}, N.,  \& {Smith-Moore}, M. 1988, {Michigan Catalogue of Two-dimensional
  Spectral Types for the HD Stars. Volume 4, Declinations -26$^\circ$.0 to
  -12$^\circ$.0.}

\bibitem[\protect\citeauthoryear{{Howard} et~al.}{{Howard}
  et~al.}{2010}]{Howard+Johnson+Marcy+etal_2010}
{Howard}, A.~W., {Johnson}, J.~A., {Marcy}, G.~W., et~al. 2010, \apj, 721, 1467

\bibitem[\protect\citeauthoryear{{Janson} et~al.}{{Janson}
  et~al.}{2011}]{Janson+Carson+Thalmann+etal_2011}
{Janson}, M., {Carson}, J., {Thalmann}, C., et~al. 2011, \apj, 728, 85

\bibitem[\protect\citeauthoryear{King et~al.}{King
  et~al.}{2010}]{2010A&A...510A..99K}
King, R.~R., {McCaughrean}, M.~J., {Homeier}, D., et~al. 2010, \aap, 510, A99

\bibitem[\protect\citeauthoryear{{Konopacky} et~al.}{{Konopacky}
  et~al.}{2010}]{2010ApJ...711.1087K}
{Konopacky}, Q.~M., {Ghez}, A.~M., {Barman}, T.~S., et~al. 2010, \apj, 711,
  1087

\bibitem[\protect\citeauthoryear{{Lafreni{\`e}re} et~al.}{{Lafreni{\`e}re}
  et~al.}{2007}]{Lafreniere+Marois+Doyon+etal_2007}
{Lafreni{\`e}re}, D., {Marois}, C., {Doyon}, R., {Nadeau}, D.,  \& {Artigau},
  {\'E}. 2007, \apj, 660, 770

\bibitem[\protect\citeauthoryear{{Lagrange} et~al.}{{Lagrange}
  et~al.}{2006}]{Lagrange+Beust+Udry+etal_2006}
{Lagrange}, A.-M., {Beust}, H., {Udry}, S., {Chauvin}, G.,  \& {Mayor}, M.
  2006, \aap, 459, 955

\bibitem[\protect\citeauthoryear{{Lenzen} et~al.}{{Lenzen}
  et~al.}{2003}]{Lenzen+Hartung+Brandner+etal_2003}
{Lenzen}, R., {Hartung}, M., {Brandner}, W., et~al. 2003, in \procspie, Vol.
  4841, Instrument Design and Performance for Optical/Infrared Ground-based
  Telescopes, ed. M.~{Iye} \& A.~F.~M. {Moorwood}, 944

\bibitem[\protect\citeauthoryear{{Lindegren} et~al.}{{Lindegren}
  et~al.}{2018}]{Gaia_Astrometry_2018}
{Lindegren}, L., {Hernandez}, J., {Bombrun}, A., et~al. 2018, arxiv, 1804.09366

\bibitem[\protect\citeauthoryear{{Liu} et~al.}{{Liu}
  et~al.}{2002}]{Liu+Fischer+Graham+etal_2002}
{Liu}, M.~C., {Fischer}, D.~A., {Graham}, J.~R., et~al. 2002, \apj, 571, 519

\bibitem[\protect\citeauthoryear{{Ma} et~al.}{{Ma}
  et~al.}{1998}]{Ma+Arias+Eubanks+etal_1998}
{Ma}, C., {Arias}, E.~F., {Eubanks}, T.~M., et~al. 1998, \aj, 116, 516

\bibitem[\protect\citeauthoryear{{Mamajek} \& {Hillenbrand}}{{Mamajek} \&
  {Hillenbrand}}{2008}]{Mamajek+Hillenbrand_2008}
{Mamajek}, E.~E.,  \& {Hillenbrand}, L.~A. 2008, \apj, 687, 1264

\bibitem[\protect\citeauthoryear{{Marleau} \& {Cumming}}{{Marleau} \&
  {Cumming}}{2014}]{Marleau+Cumming_2014}
{Marleau}, G.-D.,  \& {Cumming}, A. 2014, \mnras, 437, 1378

\bibitem[\protect\citeauthoryear{{Marley} et~al.}{{Marley}
  et~al.}{2007}]{Marley+Fortney+Hubickyj_2007}
{Marley}, M.~S., {Fortney}, J.~J., {Hubickyj}, O., {Bodenheimer}, P.,  \&
  {Lissauer}, J.~J. 2007, \apj, 655, 541

\bibitem[\protect\citeauthoryear{{Marois} et~al.}{{Marois}
  et~al.}{2006}]{Marois+Lafreniere+Doyon+etal_2006}
{Marois}, C., {Lafreni{\`e}re}, D., {Doyon}, R., {Macintosh}, B.,  \& {Nadeau},
  D. 2006, \apj, 641, 556

\bibitem[\protect\citeauthoryear{{Mestel}}{{Mestel}}{1968}]{Mestel_1968}
{Mestel}, L. 1968, \mnras, 138, 359

\bibitem[\protect\citeauthoryear{{Mishenina} et~al.}{{Mishenina}
  et~al.}{2004}]{Mishenina+Soubiran+Kovtyukh+etal_2004}
{Mishenina}, T.~V., {Soubiran}, C., {Kovtyukh}, V.~V.,  \& {Korotin}, S.~A.
  2004, \aap, 418, 551

\bibitem[\protect\citeauthoryear{{Montet} et~al.}{{Montet}
  et~al.}{2015}]{Montet+Bowler+Shkolnik+etal_2015}
{Montet}, B.~T., {Bowler}, B.~P., {Shkolnik}, E.~L., et~al. 2015, \apjl, 813,
  L11

\bibitem[\protect\citeauthoryear{{Mugrauer} \& {Neuh{\"a}user}}{{Mugrauer} \&
  {Neuh{\"a}user}}{2005}]{Mugrauer+Neuhauser_2005}
{Mugrauer}, M.,  \& {Neuh{\"a}user}, R. 2005, \mnras, 361, L15

\bibitem[\protect\citeauthoryear{{Noyes} et~al.}{{Noyes}
  et~al.}{1984}]{Noyes+Hartmann+Baliunas+etal_1984}
{Noyes}, R.~W., {Hartmann}, L.~W., {Baliunas}, S.~L., {Duncan}, D.~K.,  \&
  {Vaughan}, A.~H. 1984, \apj, 279, 763

\bibitem[\protect\citeauthoryear{{Pace}}{{Pace}}{2013}]{Pace_2013}
{Pace}, G. 2013, \aap, 551, L8

\bibitem[\protect\citeauthoryear{{Peretti} et~al.}{{Peretti}
  et~al.}{2018}]{Peretti+Segransan+Lavie+etal_2018}
{Peretti}, S., {S{\'e}gransan}, D., {Lavie}, B., et~al. 2018, ArXiv e-prints,
  1805.05645

\bibitem[\protect\citeauthoryear{{Plewa} et~al.}{{Plewa}
  et~al.}{2015}]{Plewa+Gillessen+Eisenhauer_2015}
{Plewa}, P.~M., {Gillessen}, S., {Eisenhauer}, F., et~al. 2015, \mnras, 453,
  3234

\bibitem[\protect\citeauthoryear{{Pourbaix} \& {Arenou}}{{Pourbaix} \&
  {Arenou}}{2001}]{Pourbaix+Arenou_2001}
{Pourbaix}, D.,  \& {Arenou}, F. 2001, \aap, 372, 935

\bibitem[\protect\citeauthoryear{{Queloz} et~al.}{{Queloz}
  et~al.}{2000}]{Queloz+Mayor+Weber+etal_2000}
{Queloz}, D., {Mayor}, M., {Weber}, L., et~al. 2000, \aap, 354, 99

\bibitem[\protect\citeauthoryear{{Ram{\'\i}rez}, {Allende Prieto}, \&
  {Lambert}}{{Ram{\'\i}rez} et~al.}{2007}]{Ramirez+Allende_Prieto+Lambert_2007}
{Ram{\'\i}rez}, I., {Allende Prieto}, C.,  \& {Lambert}, D.~L. 2007, \aap, 465,
  271

\bibitem[\protect\citeauthoryear{{Ram{\'\i}rez}, {Allende Prieto}, \&
  {Lambert}}{{Ram{\'\i}rez} et~al.}{2013}]{Ramirez+Allende_Prieto+Lambert_2013}
{Ram{\'\i}rez}, I., {Allende Prieto}, C.,  \& {Lambert}, D.~L. 2013, \apj, 764,
  78

\bibitem[\protect\citeauthoryear{{Rocha-Pinto}, {Castilho}, \&
  {Maciel}}{{Rocha-Pinto} et~al.}{2002}]{Rocha-Pinto+Castilho+Maciel_2002}
{Rocha-Pinto}, H.~J., {Castilho}, B.~V.,  \& {Maciel}, W.~J. 2002, \aap, 384,
  912

\bibitem[\protect\citeauthoryear{{Rousset} et~al.}{{Rousset}
  et~al.}{2003}]{Rousset+Lacombe+Puget+etal_2003}
{Rousset}, G., {Lacombe}, F., {Puget}, P., et~al. 2003, in \procspie, Vol.
  4839, Adaptive Optical System Technologies II, ed. P.~L. {Wizinowich} \&
  D.~{Bonaccini}, 140

\bibitem[\protect\citeauthoryear{{Sahlmann} et~al.}{{Sahlmann}
  et~al.}{2011}]{Sahlmann+Segransan+Queloz_2011}
{Sahlmann}, J., {S{\'e}gransan}, D., {Queloz}, D., et~al. 2011, \aap, 525, A95

\bibitem[\protect\citeauthoryear{{Salpeter}}{{Salpeter}}{1955}]{Salpeter_1955}
{Salpeter}, E.~E. 1955, \apj, 121, 161

\bibitem[\protect\citeauthoryear{{Saumon} \& {Marley}}{{Saumon} \&
  {Marley}}{2008}]{Saumon+Marley_2008}
{Saumon}, D.,  \& {Marley}, M.~S. 2008, \apj, 689, 1327

\bibitem[\protect\citeauthoryear{Saumon \& {Marley}}{Saumon \&
  {Marley}}{2008}]{2008ApJ...689.1327S}
Saumon, D.,  \& {Marley}, M.~S. 2008, \apj, 689, 1327

\bibitem[\protect\citeauthoryear{{Serabyn} et~al.}{{Serabyn}
  et~al.}{2009}]{Serabyn+Mawet+Bloemhof+etal_2009}
{Serabyn}, E., {Mawet}, D., {Bloemhof}, E., et~al. 2009, \apj, 696, 40

\bibitem[\protect\citeauthoryear{{Service} et~al.}{{Service}
  et~al.}{2016}]{Service+Lu+Campbell+etal_2016}
{Service}, M., {Lu}, J.~R., {Campbell}, R., et~al. 2016, \pasp, 128, 095004

\bibitem[\protect\citeauthoryear{{Snellen} \& {Brown}}{{Snellen} \&
  {Brown}}{2018}]{Snellen+Brown_2018}
{Snellen}, I.~A.~G.,  \& {Brown}, A.~G.~A. 2018, Nature Astronomy, 1808.06257

\bibitem[\protect\citeauthoryear{{Soderblom}}{{Soderblom}}{2010}]{Soderblom_2010}
{Soderblom}, D.~R. 2010, \araa, 48, 581

\bibitem[\protect\citeauthoryear{{Soubiran} et~al.}{{Soubiran}
  et~al.}{2016}]{Soubiran+Campion+Brouillet+etal_2016}
{Soubiran}, C., {Le Campion}, J.-F., {Brouillet}, N.,  \& {Chemin}, L. 2016,
  \aap, 591, A118

\bibitem[\protect\citeauthoryear{{Sozzetti} \& {Desidera}}{{Sozzetti} \&
  {Desidera}}{2010}]{Sozzetti+Desidera_2010}
{Sozzetti}, A.,  \& {Desidera}, S. 2010, \aap, 509, A103

\bibitem[\protect\citeauthoryear{{Spada} et~al.}{{Spada}
  et~al.}{2013}]{Spada+Demarque+Kim+etal_2013}
{Spada}, F., {Demarque}, P., {Kim}, Y.-C.,  \& {Sills}, A. 2013, \apj, 776, 87

\bibitem[\protect\citeauthoryear{{Spiegel} \& {Burrows}}{{Spiegel} \&
  {Burrows}}{2012}]{Spiegel+Burrows_2012}
{Spiegel}, D.~S.,  \& {Burrows}, A. 2012, \apj, 745, 174

\bibitem[\protect\citeauthoryear{{Stassun}, {Mathieu}, \& {Valenti}}{{Stassun}
  et~al.}{2006}]{2006Natur.440..311S}
{Stassun}, K.~G., {Mathieu}, R.~D.,  \& {Valenti}, J.~A. 2006, \nat, 440, 311

\bibitem[\protect\citeauthoryear{{Takeda} et~al.}{{Takeda}
  et~al.}{2007}]{Takeda+Kawanomoto+Honda+etal_2007}
{Takeda}, Y., {Kawanomoto}, S., {Honda}, S., {Ando}, H.,  \& {Sakurai}, T.
  2007, \aap, 468, 663

\bibitem[\protect\citeauthoryear{{Thalmann} et~al.}{{Thalmann}
  et~al.}{2009}]{Thalmann+Carson+Janson+etal_2009}
{Thalmann}, C., {Carson}, J., {Janson}, M., et~al. 2009, \apjl, 707, L123

\bibitem[\protect\citeauthoryear{{Troy} et~al.}{{Troy}
  et~al.}{2000}]{Troy+Dekany+Brack+etal_2000}
{Troy}, M., {Dekany}, R.~G., {Brack}, G., et~al. 2000, in \procspie, Vol. 4007,
  Adaptive Optical Systems Technology, ed. P.~L. {Wizinowich}, 31

\bibitem[\protect\citeauthoryear{{Tull} et~al.}{{Tull}
  et~al.}{1995}]{Tull+MacQueen+Sneden+etal_1995}
{Tull}, R.~G., {MacQueen}, P.~J., {Sneden}, C.,  \& {Lambert}, D.~L. 1995,
  \pasp, 107, 251

\bibitem[\protect\citeauthoryear{{Valenti} \& {Fischer}}{{Valenti} \&
  {Fischer}}{2005}]{Valenti+Fischer_2005}
{Valenti}, J.~A.,  \& {Fischer}, D.~A. 2005, The Astrophysical Journal
  Supplement Series, 159, 141

\bibitem[\protect\citeauthoryear{{van Leeuwen}}{{van
  Leeuwen}}{2007}]{vanLeeuwen_2007}
{van Leeuwen}, F. 2007, \aap, 474, 653

\bibitem[\protect\citeauthoryear{{van Saders} et~al.}{{van Saders}
  et~al.}{2016}]{vanSaders+Ceillier+Metcalfe+etal_2016}
{van Saders}, J.~L., {Ceillier}, T., {Metcalfe}, T.~S., et~al. 2016, \nat, 529,
  181

\bibitem[\protect\citeauthoryear{{Vigan} et~al.}{{Vigan}
  et~al.}{2016}]{Vigan+Bonnefoy+Ginski+etal_2016}
{Vigan}, A., {Bonnefoy}, M., {Ginski}, C., et~al. 2016, \aap, 587, A55

\bibitem[\protect\citeauthoryear{{Voges} et~al.}{{Voges}
  et~al.}{1999}]{Voges+Aschenbach+Boller+etal_1999}
{Voges}, W., {Aschenbach}, B., {Boller}, T., et~al. 1999, \aap, 349, 389

\bibitem[\protect\citeauthoryear{{Vogt} et~al.}{{Vogt}
  et~al.}{1994}]{Vogt+Allen+Bigelow+etal_1994}
{Vogt}, S.~S., {Allen}, S.~L., {Bigelow}, B.~C., et~al. 1994, in \procspie,
  Vol. 2198, Instrumentation in Astronomy VIII, ed. D.~L. {Crawford} \& E.~R.
  {Craine}, 362

\bibitem[\protect\citeauthoryear{{Vogt} et~al.}{{Vogt}
  et~al.}{2014}]{Vogt+Radovan+Kibrick+etal_2014}
{Vogt}, S.~S., {Radovan}, M., {Kibrick}, R., et~al. 2014, \pasp, 126, 359

\bibitem[\protect\citeauthoryear{{Wizinowich} et~al.}{{Wizinowich}
  et~al.}{2000}]{Wizinowich+Acton+Shelton+etal_2000}
{Wizinowich}, P., {Acton}, D.~S., {Shelton}, C., et~al. 2000, \pasp, 112, 315

\bibitem[\protect\citeauthoryear{{Wright} et~al.}{{Wright}
  et~al.}{2011}]{Wright+Drake+Mamajek+etal_2011}
{Wright}, N.~J., {Drake}, J.~J., {Mamajek}, E.~E.,  \& {Henry}, G.~W. 2011,
  \apj, 743, 48

\bibitem[\protect\citeauthoryear{{Yelda} et~al.}{{Yelda}
  et~al.}{2010}]{Yelda+Lu+Ghez+etal_2010}
{Yelda}, S., {Lu}, J.~R., {Ghez}, A.~M., et~al. 2010, \apj, 725, 331

\end{thebibliography}

\end{document}